\documentclass{aa}
\usepackage[varg]{txfonts}
\usepackage{amsmath}
\usepackage{amssymb}
\usepackage{graphicx}
\usepackage{booktabs}
\usepackage{marvosym}
\usepackage{float}
\usepackage{lipsum}
\usepackage{times}
\usepackage{multirow}
\usepackage{rotating}
\usepackage{longtable}
\usepackage{lscape}
\usepackage{epstopdf}
\usepackage{hyperref}
\usepackage[svgnames]{xcolor}
\hypersetup{colorlinks,citecolor=Blue}

\def\comm#1   {{\tt (COMMENT: #1) }}

\newcommand{\smo}{Smol\v{c}i\'{c} }

\def\nTotCpts{8,035}
\def\nCOSMOSCpts{7,729}
\def\nibandCpts{97}
\def\nIRACCpts{209}
 
\def\nTotCptsMulti{48}
\def\nCOSMOSCptsMulti{48}
\def\nibandCptsMulti{0}
\def\nIRACCptsMulti{0}
 
\def\nCOSMOSCptsSingle{7,681}
\def\nibandCptsSingle{97}
\def\nIRACCptsSingle{209}
\def\nCOSMOSCptsSinglePfalseBig{57}
\def\nibandCptsSinglePfalseBig{39}

\def\nherscheltot{4,879}
\def\nherschelCOSMOS{4,836}
\def\nherscheliband{43}

\def\nXrayCpetnaest{906}
\def\nXrayiband{5}
\def\nXrayIRAC{16}
\def\nXrayCptTot{927}

\def\nz{7,778}
\def\nzspec{2,740}
\def\nzphot{5,123}
\def\nznone{124}

\def\nTotCptsXAGN{859}
\def\nTotCptsSEDAGN{1,178}
\def\nTotCptsMIRAGN{466}
\def\nTotCptsMIRAGNnoX{237}
\def\nTotCptsMIRAGNandX{229}
 
\def\nTotCptsREXXAGN{277}
\def\nTotCptsREXSEDAGN{337}
\def\nTotCptsREXMIRAGN{120}
\def\nTotCptsREXHLAGN{486}
\def\nTotCptsREXSFG{855}
\def\nTotCptsREXQAGN{505}
\def\nTotCptsCOSMOSibandAeff{7,826}
\def\nTotCptsQREXAGNoverlap{505}
\def\nTotCptsSED_XMIRAGNoverlap{651}

\def\nTotCptsHLAGN{1,623}
\def\nTotCptsMLAGN{1,648}
\def\nTotCptsSFGpure{4,555}
 
\def\nTotCptsQMLAGN{793}
\def\nTotCptsREXMLAGN{1,360}
\def\nTotCptsSFG{5,410}
 
\def\nTotCptsREX{1,846}

\titlerunning{Composition of the VLA-COSMOS 3 GHz radio population}
\authorrunning{V.~\smo \ et al.}

\title{The VLA-COSMOS 3 GHz Large Project: \\ Multiwavelength counterparts and the composition of the faint radio population}
\author{ V.~\smo\inst{1}\thanks{\emph{vs@phy.hr}} 
	\and I.~Delvecchio\inst{1}\thanks{\emph{ivand@phy.hr}}
	\and G.~Zamorani \inst{2}
	\and N. Baran\inst{1}
	\and M.~Novak \inst{1}
	\and J.~Delhaize \inst{1}
	\and E.~Schinnerer \inst{3}
	\and S.~Berta \inst{1}\thanks{\emph{Visiting scientist}}
	\and M.~Bondi \inst{4}
	\and P.~Ciliegi \inst{2}
	\and P.~Capak \inst{5}
	\and F.~Civano \inst{6}
	\and A.~Karim \inst{7}
	\and O.~Le~Fevre \inst{8}
	\and O.~Ilbert \inst{8}
	\and C.~Laigle \inst{9}
	\and S.~Marchesi \inst{10}
	\and H.~J.~McCracken \inst{9}
	\and L.~Tasca \inst{8}
	\and M.~Salvato \inst{11}
	\and E.~Vardoulaki \inst{7}
}
\institute{University of Zagreb, Physics Department, Bijeni\v{c}ka cesta 32, 
	10002 Zagreb, Croatia
	\and INAF - Osservatorio Astronomico di Bologna, Via Ranzani 1, I-40127, Bologna, Italy
	\and Max Planck Institute for Astronomy, K\"onigstuhl 17, D-69117 Heidelberg, Germany
	\and Istituto di Radioastronomia di Bologna - INAF, via P. Gobetti, 101, 40129, Bologna, Italy
	\and Spitzer Science Center, California Institute of Technology, 220-6, Pasadena, CA, USA 91125
	\and Harvard-Smithsonian Center for Astrophysics, 60 Garden Street, Cambridge, MA 02138, USA
	\and Argelander-Institut für Astronomie, University of Bonn, Auf dem H\"ugel 71, 53121, Bonn, Germany 
	\and Aix Marseille Universit\'{e}, Laboratoire d'Astrophysique de Marseille, 38 rue Frederic Joliot-Curie, 13388 Marseille, France 
	\and Institut d' Astrophysique de Paris, Sorbonne Universit\'es, UPMC, Univ Paris 06 et CNRS, UMR 7095, 98 bis bd Arago, 75014 Paris, France 
	\and Department of Physics \& Astronomy, Clemson University, Clemson, SC 29634, USA
	\and Max Planck Institute for Extraterrestrial Physics, Giessenbachstr. 1, D-85748 Garching, Germany 
	}

\begin{document}
\abstract{We study the composition of the faint radio population selected from the VLA-COSMOS 3~GHz Large Project, a radio continuum survey performed at 10~cm wavelength. The survey covers a 2.6 square degree area with a mean $rms$ of $\sim2.3~\mu$Jy/beam, cataloging 10,830 sources above $5\sigma$, and enclosing the full 2 square degree COSMOS field. By combining these radio data with  optical, near-infrared (UltraVISTA), and  mid-infrared (Spitzer/IRAC) data, as well as X-ray data (\textit{Chandra}), we find counterparts to  radio sources for 
$\sim$93\% of the total radio sample (in the 
unmasked areas of the COSMOS field, i.e., those not affected by saturated or bright sources in the optical to NIR bands), reaching out to $z\lesssim6$. 
We further classify the sources as star forming galaxies or AGN based on various criteria, such as X-ray luminosity, observed MIR color, UV-FIR spectral-energy distribution, rest-frame NUV-optical color corrected for dust extinction, and radio-excess relative to that expected from the hosts'  star-formation rate. 
We separate the AGN
into sub-samples dominated by low-to-moderate and moderate-to-high radiative luminosity AGN, candidates for high-redshift analogues to local low- and high-excitation emission line AGN, respectively. We study the fractional contributions of these sub-populations down to radio flux levels of $\sim$11~$\mu$Jy at 3~GHz (or  $\sim$20~$\mu$Jy at 1.4~GHz assuming a spectral index of -0.7). We find that the dominant fraction at 1.4~GHz flux densities above $\sim$200~$\mu$Jy is constituted of   low-to-moderate radiative luminosity AGN. Below densities of $\sim$100~$\mu$Jy the fraction of star-forming galaxies increases to $\sim60$\%, followed by the moderate-to-high radiative luminosity AGN ($\sim20$\%), and low-to-moderate radiative luminosity AGN ($\sim20$\%). Based on this observational evidence, we extrapolate the fractions down to sensitivities of the Square Kilometer Array (SKA). Our estimates suggest that at the faint flux limits to be reached by the (Wide, Deep, and UltraDeep) SKA1 surveys, a selection based only 
on radio flux limits can provide a simple tool to efficiently identify samples  highly ($>75\%$) dominated by star-forming galaxies.
}

\maketitle 	
\makeatother

\section{Introduction\label{sec:intro}}

Understanding how galaxies form in the early universe and their subsequent evolution through cosmic time
is a major goal of modern astrophysics. Panchromatic look-back sky surveys significantly advanced the field
in the past decade, demonstrating that a multi-wavelength X-ray to radio approach is key for such studies. With the major upgrade of existing and the onset of new radio facilities, such as the Karl G.\ Jansky Very Large Array (VLA), Australia Telescope Compact Array (ATCA) and Atacama Large Millimeter/submillimeter
Array (ALMA), delivering now an order of magnitude increase in sensitivity, radio-astronomy has entered its 'golden age', and has moved towards the forefront of multi-wavelength research (see \citealt{padovani16} for a recent review). 
For an overall assessment of the evolution of radio sources a combination of several areal and sensitivity (as well as spectroscopic) coverages (a 'wedding-cake' approach; see e.g.~Fig.~1 in \citealt{smo17}, and references therein) is optimum, allowing to construct statistically significant samples, even large enough  when split simultaneously by various parameters such as redshift, galaxy type, stellar mass, star formation rate etc. 
In this context, the  Cosmic Evolution Survey (COSMOS;  \citealt{scoville07})\footnote{\url{http://cosmos.astro.caltech.edu/}} contains one of the richest multi-wavelength (X-ray to radio) data-sets over a 2 square degree area on the sky, ideal for the exploration of galaxy evolution. 

The COSMOS field has been observed with the VLA at 1.4~GHz (rms~$\sim10-15~\mu$Jy/beam; \citealt{schinnerer04,schinnerer07,schinnerer10}), and at 3~GHz with the upgraded system to substantially better sensitivity (rms~$\sim2.3~\mu$Jy/beam), yielding about four times more radio sources compared to the 1.4~GHz data \citep{smo17}. This makes the VLA-COSMOS 3~GHz Large Project to-date the deepest radio continuum survey over a relatively large field. Combined with the rich COSMOS multi-wavelength data it, thus, yields a unique data-set to test the composition (i.e., fractions of various galaxy types) of the  faintest  radio source populations that can currently be probed, and it allows us to use these findings to make predictions for the populations to be detected by future surveys, such as those planned with the Square Kilometre Array (SKA), and its precursors \citep{prandoni15}.

Past research has shown that radio-detected galaxy populations are dominated by two populations: star-forming and AGN galaxies (e.g., \citealt{miley80}, \citealt{condon92}). The observed radio emission is dominated in both galaxy types by synchrotron radiation (at GHz wavelengths), arising either from supernovae remnants (thus tracing star-formation in galaxies) or near supermassive black holes, by ejection of relativistic jets of plasma \citep{burbrige56, sadler89, condon92}. 
Furthermore, 
AGN detected in the radio band  have been found to separate into two, physically distinct populations. These can be classified via the host galaxies' optical spectroscopic emission line properties as high- and low-excitation radio AGN (e.g., \citealt{hine79, hardcastle06, hardcastle07, allen06, smolcic09, heckman14, smolcic15}). The hosts of  local high-excitation radio AGN are shown to have green optical colours (e.g., \citealt{smolcic09}). These types of objects can also be identified as AGN in X-ray and/or mid-infrared (MIR) wavelengths (e.g., \citealt{hardcastle13}), and they fit into the classical unified model of AGN, where the central supermassive black hole is thought to be accreting radiatively efficiently from a geometrically thin, but optically thick accretion disk, encompassed by a dusty torus (\citealt{shakura73}). Low-excitation radio AGN are  found to be hosted by massive, red, quiescent galaxies, and identifiable only in the radio band (see \citealt{hine79, laing94, smolcic09}). 
They  
do not display properties of a classical, unified-model AGN as they are thought to be accreting radiatively inefficiently, possibly through advection dominated accretion flows, associated with a puffed-up, geometrically thick, but optically thin accretion structure (e.g., \citealt{heckman14}). Since their radio emission far exceeds the level expected from the star forming activity in the host galaxy, they are also often referred to as radio-loud or radio-excess objects (e.g., \citealt{condon92, bonzini13, padovani15, delvecchio17}).

Identifying the above mentioned galaxy types poses a challenge for radio continuum surveys as i) multi-wavelength (X-ray to  FIR) photometric and/or optical spectroscopic data is required for the identification, and ii) the multi-wavelength signatures of star-formation and AGN activity may not be in one-to-one correlation with those in the radio band, reflecting a composite nature of galaxies. Largely for these reasons, the cause of the upturn in the observed (Euclidean-normalized) radio counts at $\sim1$~mJy  was long debated in the literature (e.g., \citealt{georgakakis99,gruppioni99,jarvis04,cowie04,huynh05,afonso05,simpson06}). 
Although the popular interpretation for years was that star forming galaxies dominated the submilliJansky radio population, \citet{gruppioni99}, studying a sample of 68 faint radio sources ($S > 0.2$~mJy) in the Marano Field concluded that 
star forming galaxies do not constitute the main population, as even at sub-mJy level the majority of their radio sources were identified with early-type galaxies.
Using a robust classification of $\sim$65\% of the 1.4~GHz VLA-COSMOS radio sources \citet{smolcic09b} found a fairly constant fraction ($\sim$30-40\%) of star forming galaxies  in the flux density range between $\sim$50$~\mu$Jy and 0.7~mJy. In a more recent study, \citet{padovani15} used a sample of 680 radio sources detected at 1.4~GHz in the Extended \textit{Chandra} Deep Field South (E-CDFS). They show that AGN and star forming galaxies are approximately equally numerous between 32~$\mu$Jy and 1~mJy, with star forming galaxies becoming the dominant population only below $\sim100~\mu$Jy at 1.4~GHz. This is qualitatively consistent with the Square Kilometre Array Design Study (SKADS) semi-empirical simulation of the radio source counts predicting that star forming galaxies will only start to dominate the counts at 1.4~GHz fluxes below $100~\mu$Jy  \citep{wilman08}. 

Given the onset of the upcoming, revolutionary radio facilities such as the Square Kilometre Array (SKA)\footnote{\url{http://www.skatelescope.org}} \citep{norris13, prandoni15} and the  Next Generation Very Large Array (ngVLA)\footnote{\url{https://science.nrao.edu/futures/ngvla}} \citep{hughes15}, to be operational at the same time as the The Large Synoptic Survey Telescope (LSST)\footnote{\url{https://www.lsst.org/}} and Euclid\footnote{\url{www.euclid-ec.org}}, it becomes even more important i) to study the composition of the radio source population as a function of radio flux density, probing the to-date faintest achievable flux limits over significant areas, and accounting for the possibility of a composite  nature of galaxies (containing both AGN and star-formation activity), and ii) to make predictions applicable to the future radio surveys, facilitating the identification of the various galaxy populations within these surveys.
 The COSMOS project \citep{scoville07} is optimal for such research as it is currently one of the most advanced panchromatic surveys covering a 2 square degree area and sampling galaxies and AGN out to high ($z\lesssim6$) redshift. It incorporates one of the deepest radio datasets ever obtained with the  VLA at 3~GHz 
 detecting 10,830 sources down to $\approx11.5~\mu$Jy ($\approx5\sigma$ at an angular resolution of $0.75\arcsec$;
 \smo et al., accepted). Here we combine these  radio data with the COSMOS X-ray to FIR multi-wavelength datasets to search for counterparts of the radio sources, and analyze the composition of the faint, microJansky radio population, extrapolating this down to the flux limits that will be achieved with the SKA in Phase 1 (SKA1; \citealt{prandoni15}).

In Section \ref{sec:data} we describe the datasets used in the analysis. Section \ref{sec:sample} summarizes the method of association of multiwavelength, optical-MIR counterparts to our radio sources (the details of which are given in Appendix~\ref{sec:method}). In Section \ref{sec:X} we introduce the X-ray counterparts to our radio sources. In Section \ref{sec:redshiftsSample} we analyze the redshifts of the counterparts.  In Section \ref{sec:galpop} we present a multi-wavelength assessment of the galaxy populations present within our radio source sample. In Sec.~\ref{sec:catalog} we present the counterpart, and source population catalog released.
The composition of the microJansky radio source counts in the context of different galaxy populations, and emitting-mechanisms in the radio are discussed in Section \ref{sec:millijansky}. We summarize in Sec.~\ref{sec:summary}. Throughout this paper, magnitudes are given in the AB system, coordinates in the J2000 epoch, and we use the following cosmological parameters, $\Omega_{M}=0.3$, $\Omega_{\Lambda}=0.7$ and $H_{0}=70~\rm km~s^{-1}~Mpc^{-1}$. 
We define the radio spectral index, $\alpha$, via $S_\nu\propto\nu^\alpha$, where $S_\nu$ is the flux density at frequency $\nu$. A \citet{chabrier03} initial mass function (IMF) is used.

\section{Data\label{sec:data}}

In this section we describe the multi-wavelength data used in the analysis.

\subsection{Radio data}

\begin{figure} 
\centering
\includegraphics[bb = 0 35 865 866, width=\columnwidth]{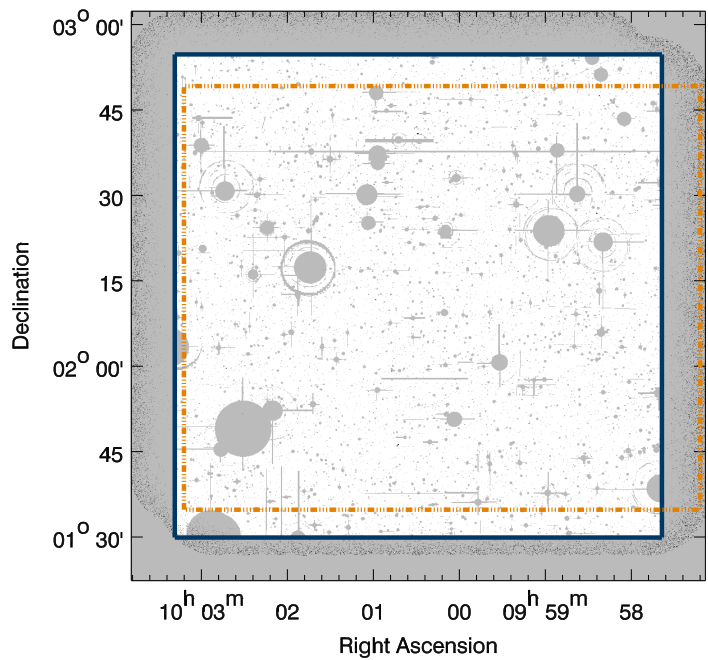}
\caption{
VLA-COSMOS 3~GHz Large Project mosaic with grayed-out regions masked in the COSMOS2015 catalog due to the presence of saturated or bright sources in the optical to NIR bands.
 Also indicated are the areas covered by Subaru  Suprime Cam (blue full line, corresponding to the outer edge of the unmasked area), and UltraVISTA (Ultra Deep Survey with the VISTA telescope; orange dash-dotted line; see \citealt{laigle16} and references therein). The masked regions reduce the effective area to 1.77~deg$^2$, with 8,696 (80\%) out of 10,830 radio sources outside the masked areas. 
  }
\label{fig:counterpartsRADEC}
\end{figure}

The VLA-COSMOS 3~GHz Large Project entails 384~hours of observations with the VLA in S-band (centered at 3~GHz with 2,048~MHz bandwidth) over a total area of 2.6 square degrees  (see \smo et al., accepted, for details). 
 The VLA S-band was chosen for the observations as the combination of a large effective bandwidth ($\sim2$~GHz) and fairly large field of view (half-power beam width $15\arcmin$)\footnote{
 Note that the primary beam half-power beam width is a function
of frequency and varies by a factor of 2 between the lower and upper frequency
edge of the S-band. At 3 GHz frequency it corresponds to $15\arcmin$ allowing to cover a 2 square degree field with $\sim64$ pointings. 
}
allow for a time-efficient coverage of a field like COSMOS ($\sim2$ square degrees) to 
 a high sensitivity, even though the sources detected  are typically fainter at this observing frequency compared to the commonly used 1.4 GHz  observing frequency (L-band)\footnote{
  For a typical spectral index of $\alpha=-0.7$ the flux density at 3~GHz observing frequency (S-band) is a factor of $1.67$ lower than at 1.4~GHz observing frequency (L-band). 
Assuming an effective bandwidth of 1.664 GHz in S-band, and 400-600 MHz in L-band, to reach an $rms$ of 2.3~$\mu$Jy/beam at 3 GHz observing frequency, which is equivalent to an rms of $3.9$ $\mu$Jy/beam at 1.4 GHz observing frequency for $\alpha=-0.7$ requires per pointing a factor of about $1.4 - 1.9$ more observing time in  L-band, compared to  S-band. On the other hand, to reach a uniform $rms$ over a 2 square degree field 23 pointings are required in the L-band, while 64 are required in the S-band. This contributes to a slightly higher overhead time in the S-band due to longer telescope slewing times. We also note that the S-band offers further advantages over the L-band, such as a (factor of $\sim2$) higher angular resolution, as well as the large bandwidth which can simultaneously provide spectral shape information for bright sources in the field. }.   
The observing layout was such that a uniform $rms$, with a median $rms\sim2.3~\mu$Jy/beam (at a resolution of $0.75\arcsec$), was reached over the inner 2 square degrees coincident with the (Subaru Suprime Cam) COSMOS field (see Fig.~\ref{fig:counterpartsRADEC}).  Over the full 2.6 square degrees the survey yielded a total of 10,830 cataloged sources.  
Of these, 67 were found to be multi-component sources, i.e., objects that are composed of two or more detected radio components, completely separated from each other, but belonging to the same source (see \citealt{smo17}; Vardoulaki et al., in prep.). The combination of the multiple components into one reported source was performed by visual inspection using the auxiliary multi-wavelength data available in the COSMOS field (see \citealt{smo17}). 
These sources are mainly radio galaxies with resolved core/jet/lobe structures (see Fig.~7 in \smo et al., accepted), but can also be star-forming galaxies where the radio emission follows the disk morphology. The remaining 10,763 detected sources are referred to as single-component sources.

Prior to the VLA-COSMOS 3~GHz Large Project, the field was observed with the VLA at 1.4~GHz frequency within the VLA-COSMOS 1.4~GHz Pilot, Large and Deep Projects \citep{schinnerer04,schinnerer07,schinnerer10}. Within the 1.4 GHz Large Project a  uniform $rms$ of 10-15~$\mu$Jy/beam was reached over a resolution element of $1.5\arcsec$, across the 2 square degree field. The Deep Project added further observations towards the inner square degree, yielding an $rms$ of 12~$\mu$Jy/beam over a resolution elements of $2.5\arcsec$. A Joint catalog was generated combining the sources detected at $\geq5\sigma$ at $1.5\arcsec$ and/or $2.5\arcsec$ resolution in the Large and/or Deep Projects, and lists a total of 2,864 sources \citep{schinnerer10}. Matching the 3 GHz catalog with the Joint 1.4 GHz catalog using  a radius  of $1.3\arcsec$ (corresponding to approximately half the synthesized beam size at 1.4~GHz) yields 2,530 matches (see \citealt{smo17}). Hence, for 2,530 out of 10,830 sources in the 3 GHz 
catalog a 
spectral index can be calculated based on the two frequencies. For the remainder of the sources a spectral index of -0.7 is assumed, consistent with the average value derived for the full 3 GHz population taking into account also limits of the spectral indices for sources detected at 3~GHz, but not detected at 1.4~GHz (for more details see Sec.~4.2., and Fig.~14 in \smo et al., accepted).
It is widespread in the literature to assume a single spectral index for the radio spectral energy distribution (usually taken to be $\alpha=-0.7$ or $\alpha=-0.8$). The usually observed spread in spectral indices is $\sigma\approx0.35$ (e.g., \smo et al., accepted), and we note that 
 the uncertainty of the spectral index can induce significant errors in the derived radio luminosity for a single object. However, on statistical basis the symmetry of the spread is expected to cancel out the variations yielding a valid average luminosity for the given population (see also \citealt{novak17,delhaize17}, \smo \ et al., submitted, for more specific discussions on this).

\subsection{Near-ultraviolet -- mid-infrared data and photometric redshifts}

To complement our dataset with optical, near-infared (NIR) and mid-infrared (MIR) wavelengths we use 
\textit{i)} the latest photometry and photometric redshift catalogs (COSMOS2015 hereafter) \citep{laigle16},
\textit{ii)} the $i$-band selected catalog (version 1.8, \citealt{capak07}),
\textit{iii)} the \textit{Spitzer} - COSMOS (S-COSMOS) InfraRed Array Camera\footnote{https://www.cfa.harvard.edu/irac/} 3.6~$\mu$m selected catalog (IRAC hereafter) \citep{sanders07}. 

The COSMOS2015 catalog lists optical and NIR photometry in over 30 bands for 1,182,108 sources identified either in the $z^{++}YJHK_S$\footnote{
The  observations of the COSMOS field in $z$-band were performed with Subaru Suprime-Cam \citep{taniguchi07,taniguchi15}. The initial COSMOS z-band ($z^+$) data are superseded by deeper $z$-band observations obtained with upgraded CCDs and a different filter ($z^{++}$). Data in $YJHK_S$ bands refers to that obtained by the Ultra Deep Survey with the VISTA telescope (see also \citealt{laigle16}).
} stacked detection image 
 (within the area encompassed by the Ultra Deep Survey with the VISTA telescope, UltraVISTA, where $YJHK_S$ are taken from UltraVISTA DR2\footnote{http://www.eso.org/sci/observing/phase3/data\_releases/uvista\_dr2.pdf}), or the $z^{++}$ band, outside the UltraVISTA footprint (see Fig.~\ref{fig:counterpartsRADEC}). 
The total area covered by the catalog used is $\sim2.3$ square degrees, which reduces to an effective area of 1.77 square degrees if masked regions are excluded. The cataloged fluxes were extracted as described in detail by  \citet{laigle16}, using \textit{SExtractor}  on the positions of the sources detected in the stacked images 
separately for $2\arcsec$, and $3\arcsec$ diameter apertures, and corrections to total magnitudes are given for each source, as well as corrections for galactic dust extinction from \citet{schlegel98}. 
The point spread functions (PSFs) were homogenized in the optical/NIR maps to a resolution of $\theta\sim$0\farcs8 prior to extraction of the COSMOS2015 catalog.
The $3\sigma$ limits for $2\arcsec$ ($3\arcsec$) diameter apertures are 25.9 (26.4) for $z^{++}$, and 24.8 (25.3), 24.7 (25.2), 24.3 (24.9), 24.0 (24.5) for the $Y$, $J$, $H$, $K_S$ Deep bands, respectively (see Tab.~1 and Figs. 1-3 in \citealt{laigle16} for more details).
The above mentioned procedure was also used to extract the four IRAC band photometry, where particular care was taken to properly deblend the IRAC photometry using IRACLEAN (see \citealt{laigle16} for details). We note that the IRAC channel 1 and 2 data were drawn from the SPLASH survey (PI: P.~Capak), while channel 3 and 4 were taken from the S-COSMOS survey (see also below). The 3$\sigma$ depth for $3\arcsec$ apertures in IRAC channel 1, 2, 3 and 4 is 25.5, 25.5, 23.0 and 22.9 ($m_{AB}$), respectively.  In addition, this dataset also contains 24$~\mu$m photometry from \citep{LeFloch09} from the Multi-Band Imaging Photometer for Spitzer (MIPS) down to a magnitude limit of 19.45.
\citet{laigle16} have computed photometric redshifts for the sources listed in the COSMOS2015 catalog using \textsc{LePhare} (see also \citealt{ilbert09} and \citealt{ilbert13}). They report a photometric redshift accuracy\footnote{Normalized median absolute deviation \citep{hoaglin83} defined as 1.48 times the median value of $ | z_\mathrm{phot}-z_\mathrm{spec}  | / (1 + z_\mathrm{spec})$ } within the UltraVISTA-COSMOS area of $\sigma_{\Delta z/(1+z_s)}~\sim~0.01$ for $22\textless~i^+\textless23$, and $\sigma_{\Delta z/(1+z_s)}~\textless~0.034$ for $24\textless~i^+\textless25$ based on comparison with a spectroscopic redshift sample in the COSMOS field (cf. Table 4 in \citealt{laigle16}). 

We also use the  $i$-band selected COSMOS photometric catalog version 1.8 (initially described in \citealt{capak07}). This catalog contains 2,017,800 $i$-band selected sources, and 15 photometric bands ranging from 0.3~$\mu$m to 2.4~$\mu$m, measured in $3\arcsec$ diameter apertures. The total area covered is $\sim3.1$ square degrees. The 5$\sigma$ depth in a $3\arcsec$ aperture is $i^+_{AB}$~<~26.2.  The catalog has been cross-matched with MIPS 24$~\mu$m data from \citep{LeFloch09}, and also contains photometric redshifts derived by \citet{ilbert09} using \textsc{LePhare} with an accuracy similar to that reached for the COSMOS2015 sources. The PSF-homogenized images have a resolution of $\theta\sim$0\farcs8, as in the COSMOS2015 catalog. In comparison to the  $i$-band selected COSMOS catalog \citep{capak07} within the 1.5 square degree area of the COSMOS field covered by UltraVISTA, \citet{laigle16} show that 96.5\% (83.9\%) of sources brighter than Subaru $i^+$~=~25.5 (26.1) are detected also in 
the COSMOS2015 catalog.  Thus, given that some of the $i$-band selected sources are not recovered in the COSMOS2015 catalog, we here also make use of the $i$-band selected COSMOS catalog.

Additionally, we use the \textit{Spitzer} IRAC catalog \citep{sanders07} of 345,512 sources detected ($\gtrsim$1~$\mu$Jy) at 3.6~$\mu$m, over an area of $\sim2.8$ square degrees. The catalog includes photometry in the 4 IRAC channels at 3.6, 4.5, 5.6, and 8.0~$\mu$m, for sources detected in the 3.6~$\mu$m image. Their respective 3$\sigma$ depths are 24.6, 23.8, 21.8, 21.6. 
For each channel, fluxes extracted within four apertures are given. We use the 1\farcs9 aperture photometry given in the catalog. The  3.6~$\mu$m image resolution is $\theta=$1\farcs66.

The combined use of the three catalogs increases the likelihood of finding counterparts to radio sources, especially those  (either very blue or very red) counterparts whose spectral energy distribution makes them undetectable in the stacked  $z^{++}YJHK_S$ map of COSMOS2015 but that are detected in the $i$-band or IRAC. Furthermore, highly obscured and/or high redshift sources might be detectable only in  the 3.6~$\mu$m IRAC  band.

 \subsection{Far-infrared and (sub-)millimetre photometry}
\label{subsec:mmdata}
To derive accurate estimates of the star-formation rate (SFR) of each radio source, we use far-infrared photometry from the \textit{Herschel Space Observatory}, that encompasses the full 2 square degrees of the COSMOS field. \textit{Herschel} imaging is taken from the \textit{Photoconductor Array Camera and Spectrometer} (PACS, 100 and 160$~\mu$m, \citealt{Poglitsch10}) and Spectral and Photometric Imaging Receiver (SPIRE, 250, 350, and 500$~\mu$m, \citealt{Griffin10}), which are part of the \textit{PACS Evolutionary Probe} (PEP, \citealt{Lutz11}) and the \textit{Herschel Multi-tiered Extragalactic Survey} (HerMES, \citealt{Oliver12}) projects. The COSMOS2015 catalog (\citealt{laigle16}) reports the de-blended \textit{Herschel} fluxes extracted by using 24$~\mu$m positions as priors (from \citealt{LeFloch09}), and unambiguously associated to the corresponding optical/NIR counterpart via 24$~\mu$m IDs reported in both catalogs. A similar association technique is applied to assign \textit{Herschel} 
counterparts to i-band selected 
sources (\citealt{capak07}). 

In addition to far-infrared data, we retrieve also the available (sub)-millimetre photometry from at least one of the observational campaigns conducted over the COSMOS field: JCMT/SCUBA-2 at 450 and 850$~\mu$m (\citealt{Casey13}), LABOCA at 870$~\mu$m (F. Navarrete et al. priv. Comm.), Bolocam (PI: J. Aguirre), JCMT/AzTEC (\citealt{Scott08}) and ASTE/AzTEC (\citealt{Aretxaga11}) at 1.1~mm, MAMBO at 1.2~mm (\citealt{Bertoldi07}), and interferometric observations at 1.3~mm with PdBI (\citealt{Smolcic12}; \citealt{Miettinen15}) and ALMA (PI: M. Aravena, M. Aravena et al. in prep.).

\subsection{X-ray data}
\label{subsec:xraydata}
We use the most recent COSMOS X-ray catalog of point sources \citep{marchesi16}, drawn from the \textit{Chandra} COSMOS-Legacy survey \citep{civano16}. The \textit{Chandra} COSMOS-Legacy survey combines the \textit{Chandra} COSMOS \citep{elvis09} data with the new \textit{Chandra} ACIS-I data (2.8~Ms observing time) resulting in a total exposure time of 4.6~Ms over 2.15~deg$^2$ area,  reaching (at best) a limiting depth of 2.2$\times$10$^{-16}$~erg~s$^{-1}$~cm$^{-2}$ at [0.5~-~2]~keV, 1.5$\times$10$^{-15}$~erg~s$^{-1}$~cm$^{-2}$ at [2~-~10]~keV, and 8.9$\times$10$^{-16}$~erg~s$^{-1}$~cm$^{-2}$ at [0.5~-~10]~keV \citep{civano16}. The X-ray observations cover the central 1.5 square degrees of the COSMOS field, with an average effective exposure time of 160~ks, while in the outer regions it reduces down to 80~ks.
The catalog contains 4,016 X-ray point sources in the \textit{Chandra} COSMOS-Legacy survey, out of which 3,877 have optical--MIR counterparts, matched using the likelihood ratio technique \citep{marchesi16}. The counterparts were searched for in three different bands. The $i$-band (760~nm) was taken from Subaru where the data was unsaturated ($i_{AB}\textgreater20$, \citealt{capak07}), and CFHT and SDSS otherwise. 
Matching was also done with the \textit{Spitzer} IRAC catalog \citep{sanders07}, and the SPLASH IRAC magnitudes from \citet{laigle16}, in which the photometry was extracted at positions with a detection in the $z^{++}YHJK_S$ stacked image, as described in more detail in the previous Section. The catalog lists X-ray fluxes, and intrinsic (i.e. unobscured) X-ray luminosities in the full [0.5~-~10]~keV, soft [0.5~-~2]~keV, and hard [2~-~10]~keV X-ray bands.  Later on, X-ray fluxes and luminosities will be given up to 8~keV, by assuming a simple power-law spectrum with photon index $\Gamma$=1.4 (\citealt{marchesi16}). Also, spectroscopic, or photometric redshifts of the X-ray sources are available in the catalog (see \citealt{civano16}).

\subsection{Spectroscopic redshifts}
\label{sec:redshifts}
We use the most up-to-date spectroscopic redshift catalog available to the COSMOS team (see Salvato et al.\ in prep.). It contains 97,102 sources with spectroscopic redshifts. The catalog has been compiled from a number of spectroscopic surveys of the COSMOS field, including SDSS (DR12), zCOSMOS \citep{lilly07,lilly09}, VIMOS Ultra Deep Survey (VUDS) \citep{lefevre15}, MOSDEF \citep{kriek15}, and DEIMOS\footnote{ The DEIMOS data were taken from various independent observational campaigns led by: Nick Scoville, Peter Capak, David Sanders, Guenther Hasinger, Jeyhan Kartaltepe, Mara Salvato, and Caitlin Casey.}.

\section{Optical-MIR counterparts}
\label{sec:sample}

In this section we describe the cross-correlation of our radio sources with optical ($i$-band-selected), NIR ($z^{++}$-, or $z^{++}YJHK_S$-stack-selected, COSMOS2015), and MIR ($3.6~\mu$m-selected, IRAC) sources.  We here report the numbers of optical-MIR counterparts associated to our radio sources within the same unmasked area  of 1.77 deg$^2$ (see Fig.~\ref{fig:counterpartsRADEC}), while in Appendix~\ref{sec:fullcat} we present the full counterpart catalog that includes also sources outside the  1.77 deg$^2$ area. We separately associate the multi-component and single-component radio sources with their optical-MIR counterparts, as  detailed in Secs.~\ref{sec:cptMulti}, and \ref{sec:cptSingle}, respectively. We present an assessment of the optical-MIR counterparts of our radio sources in Sec.~\ref{sec:optMIRcpts_assesment}.

\subsection{
Counterparts of multi-component radio sources 
}
\label{sec:cptMulti}
 Out of the 67 multi-component radio sources identified in our 3~GHz radio catalog, 48 lie within the optical-MIR unmasked area of 1.77 deg$^2$. By visual inspection we find a match for all 48 sources in the COSMOS2015 catalog, accounting for the full sample of multi-component radio sources (see Table \ref{tab:cpts}).

\begin{figure} 
	\centering	\includegraphics[bb = 0 0 432 432,width=0.8\linewidth]{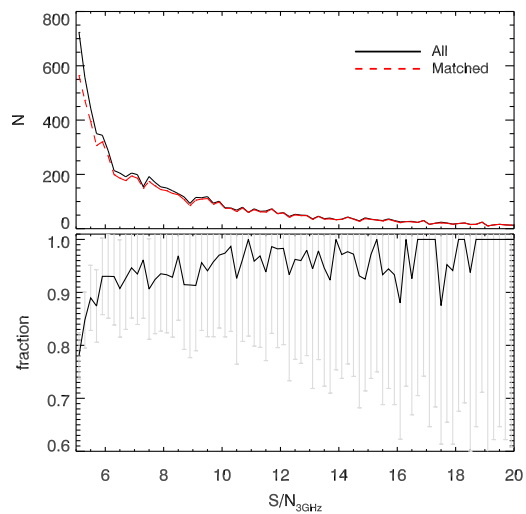}
	\caption{\textit{Top: } Signal-to-noise ratio distribution of all (black line) and matched (red dashed line) VLA 3 GHz single-component sources. 	\textit{Bottom: } The fraction of matched VLA 3 GHz (single-component) sources as a function of signal-to-noise ratio. Poisson errors are also shown.}
	\label{fig:snr}
\end{figure}

\subsection{ Counterparts of single-component radio sources}
\label{sec:cptSingle}

The optical-MIR counterpart assignment to our single-component 
3~GHz radio sources is described in detail in Appendix~\ref{sec:method}, and here we only present a brief overview.
We base the counterpart association on nearest neighbor matching, accounting for a false match probability ($p_\mathrm{false}$) for each match. 
The counterpart association proceeded as follows. 

First, prior to the positional matching of our 3 GHz sources separately to the sources from the COSMOS2015 \citep{laigle16}, $i$-band  \citep{capak07}, and IRAC  \citep{sanders07} catalogs the astrometry in each multi-wavelength catalog was corrected for observed small systematic offsets, relative to the radio positions (see Appendix \ref{sec:methodRadius}). Second, a positional matching of the 3 GHz radio sources was performed with the sources in the unmasked areas of each of the COSMOS2015, $i$-band and IRAC catalogs out to a search radius of $0\farcs8$ (COSMOS2015/i-band), or $1\farcs7$ (IRAC). Third, a false match probability was assigned to every counterpart match based on Monte Carlo simulations. For the simulations we matched sources from mock catalogs that realistically reflect the magnitude, and separation distributions expected for the counterparts of our radio sources (see Appendix~\ref{sec:methodFMP} for details). 
These steps led to three counterpart candidate catalogs, each, respectively, containing the 3~GHz sources matched to sources from the 1) COSMOS2015 (see Appendix~\ref{sec:crossmatching}), 2) $i$-band (see Appendix~\ref{sec:crossmatching_iband}), 3) IRAC catalogs  (see Appendix~\ref{sec:crossmatching_irac}), and with a false match probability computed for each match. 

In the fourth, final step, only one, final counterpart 
of a given 3~GHz source is selected from the 
three catalogs following a hierarchical line of reasoning, setting first, second and third priority to the COSMOS2015, $i$-band and IRAC matches, respectively.  The hierarchical choice was based on a combination of critetia, such as best resolution (and therefore more precise positions), and the availability of accurate photometric redshifts, computed using the to-date best available, i.e., deepest photometry.

In this last step, we associate to the 3~GHz single-component sources counterparts from the COSMOS2015 catalog if they are present  within 0\farcs8 and have $p_\mathrm{false}\leq20\%$, or if they have $p_\mathrm{false}>20\%$, but coincide with a reliable ($p_\mathrm{false}\leq20\%$) counterpart candidate from either of the other two catalogs (in  total, \nCOSMOSCptsSingle \ COSMOS2015 counterparts, \nCOSMOSCptsSinglePfalseBig \ of which belong to the second category). We associate $i$-band counterparts if they are  within 0\farcs8 and have $p_\mathrm{false}\leq20\%$  and do not coincide with a COSMOS2015 counterpart candidate, or if they have $p_\mathrm{false}>20\%$, but coincide with a reliable ($p_\mathrm{false}\leq20\%$) counterpart candidate from the IRAC catalog (in total, \nibandCptsSingle \ $i$-band counterparts, \nibandCptsSinglePfalseBig \ of which belong to the second category). Otherwise, we associate IRAC counterparts (\nIRACCptsSingle \ IRAC counterparts, in total). Additionally, we note that we 
do not allow the same multi-wavelength counterpart being associated to different 3~GHz sources following the same reasoning approach. 

 \begin{table}
 \begin{centering}
\begin{tabular}{lccc}
\hline 
Final  &  Total & Multi- &	X-ray \\ 
 counterpart  catalog   &    &  components &   \\ 
\hline
COSMOS2015 & \nCOSMOSCpts & \nCOSMOSCptsMulti \ &   \nXrayCpetnaest \\
$i$-band &  \nibandCpts & \nibandCptsMulti &   \nXrayiband\\
IRAC & \nIRACCpts &  \nIRACCptsMulti   &   \nXrayIRAC \\ 
\hline
Combined	&  \nTotCpts   &	\nTotCptsMulti  &  \nXrayCptTot \\ 
\hline\\
\end{tabular}
\caption{Summary of the cross-correlation of the VLA-COSMOS 3~GHz Large Project sources with sources in the multi-wavelength catalogs.  The last line reports the combined number of COSMOS2015, $i$-band, and IRAC counterparts selected in the common unmasked area of 1.77 square degrees. }
\label{tab:cpts}
 \end{centering}
\end{table}

\subsection{Assessment of optical-MIR counterparts}
\label{sec:optMIRcpts_assesment}

The final counterparts assigned to both our multi- and single-component 3~GHz radio sources are summarized in Tab.~\ref{tab:cpts}. Overall, 
\nCOSMOSCpts \footnote{\nCOSMOSCptsSingle \ and \nCOSMOSCptsMulti \ single- and multi-component sources, respectively.},
\nibandCpts \footnote{\nibandCptsSingle \ and \nibandCptsMulti \ single- and multi-component sources, respectively.}, and
 \nIRACCpts \footnote{\nIRACCptsSingle \ and \nIRACCptsMulti \ single- and multi-component sources, respectively.} 
3~GHz  sources are associated with COSMOS2015,  $i$-band, and IRAC counterparts, respectively. In total, we find \nTotCpts \ counterparts for our radio sources.  Summing the computed false match probabilities we estimate a total fraction of spurious matches of $\sim2\%$. Since our identification procedure is based on ground-based catalogs, it is likely that this "formal" fraction should be considered as a lower limit because of the limited spatial resolution of these data used in these catalogs.
The number of radio sources within the unmasked 
1.77 deg$^2$ area is 8,696, which yields an overall counterpart fraction of  92.4\% (i.e. \nTotCpts \ / 8,696). 

In Fig.~\ref{fig:snr} we show the 3~GHz S/N-ratio distribution for all the 8,696 and the matched \nTotCpts \ radio sources.
The fraction of identifications is  greater than $\sim$90\% for radio sources with signal-to-noise ratio $\gtrsim$6, below which it decreases to $\sim$80\% at signal-to-noise ratio of 5. Such a trend is not  surprising as the estimated fraction of spurious radio detections increases at the lowest signal-to-noise ratios (the fraction of false detections is 24\% for signal-to-noise ratios between 5.0 and 5.1, which decreases to less than 3\% beyond signal-to-noise ratios of 5.5.; see Fig. 17 in \smo et al., accepted).

\section{X-ray counterparts}
\label{sec:X}

We used the X-ray point-source catalog from \citet{marchesi16} (described in detail in Section \ref{sec:data}) to match the X-ray and radio sources. 
We matched the counterparts of the X-ray
sources to the counterparts of the radio sources. Given that the same (COSMOS2015, $i-$band, IRAC) catalogs have been
used to assign counterparts to X-ray and radio-detected sources using either a likelihood ratio
technique \citep{marchesi16} or a nearest neighbor match with false match probability
assigned as done here, the clearest way to match the X-ray and radio sources is to take their
assigned counterparts (see, e.g., \citealt{smolcic08}).

\citet{marchesi16} list the IDs of the COSMOS2015 and $i$-band counterparts of X-ray sources. Thus, after associating our radio sources with counterparts from the COSMOS2015 and the $i$-band selected catalog, we match their respective IDs to those given in the X-ray catalog. The radio sources with IRAC counterparts were cross-macthed to the X-ray catalog via a nearest neighbour matching between IRAC positions listed in both catalogs, but using a search radius of 0\farcs1 to account for machine rounding error. In total, we find \nXrayCptTot \ X-ray counterparts to our radio sources with COSMOS2015 (\nXrayCpetnaest), $i$-band (\nXrayiband) or IRAC (\nXrayIRAC) counterparts. 
Among the 927 X-ray counterparts, 592 (64\%) of the redshifts reported by  \citet{marchesi16} are  spectroscopic, 325 (35\%) are photometric, and 10 (1\%) do not have a redshift listed.

For X-ray sources with redshift values given in the catalog from \citet{marchesi16}, we use the intrinsic (i.e. unobscured) X-ray luminosities already provided in the catalog. In case an X-ray source does not have a spectroscopic or photometric redshift available in the \citet{marchesi16} catalog (about 10 in total entering our sample), we assign it a redshift by taking the most reliable spectroscopic or photometric measurement available in the literature, as described in Section~\ref{sec:redshiftsSample}. The corresponding X-ray luminosity ($L_x$) is calculated by following the same approach detailed in \citet{marchesi16}, which uses the hardness ratio as a proxy for nuclear obscuration (see also \citealt{xue10}). The unobscured $L_x$ estimates were then scaled from [0.5-10]~keV to [0.5-8]~keV by assuming a power-law spectrum with intrinsic slope $\Gamma$=1.8 (e.g. \citealt{tozzi06}).

\section{Redshifts of VLA 3~GHz source counterparts}
\label{sec:redshiftsSample}

We gathered redshift measurements of the radio source counterparts by taking the most reliable redshift available in the literature, either spectroscopic or photometric, as described in \citet{delvecchio17} for only the subset of (COSMOS2015) counterparts analyzed here (see their Section 2.3), and here extended to the full list of (COSMOS2015, $i-band$, and IRAC) radio source counterparts.

For the COSMOS2015 and $i$-band counterparts not detected in X-ray, the photometric redshifts were taken from their respective catalogs. They were derived using the \textsc{LePhare} SED-fitting code \citep{arnouts99, ilbert06}, as described in \citet{ilbert09,ilbert13}. Photometric redshift estimates for IRAC counterparts have been found by cross-matching to COSMOS2015 or $i$-band sources within the respective masked regions. For X-ray detected sources, we used a different set of photometric redshifts from Salvato et al. (in prep.), which were obtained via SED-fitting including AGN variability and AGN templates \citep{salvato09, salvato11}, thus these estimates are more suitable for AGN-dominated sources. 

For each counterpart reported in the spectroscopic sample compiled by Salvato et al. (in prep.), we replace the photometric redshift with the spectroscopic one, only if the spectroscopic measurement is considered ``secure'' or ``very secure''\footnote{The reliability of each spectroscopic redshift relies on the corresponding quality flag. In case of spectroscopic redshift from the zCOSMOS survey \citep{lilly07,lilly09}, we followed the prescription recommended in the zCOSMOS IRSA webpage: https://irsa.ipac.caltech.edu/data/COSMOS/spectra/z-cosmos/Z-COSMOS\_INFO.html. For the other surveys we selected spectroscopic redshifts with quality flag $Q_f\geq 3$ and discarded less reliable measurements from our analysis.}. Additional 28 spectroscopic redshifts were taken from the VIMOS Ultra Deep Survey (VUDS, \citealt{lefevre15,tasca16}).  For X-ray detections, in addition to the above criteria, we adopted the best available redshift reported in \citet{marchesi16}.

\begin{figure} 
	\includegraphics[bb=0 0 432 252, width=1\linewidth]{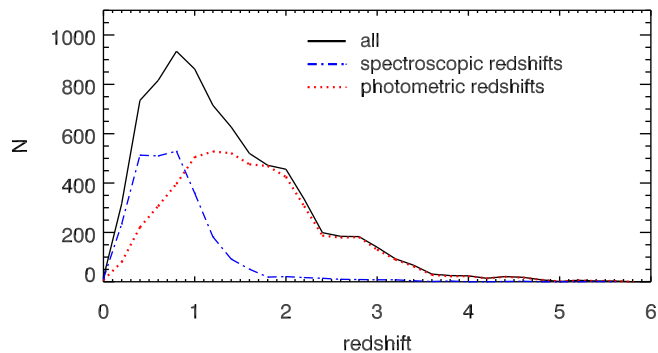}
	\caption{Redshift distribution of the matched (COSMOS2015 and $i$-band) counterparts with redshifts. The black solid line shows the distribution of the best available (spectroscopic or photometric) redshifts, the red dotted line that of the photometric redshift subsample, while the blue dot-dashed line that of the spectroscopic redshift subsample. }
	\label{fig:zdist}
\end{figure}

In total, we retrieved a redshift value for \nz \ out of \nTotCpts \ sources ($\sim$98\% of our sample), being \nzspec \ ($\sim$34\%) spectroscopic and \nzphot \ ($\sim$64\%) photometric. All of the \nznone\ out of \nTotCpts \ ($\sim$2\%) sources with no redshift measurement were associated to an IRAC-only counterpart. The mean redshift for the remaining IRAC sources is similar to the mean values of the counterparts in the COSMOS2015 and $i$-band catalogs ($z\sim1.3$). The counterpart redshift distribution is shown in Fig.~\ref{fig:zdist}. We also verified the accuracy of the photometric redshifts in our sample, based on the comparison with the available spectroscopic measurements. We found a dispersion of the $|\Delta z/(1 + z)|$ distribution of $\sigma_{ |\Delta z/(1 + z)| } =$ 0.01, peaking at 0.002 (see also \citealt{laigle16}). 

\begin{figure} 
\centering
\includegraphics[bb=0 20 432 252, width=1\linewidth]{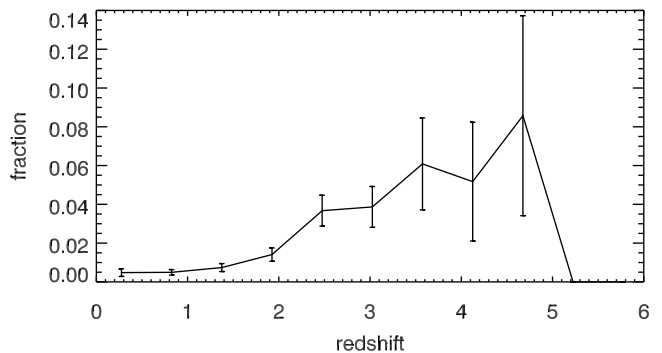}
\caption{Fraction of $i$-band only counterparts in the total (COSMOS2015 and $i$-band counterparts, combined) as a function of redshift. Poisson errors are also shown.}
\label{fig:c15iband}
\end{figure}

Having assigned redshifts to the counterparts of our 3~GHz sources, we now assess the potential incompleteness of the COSMOS2015 catalog, cross-correlated with the 3~GHz radio source catalog (c.f., Fig.~8 in \citealt{laigle16}). In Fig.~\ref{fig:c15iband} we show the fractional contribution of the $i$-band counterparts to the total (COSMOS2015 and $i$-band) counterpart sample. We find that the incompleteness (i.e. the largest fraction of $i$-band-only counterparts) rises with increasing redshift  from $\lesssim1\%$ at $z<2$ to $\sim 8\%$ close to $z=5$.

\begin{figure} 
\centering
\includegraphics[bb=0 0 793 396, width=1\linewidth]{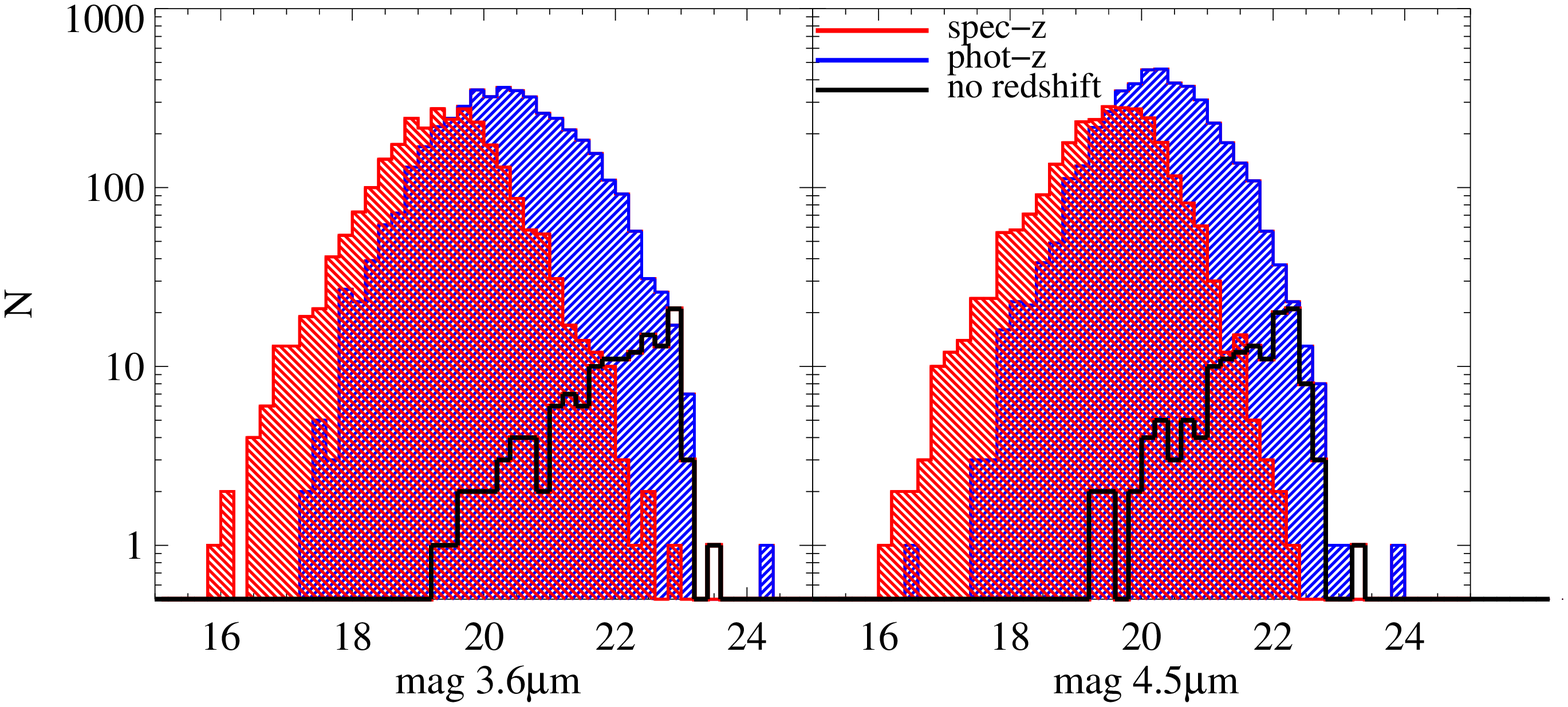}
\includegraphics[bb=0 0 793 396, width=1\linewidth]{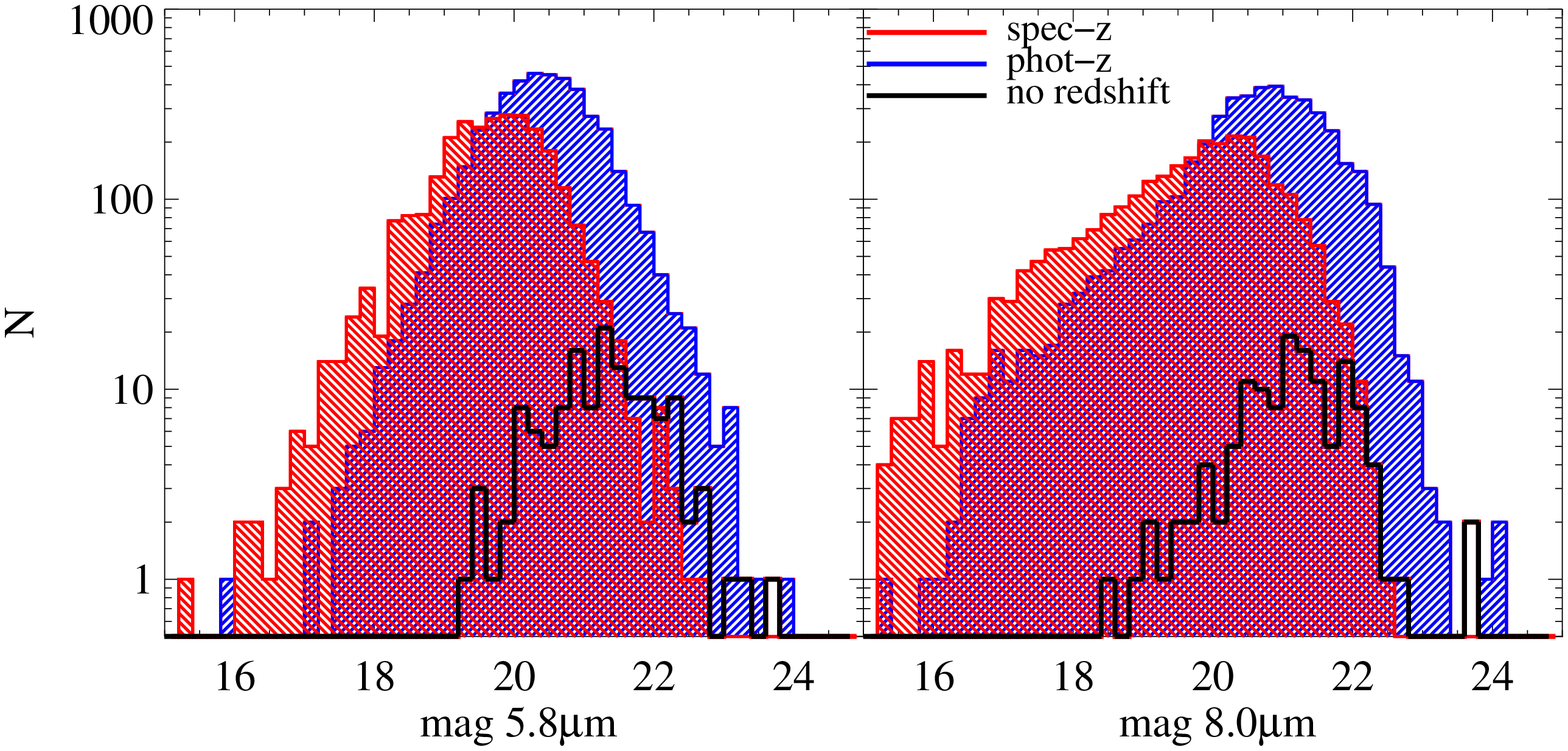}
\caption{
IRAC magnitude distributions for the 3 GHz radio sources with available IRAC magnitudes, separated into three subsamples as indicated in the right panels.
}
\label{fig:irachisto}
\end{figure}

In Fig.~\ref{fig:irachisto}  we show the distributions  for each IRAC band separately, for a subset of radio sources
with available IRAC fluxes in all channels (6915 sources), and for the various subsets of sources with 
 spectroscopic redshift (2674 sources), photometric redshift (4119 sources) and without redshift (122 sources). As expected, for all
IRAC bands these subsets lie at different magnitude ranges, with the spectroscopic redshifts
being more common for brighter sources, followed by those with photometric redshifts, and
eventually by those without redshifts at the faint tail of the magnitude distribution.

\begin{figure} 
\centering
\includegraphics[bb= 0 0 566 566, width=1\linewidth]{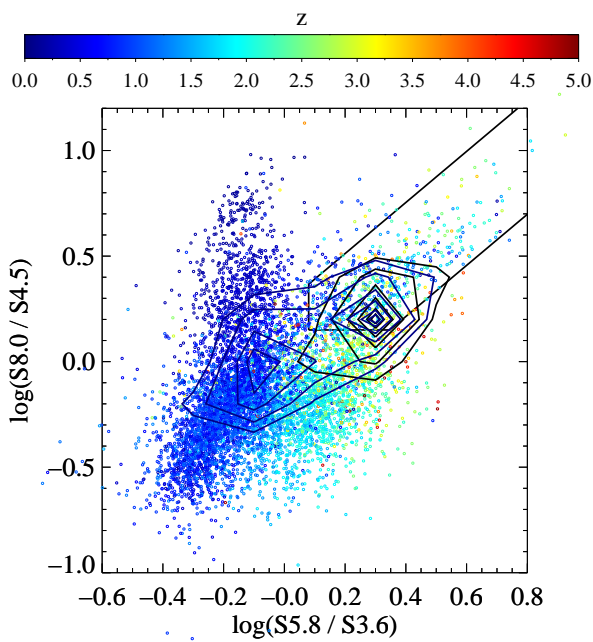}
\caption{
MIR color-color diagram for 6793 radio sources with available (spectroscopic or photmetric) redsihfts and IRAC magnitudes (symbols), color-coded according to their redshift value, as indicated in the color-bar at the top. The blue lines mark the 2D density contours for the sample with IRAC counterparts (205 sources), while the black lines represent the 2D density contours for IRAC counterparts without a redshift measurement (122 objects). The black wedge marks the Donley et al. (2012) area for the selection of MIR-AGN in this color-color plane. 
}
\label{fig:iraccolor}
\end{figure}

In Fig.~\ref{fig:iraccolor} we show the IRAC color-color distribution for the radio sources with available redshift and  fluxes in all IRAC bands (6793 sources). While the IRAC colour distribution (and average redshift) is similar between the full sample
and the subsample with IRAC-only counterparts, the 122 sources with no redshift show redder
colors and a distribution consistent with being at $z>1$. The bulk of these sources falls within
the wedge defined by \citet{donley12} for the selection of MIR-AGN in this color-color plane, suggesting a possible AGN nature of a significant fraction of them.

\section{Galaxy populations}
\label{sec:galpop}

In this Section we characterize the galaxy populations present in our 3~GHz survey. Our aim is to assess the multi-wavelength properties of our 3 GHz sources, such that various characteristics can be combined to select AGN/star-forming galaxy samples depending on the scientific application, for example, to select either i) conservative/cleaned AGN or star-forming galaxy samples, or ii)  {\em parent} populations of radio AGN/star forming galaxies with clean selection functions beyond the radio regime (but to a degree contaminated, as quantified below). For example, the latter (i.e., a selection not relying on  radio flux/luminosity) is preferrable for the purpose of, e.g., stacking in the radio map (e.g., \citealt{karim11}), or statistically assessing radio-AGN duty cycles (e.g., \citealt{smolcic09}). Furthermore, the multi-wavelength properties of our sources identified via various (non-radio and radio-based selection criteria) shed light on the composite (AGN plus star-forming) nature of our sources, over a broad wavelength range, and in particular in the radio regime. 
For sources with available redshifts (i.e., COSMOS2015 or $i$-band counterparts) we first identify AGN by using a combination of X-ray, MIR and spectral energy distribution (SED) fit criteria. For the remainder of the sample we then identify star forming galaxies and AGN based on their host galaxies' rest-frame optical colors (Sec.~\ref{sec:noradioagn}). We then analyze the properties of the galaxy samples in terms of excess of radio emission relative to the  SFRs in their host galaxies (which can be attributed to AGN activity in the radio; Sec.~\ref{sec:sfg_radineff}), and radiative bolometric luminosities of the identified AGN (Sec.~\ref{sec:Lradiative}). We categorize and summarize the samples in Sec.~\ref{sec:sample_sum}.
For the following analysis we use the (\nTotCptsCOSMOSibandAeff ) 3~GHz sources with COSMOS2015 (\nCOSMOSCpts) or $i$-band (\nibandCpts) counterparts present in the unmasked areas of the given catalogs (see Fig.~\ref{fig:counterpartsRADEC} and Table \ref{tab:cpts}), i.e., within a well-defined, unbiased effective area of 1.77 square degrees.  We also stress that the multi-wavelength assessment as presented here is statistical in nature. While the classification will not be correct for every single radio source, it will be statistically correct for the population as a whole.

\subsection{X-ray to  FIR signatures of AGN and star formation activity}
\label{sec:noradioagn}

\subsubsection{X-ray-, MIR-, and SED-selected AGN}
\label{sec:reagn}
Physical processes present in AGN imprint their signature onto the emitted radiation. Powerful X-ray emission from radio sources has been found to originate in a radiatively efficient mode of accretion onto the central black hole (e.g. \citealt{evans06}). These AGN may also be identified via emission from a warm dusty torus around the central black hole, detectable in the MIR \citep{donley12}. Thus, we identify X-ray AGN as those with a [0.5-8]~keV X-ray luminosity higher than $10^{42}~\mathrm{erg~s}^{-1}$, MIR AGN using IRAC color-color criteria, as well as those that show AGN signatures in the multi-wavelength SED using SED-fitting decomposition to disentangle the AGN emission from the host-galaxy light. Below we summarize the selection, and refer to \citet{delvecchio17} for a detailed analysis of the physical properties of such selected AGN across cosmic time.

To select X-ray AGN, we have chosen a limit in intrinsic [0.5-8]~keV (i.e. unobscured) X-ray luminosity of $L_\mathrm{X}=10^{42}~\mathrm{erg~s}^{-1}$ (e.g. \citealt{szokoly04}).  We verified that the X-ray emission expected from the IR-based SFR (obtained from SED-fitting; see below) is generally negligible (i.e. a few \%) compared to the observed X-ray emission, by assuming a canonical L$_\mathrm{X}$--SFR conversion by \citet{Symeonidis14}. This is always true for X-ray sources with $L_\mathrm{X} \geq 10^{42}~\mathrm{erg~s}^{-1}$. For this reason, we consider all sources in our sample with $L_X$ above that limit to be AGN. 
From our sample of \nTotCptsCOSMOSibandAeff \ radio sources with COSMOS2015 or $i$-band counterparts, \nTotCptsXAGN \ ($\sim$11\%) are selected as X-ray AGN. 
While all the sources with an X-ray luminosity higher than the adopted threshold are likely to be AGN, we can not exclude that AGN are present also in the sample of X-ray sources with an X-ray luminosity lower than the threshold. Indeed, out of the X-ray sources with $L_\mathrm{X}\textless10^{42}~\mathrm{erg~s}^{-1}$, $\sim33$\% have been already classified as AGN using other criteria (described below). We note that given the typical X-ray flux in the COSMOS field, this X-ray selection of AGN is progressively missing faint AGN at high redshift, where the X-ray limiting luminosity is higher than $10^{42}~\mathrm{erg~s}^{-1}$  (see, for example, Fig.~8 in \citealt{marchesi16}).

To complement the X-ray selection criterion, we have adopted the MIR selection method of \citet{donley12}. It allows for a reliable selection of sources that contain a warm dusty torus consistent with those in standard, thin-disk AGN \citep{shakura73}, and although reliable, it is not complete. This classification relies on the four IRAC bands at 3.6, 4.5, 5.8 and 8~$\mu$m. Dusty tori of AGN are found within a wedge in that diagram, but also imprint a monotonic rise of flux through the four mid-infrared bands. Sources found displaying these characteristics were recognized as MIR AGN.
Although \nTotCptsMIRAGNandX \  ($\sim$27\%) of the sources classified as X-ray AGN also satisfy the MIR AGN criteria, \nTotCptsMIRAGNnoX \ sources in our total sample of COSMOS2015 and $i$-band counterparts fit the MIR AGN criterion, while not satisfying the X-ray criterion. Thus, the MIR AGN selection can be considered complementary to the X-ray AGN selection method (see also \citealt{delvecchio17}). In total, we find \nTotCptsMIRAGN \ MIR AGN in our sample of \nTotCptsCOSMOSibandAeff \ 3~GHz sources matched with COSMOS2015 or $i$-band counterparts.

Lastly, we fit the optical to millimetre SED\footnote{
 A total of \nherscheltot \ radio sources are detected within 1.77 deg$^2$ at $\geq$3$\sigma$ in at least one \textit{Herschel} band, which are associated to optical/NIR counterparts from either the COSMOS2015 catalog (\nherschelCOSMOS) or the i-band selected catalog (\nherscheliband). For about 100 radio sources (sub-)millimetre data was also available within $1\farcs0$ of optical/NIR positions (see Sec.~\ref{subsec:mmdata})).  
}
of each radio source to identify possible evidence of AGN activity on the basis of a panchromatic SED-fitting analysis (originally developed by \citealt{dacuhna08}, and expanded by \citealt{berta13}). We adopt the fitting results already performed for the COSMOS2015 counterparts by \citet{delvecchio17}, and use the same approach to fit the SEDs of the $i$-band counterparts. The 3-component SED fitting code \textsc{SED3FIT} by \citet{berta13}\footnote{http://cosmos.astro.caltech.edu/page/other-tools} is used to disentangle the possible AGN contribution from the host-galaxy light. This approach combines the emission from stars, dust heated by star formation and a possible AGN/torus component (\citealt{feltre12}, see also \citealt{fritz06}). The dust-absorbed UV-optical stellar light is linked to the reprocessed far-IR dust emission by energy balance. To quantify the relative incidence of a possible AGN component, we followed the approach described in 
\citet{delvecchio14}, by fitting each source SED with both the {\sc magphys} code and the 3-component SED-fitting code. The fit obtained with the AGN is preferred if the reduced $\chi^2$ value of the best-fit is significantly (at $\geq$99\% confidence level, on the basis of a Fisher test) smaller than that obtained from the fit without the AGN. 
We find \nTotCptsSEDAGN \ COSMOS2015 or $i$-band counterparts fulfilling this criterion (out of these 669 are also identified as X-ray or MIR-AGN; see Fig.~2 in \citealt{delvecchio17}), which hereafter will be
referred to as SED-AGN.
In summary, combining X-ray-, MIR-, and SED-selected AGN we identify a total of \nTotCptsHLAGN \ AGN.

\subsubsection{Optically-selected AGN and star forming galaxies }
\label{sub:sfg_radineff}

Having exploited the X-ray, MIR-, and SED-based AGN selection criteria (see Sect.~\ref{sec:reagn}), we next use  a  UV/optical color-based separation method developed by \citet{ilbert10} to derive the composition of the remaining galaxy population. This method separates sources based on the rest-frame near ultraviolet ($NUV$) minus $r^+$-band colors corrected for internal dust extinction ($M_{NUV} - M_{r}$). In the work of \citet{ilbert10}, sources are considered \textit{quiescent} if $M_{NUV}~-~M_{r}>3.5$, of \textit{intermediate} (star formation) activity if $1.2<M_{NUV}~-~M_{r}<3.5$, and of \textit{high} (star formation) activity if $M_{NUV}~-~M_{r^+}<1.2$. \citet{ilbert10} have verified these color selection criteria via a comparison with other color-selections,  morphology, and specific star formation rates of their galaxies (see Figs.\ 4, 5, 6, and 9 in \citealt{ilbert10}). Following this selection here we define the 'intermediate activity' and 'high activity' galaxies as {\em star forming galaxies} (
SFGs hereafter), and we also include in this class galaxies with red ($M_{NUV}~-~M_{r^+}>3.5$) colors, when detected in the Herschel bands. The latter have properties consistent with star forming, rather than quiescent AGN galaxies (e.g., \citealt{delhaize17}). The remaining, red galaxies (not detected in the Herschel bands) are then quiescent galaxies, consistent with typical properties of radio AGN host galaxies, as verified in a number of previous studies (e.g., \citealt{best06,donoso06,sadler14,smolcic09}). Thus, they can be taken as radio-detected AGN hosted by quiescent, red galaxies (see also below). In total, we identify \nTotCptsSFG \ SFGs, and \nTotCptsQMLAGN \ red, quiescent AGN hosts.

\begin{figure}
	\centering
		\includegraphics[bb =  0 0 432 417, width=1\linewidth]{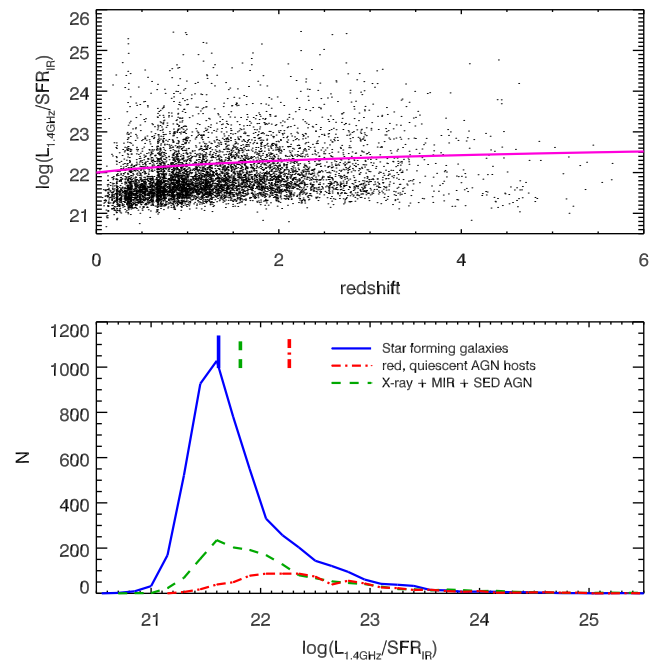}
	\caption{{\em Top panel:} Ratio of rest-frame 1.4~GHz radio luminosity and infrared star formation rate, $\log {( L_\mathrm{1.4GHz} / \mathrm{SFR_{IR}} )}$, as a function of redshift.
	The curve indicates the redshift-dependent threshold used to identify radio-excess sources (see text for details). {\em Bottom:} Distribution of $\log {( L_\mathrm{1.4GHz} / \mathrm{SFR_{IR}} )}$ for various (non-overlapping) galaxy populations as indicated in the panel. The vertical lines indicate the (bi-weighted) means of the distributions.	}
	\label{fig:ratio}
\end{figure}

\begin{figure} 
\centering
\includegraphics[bb = 0 0 432 232, width=1.\linewidth]{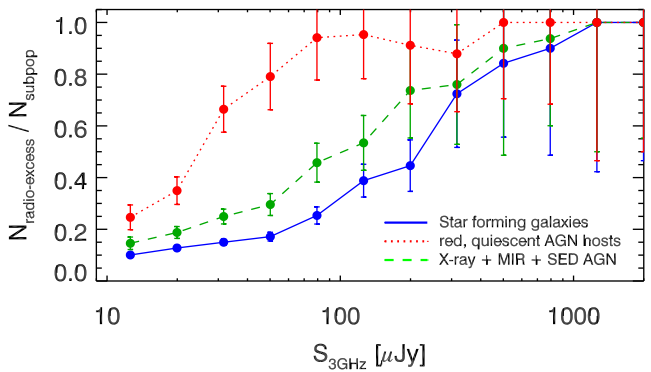} 
\caption{Fraction of radio-excess sources within the various galaxy samples (indicated in the panel) as a function of total  3~GHz radio flux. Poisson errors are also shown.}
\label{fig:radioexcessfraction}
\end{figure}

\begin{figure} 
	\centering
	\includegraphics[width=1\linewidth]{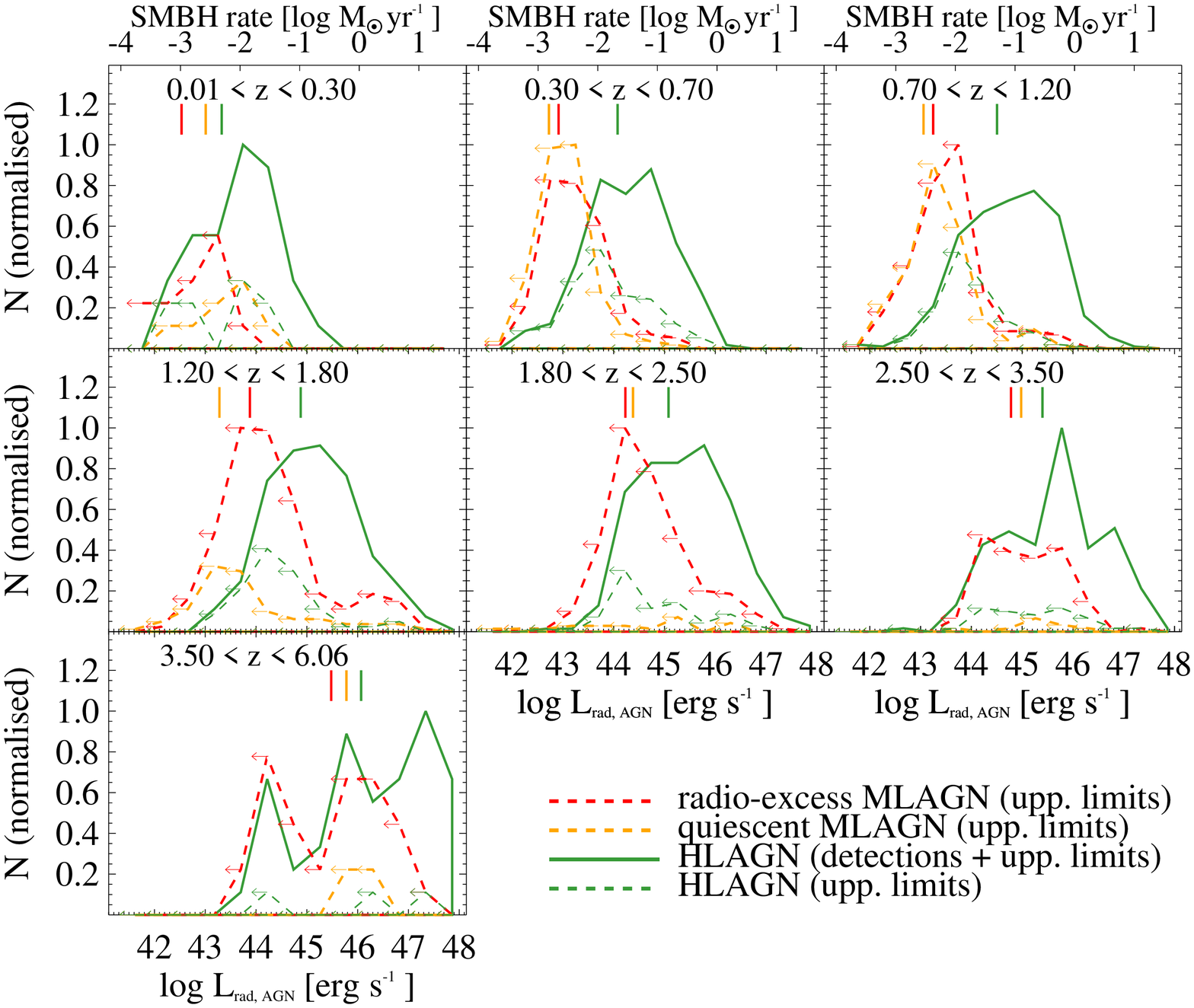}
	\caption{Normalized distribution of bolometric (radiative) AGN luminosity (L$_{\rm rad, AGN}$), as a function of redshift, for various populations, as indicated in the panels and legend, respectively.  HLAGN refer to X-ray-, MIR-, and SED-selected AGN, while MLAGN refer to radio AGN hosted by red, quiescent galaxies or showing a $>3\sigma$ radio-excess in $\log {( L_\mathrm{1.4GHz} / \mathrm{SFR_{IR}} )}$, but not selected as X-ray-, MIR-, or SED-AGN (see text for details). 
	Left-pointing arrows indicate upper limits in L$_{\rm rad, AGN}$ (see text for details). The upper $x$-axes show the supermassive black hole (SMBH) accretion rate, estimated from L$_{\rm rad, AGN}$ assuming a standard radiative efficiency of 10\%.}
	\label{fig:distlumagnred}
\end{figure}

\begin{figure*} 
	\centering
	\includegraphics[bb= 0 60 842 575, width=1\linewidth]{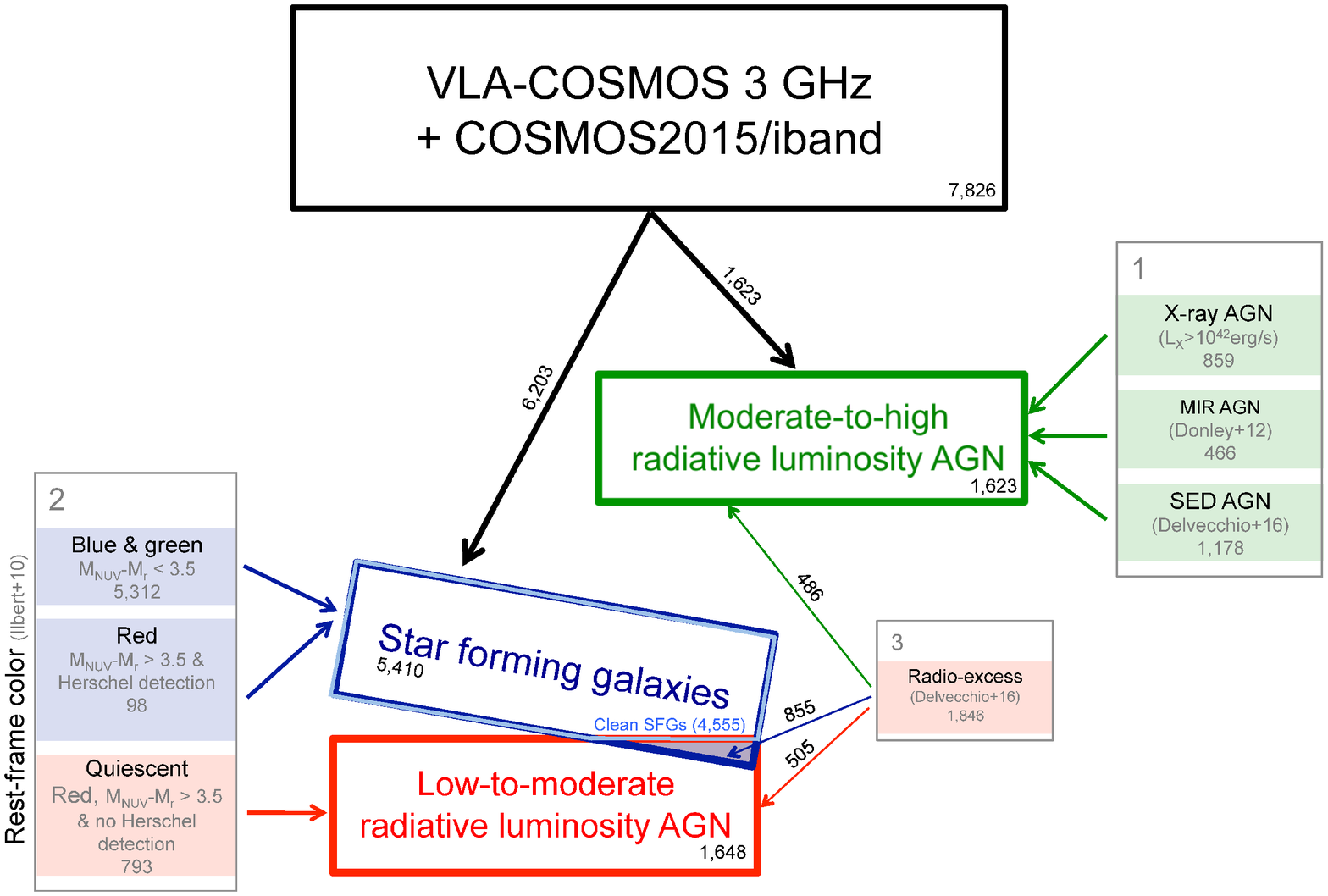}
	\caption{
		Flowchart illustrating the separation of the 3~GHz sources matched to COSMOS2015 or $i$-band counterparts (within unmasked areas of the field) into various AGN and galaxy populations (see Sec.~\ref{sec:galpop} for details).
	}
	\label{fig:flowchart}
\end{figure*}

\begin{figure}
	\centering
		\includegraphics[bb =  0 20 432 432, width=1\linewidth]{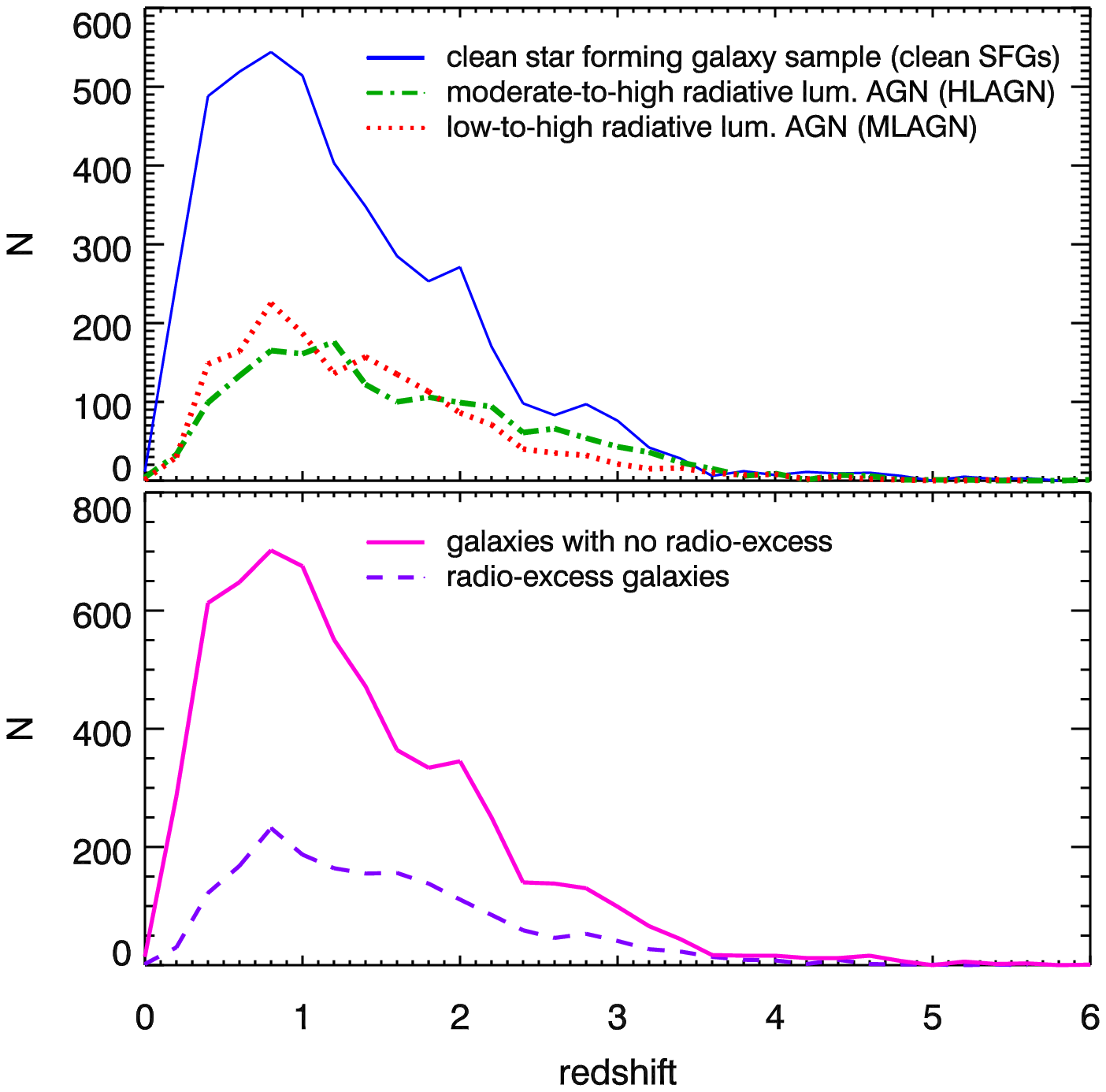}
	\caption{Redshift distribution of different (not overlapping) populations, as indicated in the panels.}
	\label{fig:riagn}
\end{figure}

\subsection{Radio-excess as AGN signature in the radio band}
\label{sec:sfg_radineff}

In the previous sections we have identified AGN and star forming galaxies in our 3~GHz radio source sample, based on their multi-wavelength (X-ray to  FIR) properties. We here investigate the excess of {\em radio} emission relative to the star formation rates in their host galaxies. This yields an insight into the AGN origin of the radio emission in these galaxies, which may not necessarily be in one-to-one correspondence with, e.g., AGN signatures in the X-ray or MIR bands, hence shedding light on the composite nature of the sources.

In Fig.~\ref{fig:ratio} we show the distribution of $\log {( L_\mathrm{1.4GHz} / \mathrm{SFR_{IR}} )}$, i.e. the logarithm of the ratio of the 1.4~GHz rest-frame radio luminosity and SFR derived from the total IR emission in the host galaxies as computed via the 3-component SED fitting procedure as described above and by \citet{delvecchio17}.
The SFRs were obtained from SED-fitting, by using the total (8-1000$~\mu$m rest-frame) infrared luminosity obtained from the best-fit galaxy template, for sources both with and without \textit{Herschel} detection, as described in \citet{delvecchio17}. This luminosity represents the fraction of the total infrared luminosity arising from star formation for HLAGN, or the total IR luminosity otherwise, and was converted to SFR via the \citet{kennicutt98} conversion factor, and scaled to a \citet{chabrier03} IMF.
As radio luminosity is an efficient star formation tracer (i.e., $\log { \mathrm{SFR} } = \log{ L_\mathrm{1.4GHz} + constant}$; e.g., \citealt{condon92}), 
an excess in $\log {( L_\mathrm{1.4GHz} / \mathrm{SFR_{IR}} )}$ can be taken as arising from non-SFR, thus AGN related processes in the radio band. 
Given a positive trend of  $\log {( L_\mathrm{1.4GHz} / \mathrm{SFR_{IR}} )}$ with redshift (see also \citealt{delhaize17}) following \citet{delvecchio17} we use a redshift-dependent threshold of the form $\log {( L_\mathrm{1.4GHz} / \mathrm{SFR_{IR}} )} = 21.984(1+z)^{0.013}$ to better quantify the excess in radio emission relative to the star formation in the host galaxies in our sample (radio-excess hereafter). For any given redshift range the dispersion ($\sigma_{\log {( L_\mathrm{1.4GHz} / \mathrm{SFR_{IR}} )}}$) is derived by fitting a Gaussian to the 
distribution obtained by mirroring the distribution at the lower end over the distribution's peak, thus eliminating the asymmetry of the distribution due to radio-excess sources. The 3$\sigma$ deviation from the peak of the Gaussian function at each redshift was chosen as the radio-excess threshold. 

We note that the reason for a redshift-dependent threshold  is purely observationally driven.
A statistically more robust and detailed analysis \citep{delhaize17}, where both the radio and the FIR upper limits are fully taken into account, also finds that the median $\log {( L_\mathrm{IR} / \mathrm{1.4GHz} )}$  for our sample of star-forming galaxies shows a significant trend with redshift, qualitatively consistent with what is shown in our Fig.~\ref{fig:ratio} (see also, e.g., \citealt{magnelli15}, Calistro Rivera et al.\, in prep.). Thus, the choice of a redshift-dependent threshold is fully consistent with the available observational data. 
We note that Delhaize et al. discuss a number of possible reasons for this observed trend. These include, for example, the uncertainty in the radio SED shape (and therefore in the K-correction), a possibly remaining AGN contribution in the SFG sample, a systematic change with redshift (and/or with the level of SFR) of the average magnetic field of star-forming galaxies.

The above described redshift-dependent threshold in $\log {( L_\mathrm{1.4GHz} / \mathrm{SFR_{IR}} )}$
 yields a total of  \nTotCptsREX \ radio-excess sources within the COSMOS2015/$i$-band counterpart sample. Dissected per selection criterion (with possible overlap between various criteria), \nTotCptsREXXAGN , \nTotCptsREXMIRAGN , \nTotCptsREXSEDAGN , \nTotCptsREXSFG , and \nTotCptsREXQAGN \ sources with radio-excess have, separately, been identified as X-ray-, MIR-, SED-AGN, star-forming, and red/quiescent galaxies, respectively. 

From  Fig.~\ref{fig:ratio} it is apparent that a large fraction of the radio AGN hosted by red, quiescent galaxies show an excess of radio emission relative to the star formation rates in their host galaxies, affirming their AGN nature at radio wavelengths. 
The star forming galaxy sample (selected based on green, and blue optical rest-frame colors) occupies the region of the $\log {( L_\mathrm{1.4GHz} / \mathrm{SFR_{IR}} )}$ distribution expected for star forming galaxies (and consistent with no radio-excess), with a tail extending towards higher radio-excess About 16\% ($=1-\nTotCptsSFGpure / \nTotCptsSFG $) of such identified star forming galaxies exhibit a $>3\sigma$ radio-excess, which is taken as the contamination of this sample by radio-AGN identified via the $>3\sigma$ radio-excess in the $\log {( L_\mathrm{1.4GHz} / \mathrm{SFR_{IR}} )}$ plane. This contamination of the star forming galaxy sample increases towards higher flux density levels, as shown in Fig.~\ref{fig:radioexcessfraction}, where we also show the radio-excess fractions as a function of radio flux for the red, quiescent hosts, and the combined, X-ray-, MIR-, and SED-AGN sample.

The $\log {( L_\mathrm{1.4GHz} / \mathrm{SFR_{IR}} )}$ distribution for the combined sample of X-ray-, MIR-, and SED-AGN peaks close to that of star forming galaxies, and exhibits an extended tail towards higher values. As discussed in detail in \citet{delvecchio17} this suggests that  a substantial amount of the radio emission in a large fraction of these AGN arises from star-formation-, rather than AGN-related processes. However, there is also a large number of such sources with significant radio-excess ($\sim30\%$ have a $>3\sigma$ radio-excess).

\subsection{Radiative AGN luminosities}
\label{sec:Lradiative}

We here investigate the distribution of the radiative (bolometric) luminosities for the AGN identified here.  We first combine the identified AGN into two categories, those that show AGN signatures i) at other than radio wavelengths (i.e., X-ray-, MIR-, and SED-selected AGN), ii) at radio wavelengths (i.e. those hosted by red, quiescent galaxies, or showing a $>3\sigma$ radio-excess in $\log {( L_\mathrm{1.4GHz} / \mathrm{SFR_{IR}} )}$, but not selected as X-ray-, MIR-, or SED-AGN). 

For each identified AGN the bolometric, radiative luminosity or its upper limit was obtained from the best fit SED template as described in detail by \citet{delvecchio17}.  Briefly, the AGN radiative luminosity is obtained from the best-fit AGN template. However, if the AGN template does not improve (at $\geq$99\% significance, based on a Fisher test) the fit to the full SED, the corresponding AGN radiative luminosity is unconstrained from SED-fitting. In this case, we report the 95${\rm th}$ percentile of the corresponding probability distribution function, which is equivalent to an upper limit at 90\% confidence level. AGN radiative luminosities derived from SED-fitting have been compared with those independently calculated from X-rays, displaying no significant systematics and a 1$\sigma$ dispersion of about 0.4~dex (see e.g. \citealt{lanzuisi15}).  In Fig.~\ref{fig:distlumagnred} we show the distribution of the AGN radiative luminosities for seven redshift bins, out to $z\sim6$. It is clear 
from this plot that the non-radio based selection of (X-ray-, MIR-, and SED-) AGN is, at every redshift, more efficient in selecting AGN with statistically higher radiative 
luminosities than AGN selection criteria mainly linked to radio-wavelengths. Given that for the latter we only  have upper limits to the radiative luminosity (left-pointing arrows in Fig.~\ref{fig:distlumagnred}), and that, thus, no unique separation threshold in radiative luminosity out to $z\sim6$ can be inferred, we hereafter abbreviate the two identified categories\footnote{consistent with the nomenclature used in \citet{delvecchio17}} as i) moderate-to-high radiative luminosity AGN (HLAGN hereafter, referring to X-ray-, MIR-, and SED-selected AGN, regardless of their radio-excess in $\log {( L_\mathrm{1.4GHz} / \mathrm{SFR_{IR}} )}$), and ii) low-to-moderate radiative luminosity AGN (MLAGN hereafter, referring to the AGN identified via quiescent, red host galaxies, or those with a $>3\sigma$ radio-excess in $\log {( L_\mathrm{1.4GHz} / \mathrm{SFR_{IR}} )}$, but not identified as X-ray-, MIR-, or SED-AGN).

\subsection{Sample summary}
\label{sec:sample_sum}
Using the multi-wavelength criteria described above the identified AGN/star-forming galaxy samples can be categorized into three main classes of objects within our 3~GHz radio sample associated with COSMOS2015 or $i$-band counterparts located inside the 1.77 square degree area unmasked in the COSMOS2015 catalog.
The selection process and relative fractions of the three classes are illustrated in Fig.~\ref{fig:flowchart}, and summarized below. 
\begin{enumerate}
	\item {\em Moderate-to-high radiative luminosity AGN (HLAGN)} were selected by using a combination of X-ray ($L_\mathrm{X}>10^{42}$ erg/s), MIR color-color \citep{donley12}, and SED-fitting \citep{delvecchio17} criteria. We identify a total of \nTotCptsHLAGN \ HLAGN in our 3 GHz sample consisting of \nTotCptsCOSMOSibandAeff \ radio sources with associated COSMOS2015 or $i$-band counterparts, and find that \nTotCptsREXHLAGN \  ($\sim30\%$) show a $>3\sigma$ radio-excess in $\log {( L_\mathrm{1.4GHz} / \mathrm{SFR_{IR}} )}$, while the radio luminosity in the remaining $\sim$70\%  is consistent (within $\pm3\sigma$) with the IR-based star-formation rates in their host galaxies.\\
	\item {\em Star forming galaxies (SFGs)} were drawn from the sample remaining after exclusion of the HLAGN by selecting galaxies with the dust-extinction corrected rest frame color i) $M_{NUV}~-~M_{r^+}<3.5$, or ii) $M_{NUV}~-~M_{r^+}>3.5$, but requiring a detection in the Herschel bands. We identify a total of \nTotCptsSFG \ star forming galaxies in our 3~GHz sample, corresponding to 69\%  of all the radio sources with associated COSMOS2015 or $i$-band counterparts. In such a selected sample we identify \nTotCptsREXSFG \ sources ($\sim16\%$) with a $>3\sigma$ radio-excess in $\log {( L_\mathrm{1.4GHz} / \mathrm{SFR_{IR}} )}$, which can be taken as the contamination of such a selected star forming galaxy sample by ($>3\sigma$ radio-excess) AGN. The sample obtained by excluding these radio-excess sources from the star forming galaxy sample is defined as the 'clean star forming galaxy sample' (i.e., {\em clean SFG sample}, hereafter). \\
	\item {\em  Low-to-moderate radiative luminosity AGN (MLAGN)} were drawn from the sample remaining after exclusion of the HLAGN. We identify a total of \nTotCptsMLAGN \ of MLAGN, and they are a combination of the following two sub-classes:\\
	\begin{itemize}
		\item[3.1.] 
		{\em Quiescent-MLAGN} were selected requiring $M_{NUV}~-~M_{r^+}>3.5$, and no detection (at $\geq$~5$\sigma$) in any of the Herschel bands. This criterion identifies \nTotCptsQMLAGN \ such sources in the COSMOS2015/$i$-band counterpart sample. \nTotCptsREXQAGN \ of these are consistent with a  $>3\sigma$ radio-excess in $\log {( L_\mathrm{1.4GHz} / \mathrm{SFR_{IR}} )}$, and their median/mean $\log {( L_\mathrm{1.4GHz} / \mathrm{SFR_{IR}} )}$ is significantly above the average for the star-forming galaxy sample. \\
		\item[3.2.] {\em Radio-excess-MLAGN}  were selected as objects with a $>3\sigma$ radio-excess in the redshift-dependent distribution of $\log {( L_\mathrm{1.4GHz} / \mathrm{SFR_{IR}} )}$. This criterion selects \nTotCptsREXMLAGN \ such AGN in the COSMOS2015/$i$-band counterpart sample, \nTotCptsQREXAGNoverlap \ of which overlap with the quiescent-MLAGN sample, and \nTotCptsREXSFG \ of which overlap with the star forming galaxy sample.
	\end{itemize}
\end{enumerate}

We note that a certain overlap exists between some of the above described populations, as stated specifically above and illustrated in Fig.~\ref{fig:flowchart}. Complete, non-overlapping samples can be formed by combining i) HLAGN, MLAGN, and clean SFG samples, or, alternatively, ii) the radio-excess and no-radio-excess samples. While the first reflects the multi-wavelength nature of AGN or star formation activity in our radio-detected sources, the latter can be taken to reflect the galaxies' AGN or star formation activity signature specifically in the radio band. Overlap between these samples (e.g., X-ray AGN with no radio-excess) gives insight into the composite (AGN plus star forming) nature of the sources across the electromagnetic spectrum. The redshift distribution of the various populations is given in Fig.~\ref{fig:riagn}.

\begin{figure} 
	\centering
	\includegraphics[bb = 0 0 432 247, width=1\linewidth]{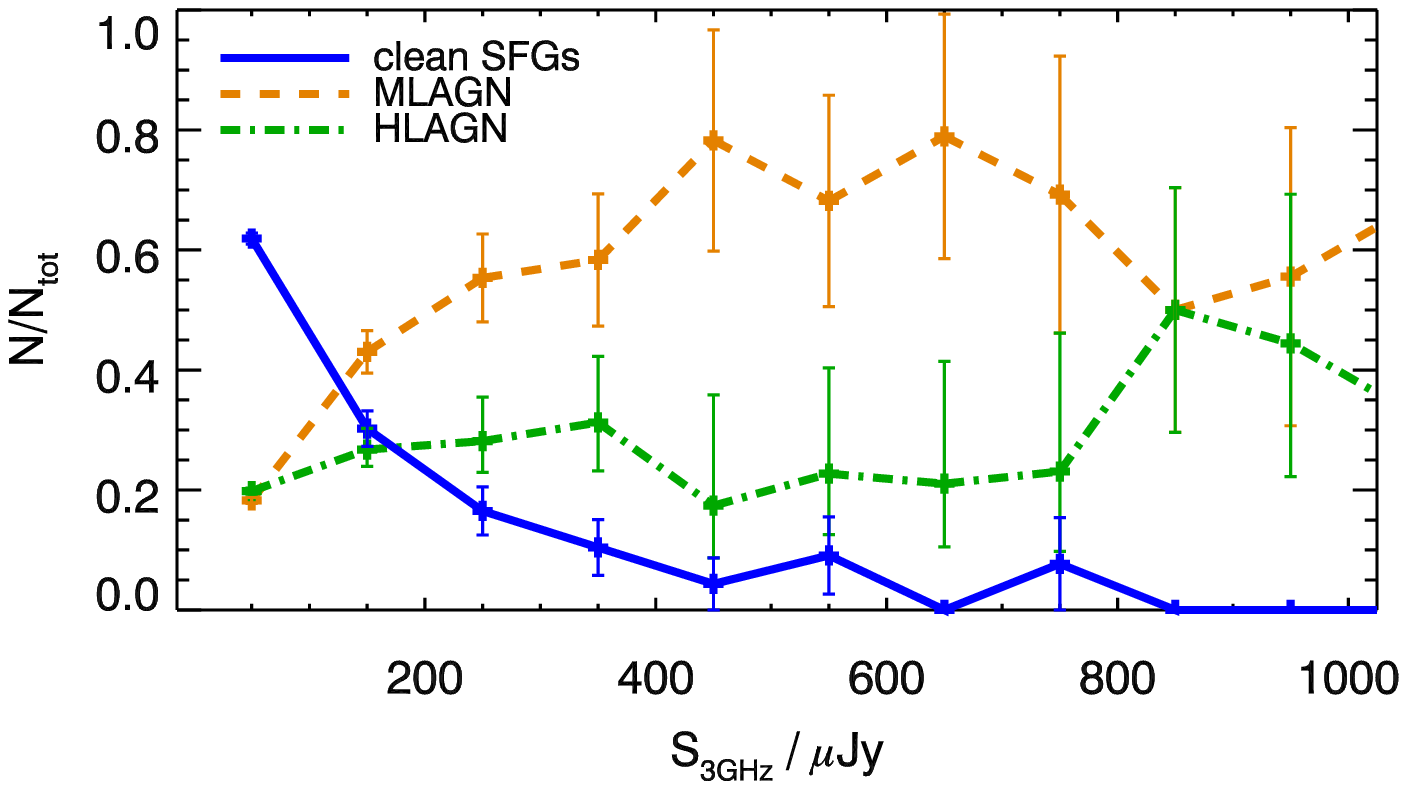}
	\includegraphics[bb = 0 0 432 247, width=1\linewidth]{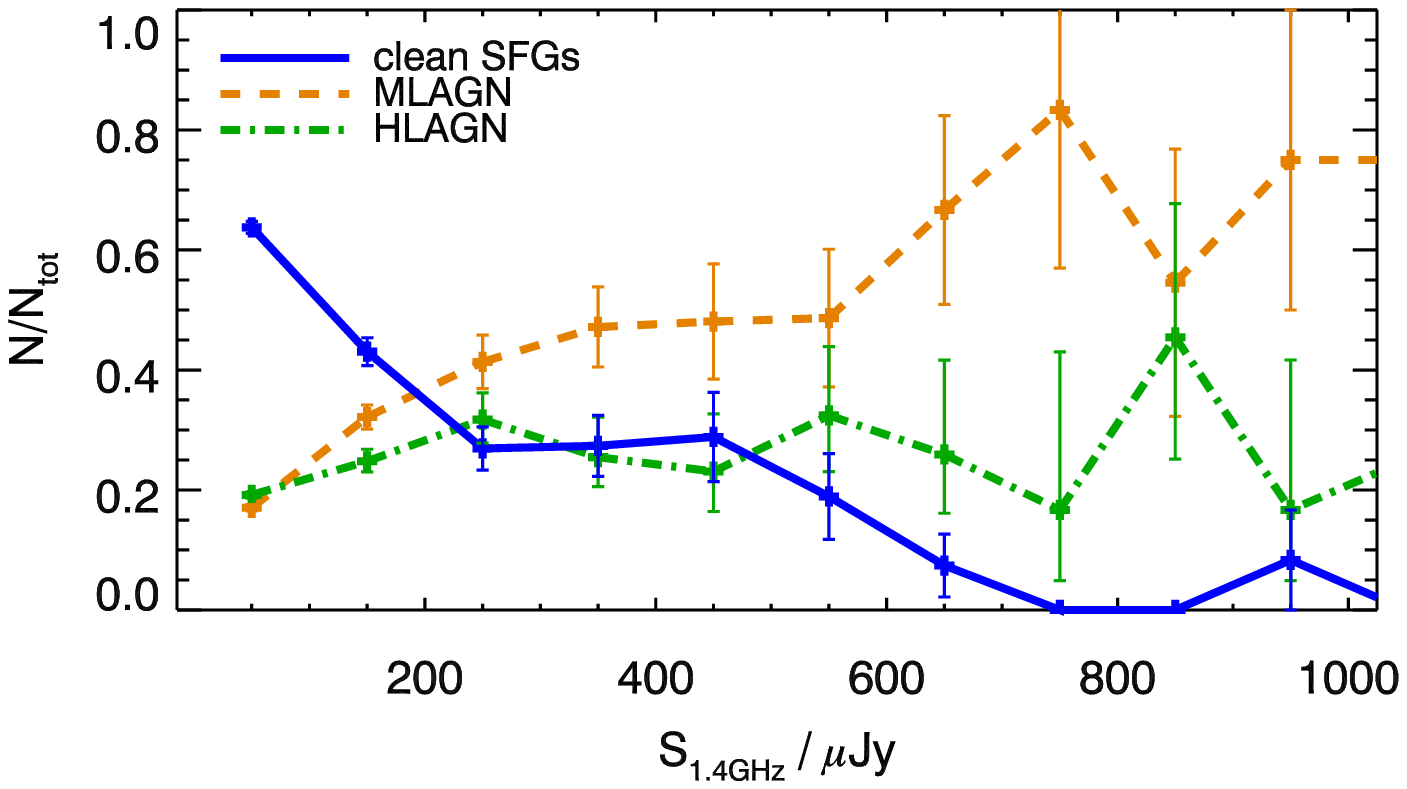}
	\includegraphics[bb = 0 20 432 247, width=1\linewidth]{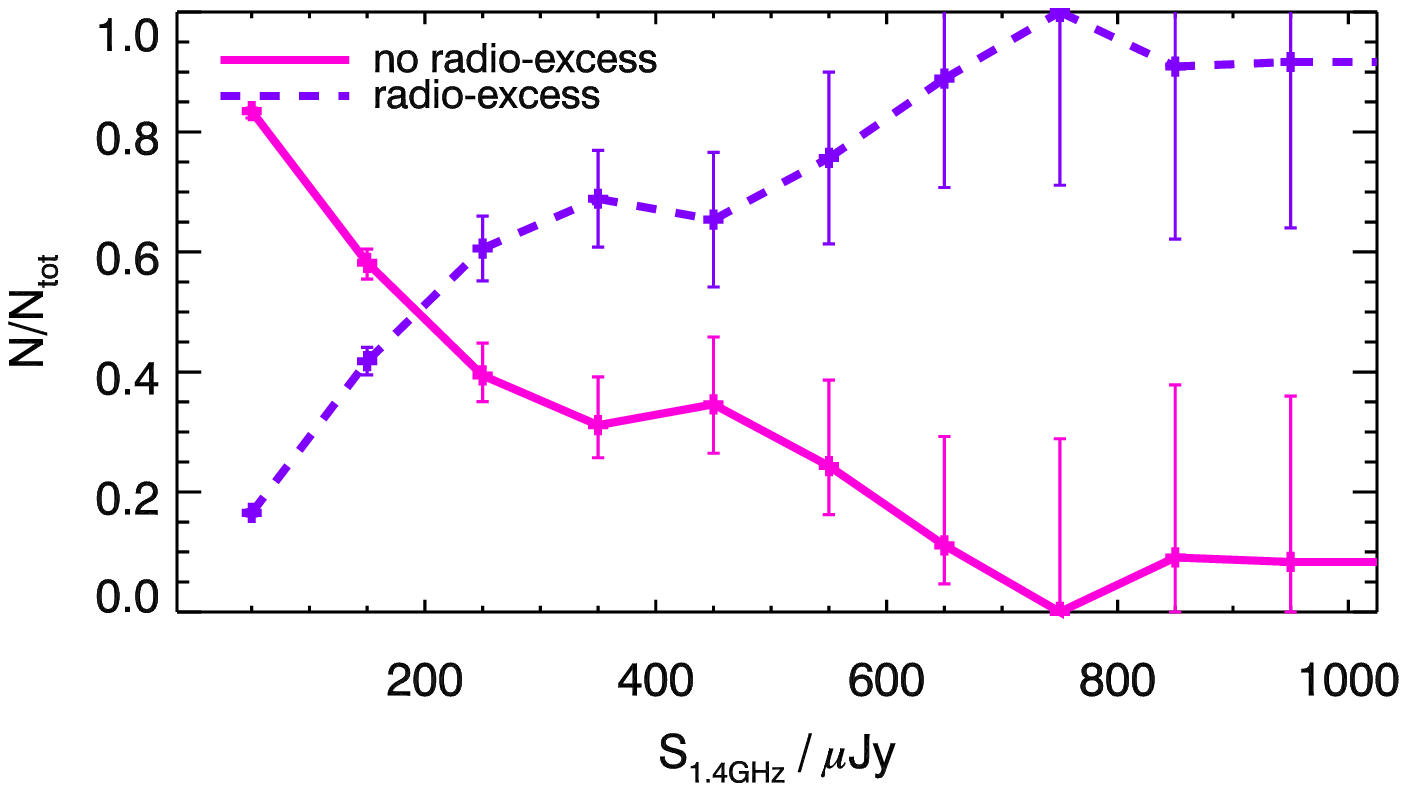}
		\caption{Fractional contributions of the various populations (indicated in the panels) as a function of 3~GHz flux (top panel), and 1.4~GHz flux (middle and bottom panels). }
	\label{fig:popfractions}
\end{figure}

\section{Final counterpart catalog}
\label{sec:catalog}

The VLA-COSMOS 3~GHz Large Project multi-wavelength counterpart catalog will be made available through the COSMOS IPAC/IRSA data-base\footnote{\url{http://irsa.ipac.caltech.edu/Missions/cosmos.html}}. It is constructed in such a way that any user can easily retrieve our classification, or adjust it to a different set of selection criteria. The catalog lists the counterpart IDs, properties, as well as the individual criteria used in this work to classify our radio sources, as follows.

\begin{enumerate}
 \item Identification number of the radio source (ID$\_$VLA).
 \item  Right ascension (J2000) of the radio source.
 \item  Declination (J2000) of the radio source.
 \item  Identifier for a single- or multi-component radio source (MULTI=0 or 1 for the first or latter, respectively).
 \item  Identification  of the counterpart catalog, either 'COSMOS2015', 'iband', or 'IRAC'.
 \item  Identification number of the counterpart (ID$\_$CPT).
 \item  Right ascension (J2000) of the counterpart, corrected for small astrometry offsets as described in the Appendix.
  \item  Declination (J2000) of the counterpart, corrected for small astrometry offsets as described in the Appendix.
  \item  Separation between the radio source and its counterpart [arcseconds].
  \item  False match probability.
 \item  Best redshift available for the source.
 \item   Integrated 3~GHz radio flux density [$\mu$Jy].
 \item   Rest-frame 3~GHz radio luminosity [log W~Hz$^{-1}$].
 \item   Rest-frame 1.4~GHz radio luminosity [log W~Hz$^{-1}$], obtained using the measured 1.4-3~GHz spectral index, if available, otherwise assuming a spectral index of -0.7.
 \item  Star-formation related infrared (8-1000$~\mu$m rest-frame) luminosity derived from SED-fitting [log $L_{\odot}$]. If the source is classified as HLAGN, this value represents the fraction of the total infrared luminosity arising from star formation, while it corresponds to the total IR luminosity otherwise (see \citealt{delvecchio17}).
 \item Star formation rate [$M_{\odot}$yr$^{-1}$] obtained from the total infrared luminosity listed in column (15), by assuming the \citet{kennicutt98} conversion factor, and scaled to a \citet{chabrier03} IMF.
 \item  X-ray-AGN: ``T'' if true, ``F'' otherwise.
 \item  MIR AGN: ``T'' if true, ``F'' otherwise.
 \item  SED-AGN: ``T'' if true, ``F'' otherwise.
 \item  Quiescent-MLAGN: ``T'' if true, ``F'' otherwise.
 \item  SFG: ``T'' if true, ``F'' otherwise.
 \item  Clean SFG: ``T'' if true, ``F'' otherwise.
 \item HLAGN: ``T'' if true, ``F'' otherwise.
 \item  MLAGN: ``T'' if true, ``F'' otherwise.
 \item  Radio-excess: ``T'' if true, ``F'' otherwise.
 \item COSMOS2015 masked area flag: ``1'' if true, ``0'' otherwise.\footnote{This flag was computed for the counterpart positions against the COSMOS2015 mosaics indicating the masked areas, and agrees to  0.17\% with the masked areas reported in the COSMOS2015 catalog.}
\end{enumerate}

 We note that the counterpart catalog presented here contains the full list of optical-MIR counterparts collected over the largest unmasked area accessible to each catalog, being 1.77, 1.73, and 2.35 deg$^2$ for COSMOS2015,  i-band, and IRAC catalogs, respectively. We note that complete, non-overlapping samples within a well defined, effective area of 1.77 square degrees (COSMOS2015 masked area flag = 0, can be formed by combining i) HLAGN, MLAGN, and clean SFG samples, or, alternatively, ii) the radio-excess and no-radio-excess samples (see previous section).

\section{Composition of the microJansky radio source population}
\label{sec:millijansky}

In this Section we analyze the composition of the faint radio population at microJansky levels. We consider non-overlapping subsamples of the total sample of radio sources with counterparts in the COSMOS2015 or $i$-band catalogs, defined in two different ways as described the previous  Sections (for a summary see Sec.~\ref{sec:sample_sum} and Fig.~\ref{fig:flowchart}): 
\begin{enumerate}
\item 
based on multi-wavelengths criteria, \\
i) moderate-to-high radiative luminosity AGN (HLAGN; selected via  X-ray-, IR-, and SED-based critera),  \\
ii) low-to-moderate radiative luminosity AGN (MLAGN; consisting of two subsamples not identified  as AGN by the X-ray, IR or SED criteria, i.e., AGN hosted by red/quiescent galaxies, and those showing a $>3\sigma$ excess in radio emission relative to the IR-based star formation rates in their host galaxies), \\ 
iii) clean star forming galaxy sample (i.e.,  rest-frame color-selected star forming galaxies, not identified by the X-ray, IR or SED criteria, and with radio-excess sources  excluded; below often referred to as the SFG sample), and

\item based on radio criteria, samples of galaxies with \\
i) radio-excess, \\
ii)  without radio-excess (defined as a $>3\sigma$ radio-excess in the redshift-dependent distribution of $\log {( L_\mathrm{1.4GHz} / \mathrm{SFR_{IR}} )}$).
\end{enumerate}

\subsection{Fractional contribution of the various populations at faint radio fluxes}

 In the top panel of Fig.~\ref{fig:popfractions} we show the fractional contributions of the various populations as a function of total 3~GHz flux ($S_\mathrm{3GHz}$). Dividing the populations into clean SFG, MLAGN, and HLAGN samples we find that the fraction of MLAGN decreases from about 75\% at $S_\mathrm{3GHz}\sim400-800~\mu$Jy down to about 50\% at $S_\mathrm{3GHz}\sim100-400~\mu$Jy, and further to about 20\% in our faintest flux bin at $S_\mathrm{3GHz}\sim50~\mu$Jy. Through the same flux ranges, the fraction of SFGs increases from  about 10\% ($S_\mathrm{3GHz}\sim400-800~\mu$Jy) to 60\% ($S_\mathrm{3GHz}\sim50~\mu$Jy), while that of HLAGN remains fairly constant in the range of 20-30\%. Hence, while the combined AGN sample (MLAGN and HLAGN) dominates the radio population at  $S_\mathrm{3GHz}\gtrsim100~\mu$Jy, it is the SFGs that form the bulk (60\%) of the population at  $S_\mathrm{3GHz}\sim50~\mu$Jy. For easier comparison with literature, in the middle panel of Fig.~\ref{fig:popfractions} we show the 
fractional contributions of the various populations as a function of total 1.4~GHz flux, computed using the 1.4~GHz detections where available and assuming a spectral index of -0.7 otherwise (see \citealt{smo17}, for details). The results are consistent with those described above. We find that only in our lowest flux bin ($S_\mathrm{1.4GHz}\sim20-100~\mu$Jy) the SFGs start dominating reaching a fraction of $\sim60\%$ at $S_\mathrm{1.4GHz}\sim50~\mu$Jy.  This is consistent with the results based on the ECDFS survey (\citealt{bonzini13}; see also below) yielding that the fraction of their SFGs is 50-60\% within $S_\mathrm{1.4GHz}\sim35-100~\mu$Jy. 
Within the total range of $S_\mathrm{1.4GHz}\sim20-700~\mu$Jy we find a median fraction of about $40\%$, $30\%$, and $30\%$ of MLAGN, HLAGN, and SFGs, respectively, 
in good agreement with the results based on the 1.4~GHz VLA-COSMOS survey \citep{smolcic08}. 

As detailed in Sec.~\ref{sec:galpop},  the radio emission of $\sim70\%$ of HLAGN is consistent with that expected from the star formation in their host galaxies (inferred from their IR emission). Hence, to set limits on the fractional contribution to the faint radio population by the star-formation- or AGN-related processes in the radio band, in the bottom panel of Fig.~\ref{fig:popfractions} we show the separation  of all of our sources into those that show ($>3\sigma$) radio-excess, and those with no radio excess relative to the host's star formation rate (see Fig.~\ref{fig:ratio}). Consistent with the above conclusions, we find that the fraction of sources with total radio luminosity consistent with that expected from the star formation in their host galaxies increases with decreasing flux density from about 10\% at  $S_\mathrm{1.4GHz}\sim700-1000~\mu$Jy to $\sim85\%$ at $S_\mathrm{1.4GHz}\sim50~\mu$Jy, while it correspondingly decreases (from $\sim90\%$ to $\sim15\%$) for sources showing 
significant ($>3\sigma$) radio-excess. The switch between the domination of the two, such selected populations, occurs at $S_\mathrm{1.4GHz}\sim200~\mu$Jy. 

In conclusion, the data used here probe the flux range of the faint radio population where a switch between the dominant AGN- or star formation-related contributions occurs. This is studied further in the context of radio source counts in the next Section.

\subsection{Euclidean-normalized radio source counts}

\begin{figure*}
	\centering
	\includegraphics[bb= 0 0 432 322, scale=0.55]{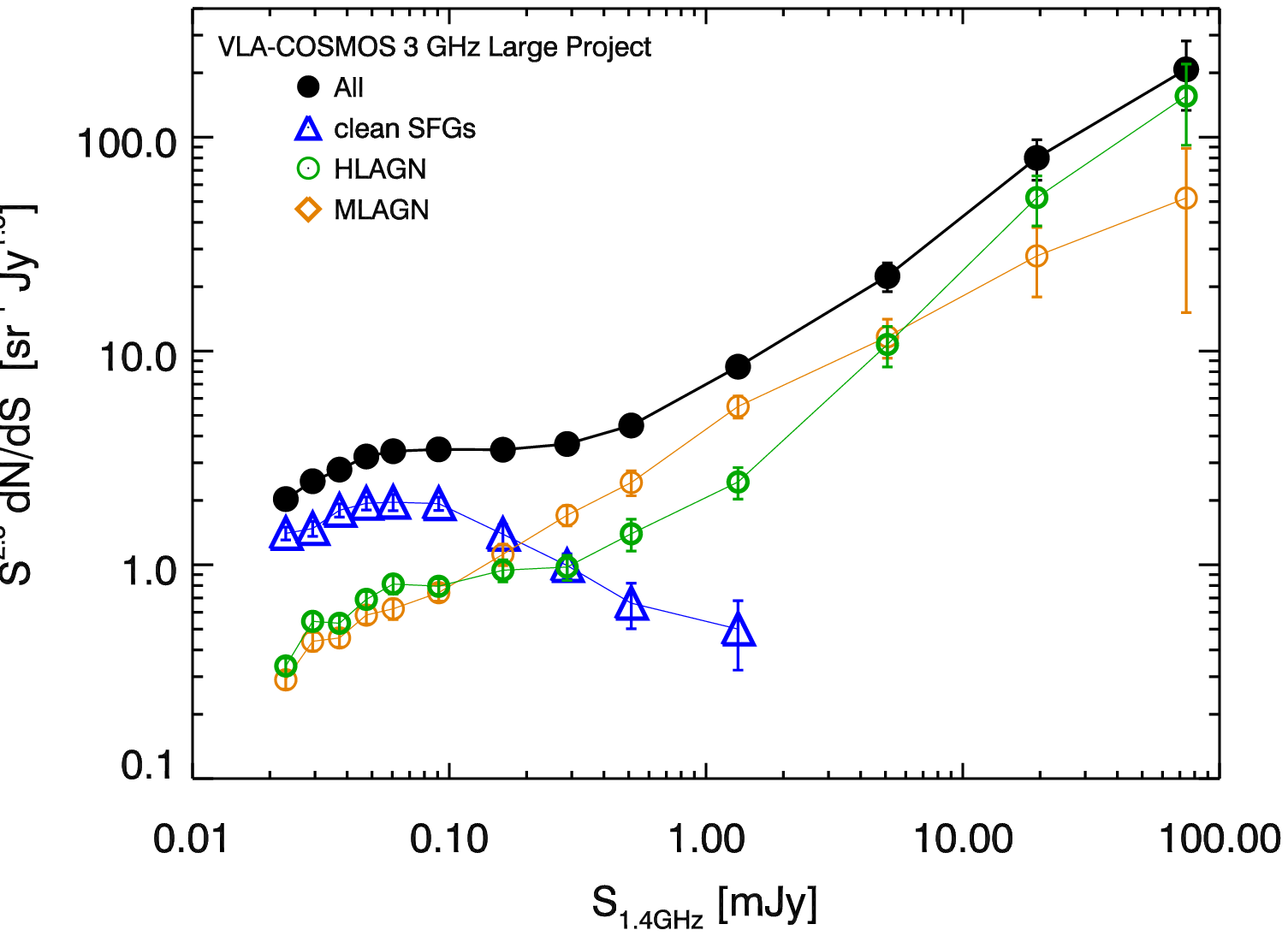}
	\includegraphics[bb= 0 0 432 322, scale=0.55]{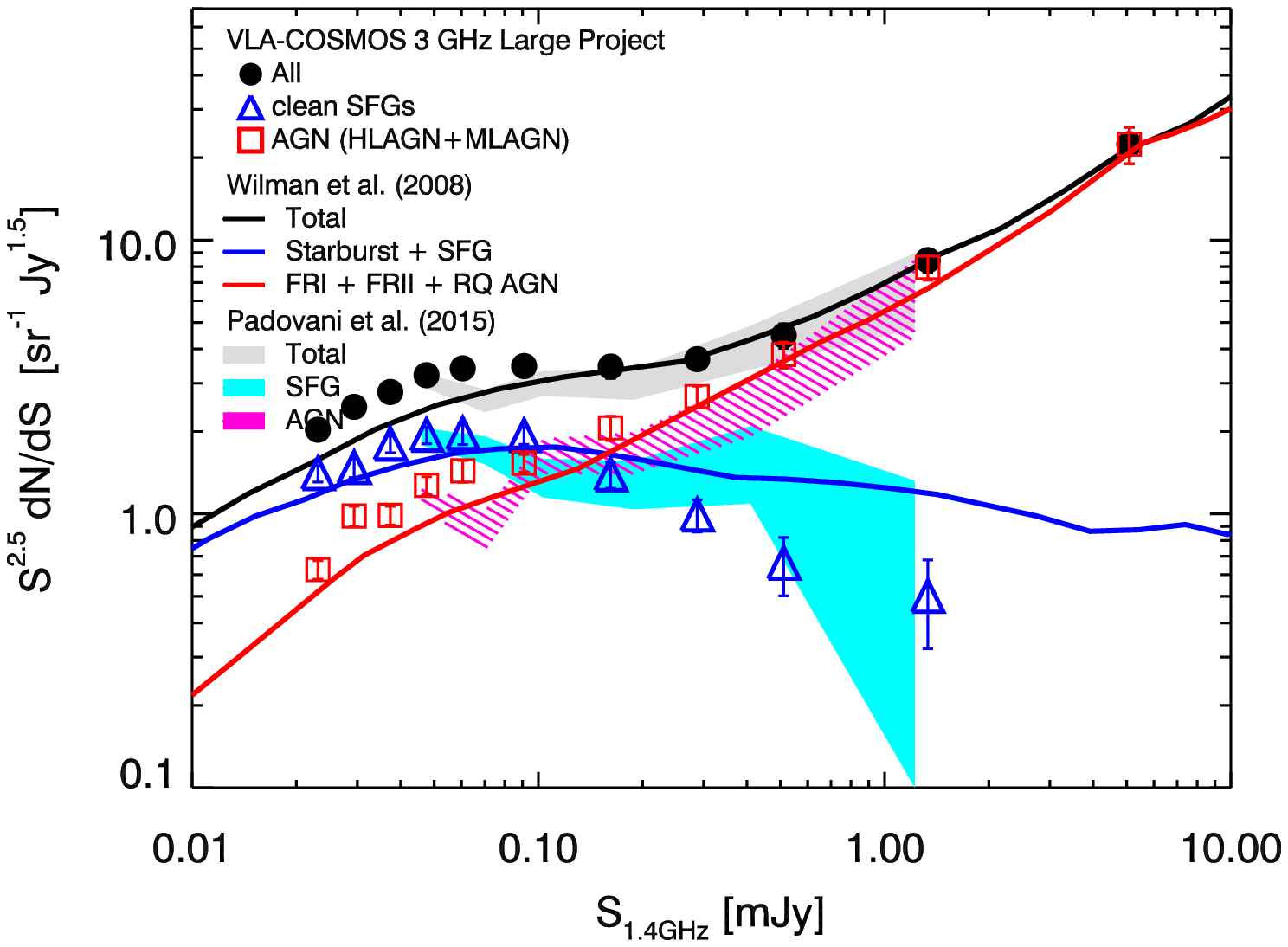}
	\includegraphics[bb= 0 10 432 322, scale=0.55]{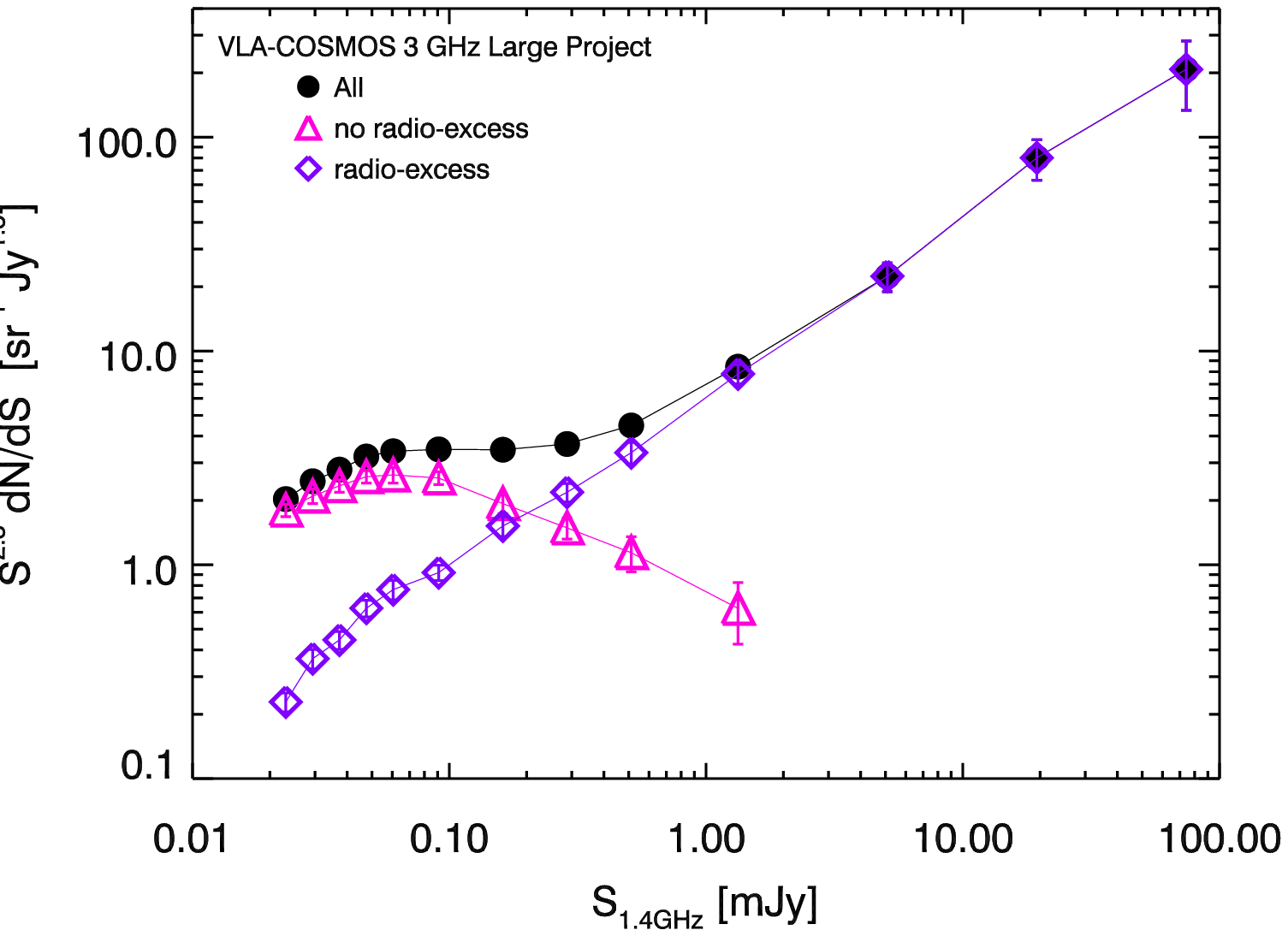}
	\includegraphics[bb= 0 10 432 322, scale=0.55]{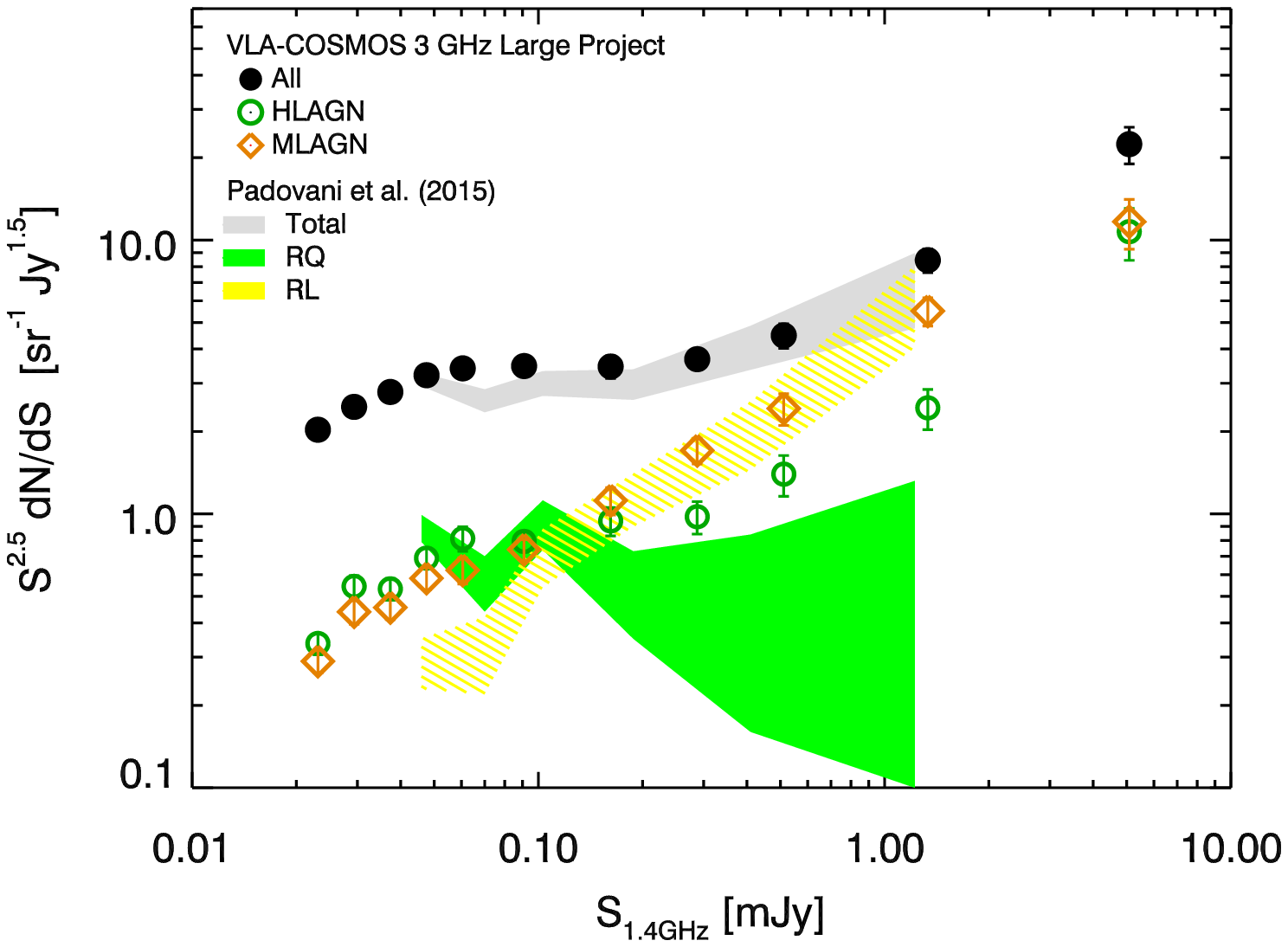}
	\caption{VLA-COSMOS 3~GHz Large Project radio source counts, separated into various (non-overlapping) populations (symbols, as indicated in the panels). The lines in the left panels connect the VLA-COSMOS 3 GHz Large project symbols. Lines and colored areas in the right hand-side panels show various source counts from the literature, as indicated in each panel. }
	\label{fig:Counts}
\end{figure*}

In Fig.~\ref{fig:Counts}, we show the 1.4~GHz source counts, normalized to Euclidean space, for all radio sources with counterparts in COSMOS2015 or $i$-band catalogs, and for each source class separately.  
The counts, {\bf tabulated in Tab.~\ref{tab:counts}}, have been corrected for completeness as described in \citet{smo17}, assuming all populations are equally complete at the same 3~GHz flux. 
From the left-hand panels in Fig.~\ref{fig:Counts}, showing the counts separated into clean SFG, MLAGN and HLAGN in the top panel and radio-excess and no-radio excess samples in the bottom panel, it is obvious that galaxies with star formation activity start dominating the counts below $S_\mathrm{1.4GHz}\sim100-150~\mu$Jy, while at higher fluxes the dominating population is related to AGN activity (see also top-right panel for the combined MLAGN and HLAGN contribution to the counts).

\begin{table*}
\begin{center}
\begin{tabular}{c|cc|cc|cc|cc|cc|cc|cccc}
& \multicolumn{2}{c|}{All} & \multicolumn{2}{c|}{AGN}& \multicolumn{2}{c|}{HLAGN}& \multicolumn{2}{c|}{MLAGN}& \multicolumn{2}{c|}{clean SFGs} & \multicolumn{2}{c|}{no radio-excess} & \multicolumn{2}{c}{radio-excess} \\
\hline
$S_\mathrm{1.4GHz}$   &count & error & count & error& count & error& count & error& count & error& count & error& count & error\\
$\mathrm{[mJy]}$ & \multicolumn{2}{c|}{[sr$^{-1}$\,Jy$^{1.5}$]} & \multicolumn{2}{c|}{[sr$^{-1}$\,Jy$^{1.5}$]}& \multicolumn{2}{c|}{[sr$^{-1}$\,Jy$^{1.5}$]}& \multicolumn{2}{c|}{[sr$^{-1}$\,Jy$^{1.5}$]}& \multicolumn{2}{c|}{[sr$^{-1}$\,Jy$^{1.5}$]} & \multicolumn{2}{c|}{[sr$^{-1}$\,Jy$^{1.5}$]} & \multicolumn{2}{c}{[sr$^{-1}$\,Jy$^{1.5}$]} \\
	\hline 
   0.02 &    2.03 &    0.13 &    0.63 &    0.05 &    0.34 &    0.03 &    0.29 &    0.03 &    1.40 &    0.10 &    1.80 &    0.12 &    0.23 &    0.02 \\ 
   0.03 &    2.46 &    0.19 &    0.98 &    0.08 &    0.54 &    0.05 &    0.44 &    0.04 &    1.48 &    0.12 &    2.10 &    0.16 &    0.36 &    0.04 \\ 
   0.04 &    2.79 &    0.19 &    0.99 &    0.08 &    0.53 &    0.05 &    0.45 &    0.04 &    1.80 &    0.13 &    2.34 &    0.16 &    0.44 &    0.04 \\ 
   0.05 &    3.21 &    0.21 &    1.27 &    0.10 &    0.69 &    0.06 &    0.58 &    0.05 &    1.94 &    0.14 &    2.58 &    0.17 &    0.63 &    0.06 \\ 
   0.06 &    3.40 &    0.28 &    1.44 &    0.13 &    0.81 &    0.08 &    0.62 &    0.07 &    1.96 &    0.17 &    2.63 &    0.22 &    0.77 &    0.08 \\ 
   0.09 &    3.47 &    0.23 &    1.53 &    0.11 &    0.79 &    0.07 &    0.74 &    0.06 &    1.93 &    0.14 &    2.55 &    0.18 &    0.92 &    0.08 \\ 
   0.16 &    3.45 &    0.33 &    2.06 &    0.21 &    0.94 &    0.11 &    1.12 &    0.13 &    1.39 &    0.15 &    1.93 &    0.20 &    1.52 &    0.16 \\ 
   0.29 &    3.67 &    0.30 &    2.68 &    0.24 &    0.98 &    0.13 &    1.70 &    0.18 &    0.99 &    0.13 &    1.49 &    0.17 &    2.19 &    0.21 \\ 
   0.51 &    4.48 &    0.46 &    3.82 &    0.42 &    1.40 &    0.24 &    2.43 &    0.32 &    0.66 &    0.16 &    1.14 &    0.21 &    3.34 &    0.39 \\ 
   1.33 &    8.44 &    0.84 &    7.94 &    0.81 &    2.44 &    0.41 &    5.50 &    0.65 &    0.50 &    0.18 &    0.63 &    0.20 &    7.82 &    0.80 \\ 
   5.09 &   22.40 &    3.42 &   22.40 &    3.42 &   10.73 &    2.30 &   11.67 &    2.41 &    -- &    -- &    -- &    -- &   22.40 &    3.42 \\ 
  19.42 &   80.08 &   17.17 &   80.08 &   17.17 &   52.22 &   13.73 &   27.85 &    9.95 &    -- &    -- &    -- &    -- &   80.08 &   17.17 \\ 
  74.16 &  207.79 &   74.20 &  207.79 &   74.20 &  155.84 &   64.10 &   51.95 &   36.82 &    -- &    -- &    -- &    -- &  207.79 &   74.20 \\ 
	\hline 
\end{tabular}
	\caption{
	{\bf
	Euclidean-normalized, and completeness-corrected 1.4 GHz radio source counts ($S^{2.5} dN/dS$ in units of sr$^{-1}$\,Jy$^{1.5}$)  for various galaxy populations, as indicated in the table (see Fig.~\ref{fig:Counts}).
	}
	}
	\label{tab:counts}
\end{center}
\end{table*}

\subsubsection{Comparison with radio source counts in the E-CDFS survey}
In the right-hand panels of Fig.~\ref{fig:Counts} we compare our  Euclidean-normalized source counts to those from \citet{padovani15}. They calculated the counts from 1.4~GHz VLA observations of the Extended \textit{Chandra} Deep Field South (E-CDFS), and in the figure the color-shaded areas correspond to their counts within the reported errors (Tab.~1 in \citealt{padovani15}). Their sample reaches a flux density limit of 32~$\mu$Jy, and covers  an area of approximately 0.3 deg$^2$. 
\citet{bonzini13} separated the $\sim800$ ($z\leq4$) radio sources within the E-CDFS into radio-loud (RL), radio-quiet (RQ) AGN, and star forming galaxies using the observed $24~\mu$m-to-1.4~GHz flux ratio ($q_\mathrm{24,obs}$). RL AGN were identified if,  at a given redshift, laying below the $2\sigma$ deviation from the average $q_\mathrm{24,obs}$, while RQ AGN were selected if being above the $2\sigma$ deviation, and at the same time fulfilling either X-ray or MIR diagnostics (similar to those used here)\footnote{Note that the $q_\mathrm{24,obs}$ ratio is defined in an inverse manner compared to the $\log {( L_\mathrm{1.4GHz} / \mathrm{SFR_{IR}} )}$ ratio used here.}. The remainder of the sample was taken as SFGs. Thus, while their classificiation scheme is largely similar to the one used here, it also contains significant differences. We refer to \citet[][their Sec.~4.3.2.]{delvecchio17}  for a detailed comparison of the two classification schemes. 

In the top-right panel of Fig.~\ref{fig:Counts} we, respectively, compare our SFG, and total AGN (MLAGN and HLAGN combined) counts with the SFG and total AGN (RQ and RL combined) counts from \citet{padovani15}. The counts show similar trends, with a consistent flux of $S_\mathrm{1.4GHz}\sim$100~$\mu$Jy below which the dominant population switches from AGN to star forming galaxies. While the SFG counts are consistent within the errors, the E-CDFS AGN counts show a slightly lower normalization  throughout the observed range. This could be due to cosmic variance given the smaller, 0.3 deg$^2$ E-CDFS area (compared to the 2 deg$^2$ COSMOS field area).
 
 In the bottom-right panel of Fig.~\ref{fig:Counts} we compare our MLAGN and HLAGN counts with the RL and RQ AGN counts from \citet{padovani15}. Due to  similarity of the AGN selection methods used here and by \citet{bonzini13}, our MLAGN (HLAGN) are expected to be a population comparable to the E-CDFS RL (RQ) AGN. 
The trends of  our MLAGN and the E-CDFS RL AGN populations are in agreement. However, while agreement also exists between the trends of our HLAGN and E-CDFS RQ AGN at $S_\mathrm{1.4GHz}<200~\mu$Jy, a discrepancy at higher fluxes is discernible, with systematically lower count values for the E-CDFS RQ AGN. This can be understood given the differences in the selections of E-CDFS RQ AGN and our HLAGN. While the first do not contain sources with radio-excess (defined by \citealt{bonzini13} as $>2\sigma$ in the redshift-dependent $q_\mathrm{24,obs}$, and using an M82 galaxy template), our sample of HLAGN contains 30\% of radio-excess sources (defined as $>3\sigma$ in the redshift-dependent distribution of $\log {( L_\mathrm{1.4GHz} / \mathrm{SFR_{IR}} )}$). Furthermore, the fraction of radio-excess sources in our HLAGN sample increases with increasing flux reaching $\sim50-100\%$ at $S_\mathrm{3GHz}\sim0.1-1$~mJy (corresponding to $S_\mathrm{1.4GHz}\sim0.17-1.7$~mJy assuming a spectral index of $-0.7$; see Fig.
~\ref{fig:radioexcessfraction}). Hence, the radio-excess sources within our HLAGN sample could explain the higher counts of our HLAGN relative to the E-CDFS RQ AGN at $S_\mathrm{1.4GHz}>0.2$~mJy.

\subsubsection{Comparison with the SKADS semi-empirical simulation}
In the top-right panel of Fig.~\ref{fig:Counts}, we compare our source counts for populations of star forming galaxies and AGN with the results of the semi-empirical simulations from \citet{wilman08}. These simulations are based on observed (or extrapolated) luminosity functions taking into account the measured large-scale clustering. \citet{wilman08} simulate classical radio-quiet quasars, FRI, and FRII sources, star forming and starburst galaxies. Since their classification is fundamentally different from ours,  we compare only the total number of AGN and star forming galaxies. We consider radio-quiet quasars, FRI and FRII sources defined by \citet{wilman08} to be AGN and compare the sum of their counts to the sum of our HLAGN and MLAGN counts. The overall agreement between the two AGN counts is good, even if \citet{wilman08} predictions are a bit lower than our AGN counts at flux densities   below $\sim500~\mu$Jy,  probably due to different population definitions, but also consistent with the overall trend of lower SKADS simulation counts, compared to ours. 
We consider the collective sample of star forming and starburst galaxies defined by \citet{wilman08} to be star forming galaxies, and compare the sum of their counts to the counts of our clean star forming galaxy sample. The prediction of \citet{wilman08}  follows the shape of our observed star forming galaxy counts below $\sim200~\mu$Jy. 

\subsection{Expectations for future surveys based on simple extrapolations}

\begin{figure*}
	\centering
	\includegraphics[bb= 0 -30 432 412, scale=0.55]{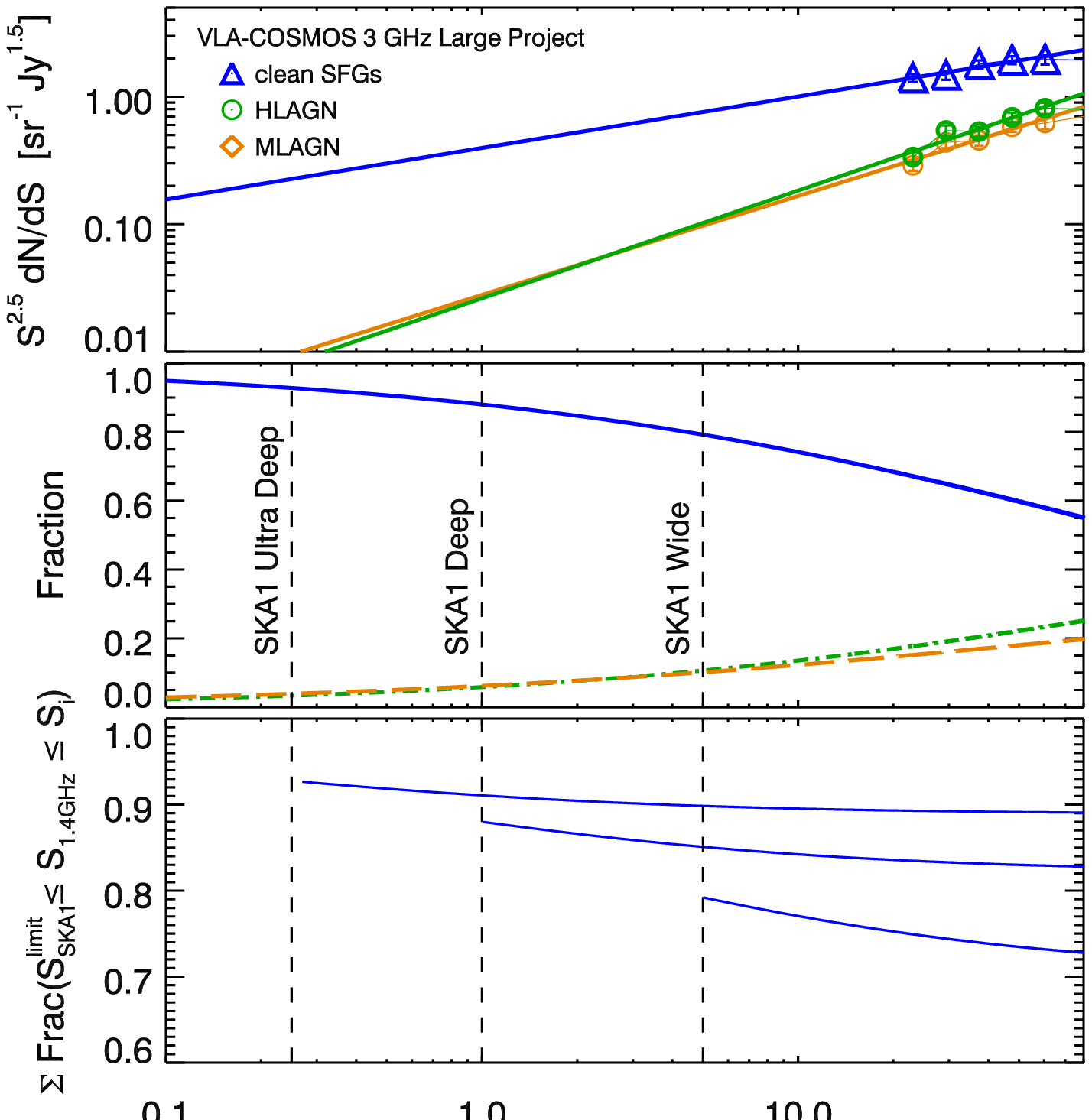}
	\includegraphics[bb= 0 -30 432 412, scale=0.55]{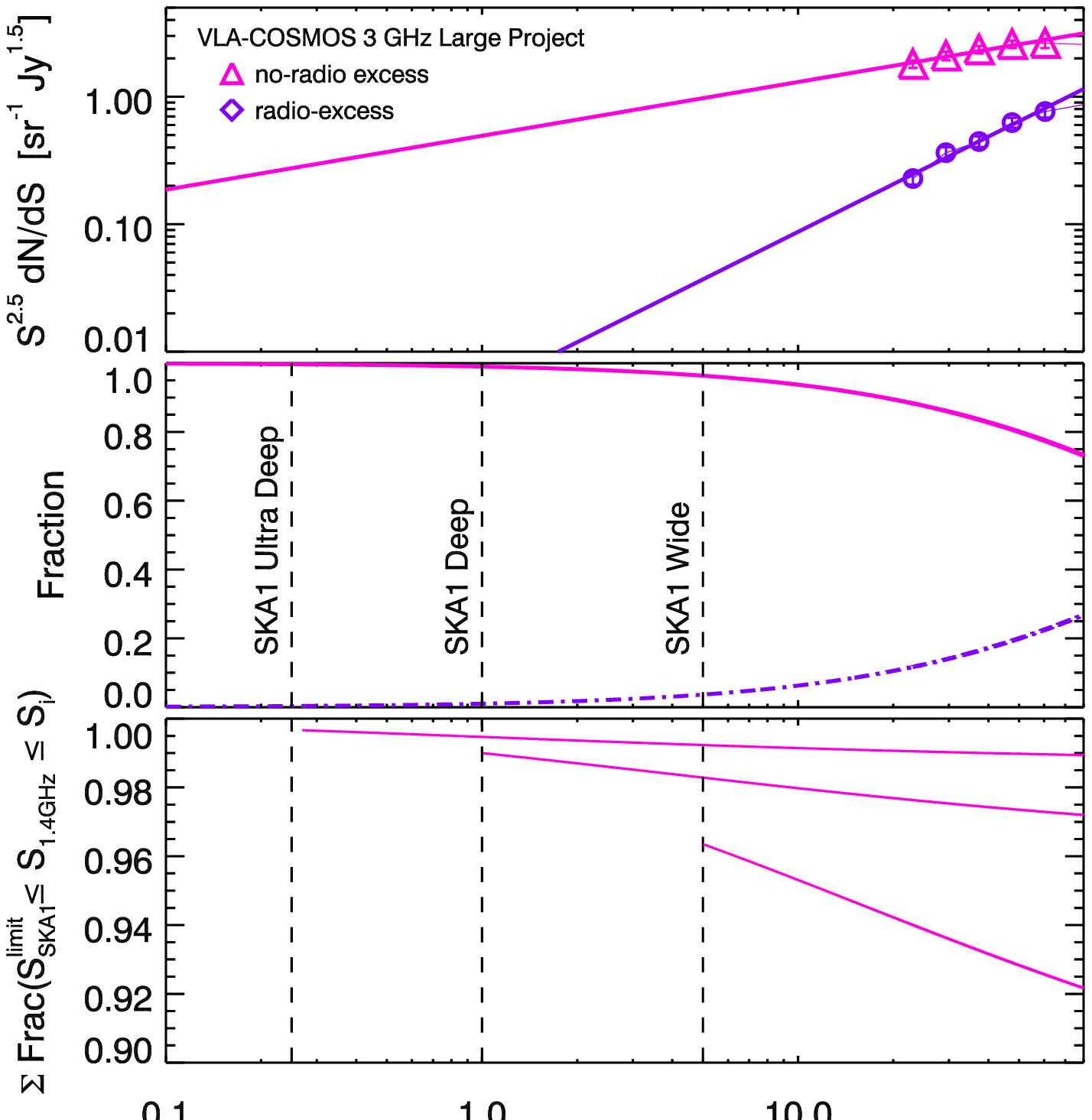}
	\caption{ {\em Top panels:} VLA-COSMOS 3~GHz Large Project radio source counts at $S_\mathrm{1.4GHz}<80~\mu$Jy for various, non-overlapping galaxy populations (symbols, as indicated in the top panels). Simple linear extrapolations (fit to the $S_\mathrm{1.4GHz}<80~\mu$Jy data) are shown by lines. {\em Middle panels: } The fractional contribution of the various populations as a function of 1.4~GHz flux, extrapolated down to the 5$\sigma$ detection limits of the SKA1 Ultra Deep, Deep, and Wide radio continuum surveys (vertical dashed lines; cf. Table 1 in \citealt{prandoni15}). {\em Bottom panels: } Cumulative fractions of clean SFGs (left), and no radio-excess sources (right) as a function of 1.4~GHz flux for each of the three SKA1 flux limits (shown by the three curves in each panel). For each given flux the $y$-axis value corresponds to the total fraction of the  population within the flux range limited at the low end by the $5\sigma$ limit of the corresponding SKA1 survey, and at the high end by the 
flux value given on the $x$-axis.
	 }
	\label{fig:CountsExtrap}
\end{figure*}

In Fig.~\ref{fig:CountsExtrap} we show the faint end of our source counts,  with two different definitions for non-overlapping populations (left and right panels), and extrapolations down to radio fluxes of $\sim$10~nJy, encompassing the  detection limits of SKA1 Ultra Deep, Deep, and Wide surveys \citep{prandoni15}. In a zero-order attempt to extrapolate the counts, not intended to be related to any physical model, we use a simple linear fit to the four faintest data points   ($S_\mathrm{1.4GHz}<80~\mu$Jy), for each of the various radio-detected subpopulations. As discernible from the middle panels in the figure, down to these fluxes the expected fractions of MLAGN and HLAGN (radio-excess sources) drop monotonously, while that of SFGs (no radio-excess sources) monotonously rises towards unity.
In the bottom panels of Fig.~\ref{fig:CountsExtrap} we show the cumulative fractions of clean SFGs, and no radio-excess sources as a function of 1.4~GHz flux for each of the three SKA1 flux limits. This was computed as the total fraction of the population within the flux range limited at the low end by the $5\sigma$ limit of the corresponding SKA1 survey, and at the high end by the given flux value. For example, selecting radio-detected sources with  $S_\mathrm{1.4GHz}<80~\mu$Jy  
(corresponding to $S_\mathrm{3GHz}<48~\mu$Jy, assuming $\alpha=-0.7$)
 will yield a sample consisting of about 72\%, 83\%, 89\% SFGs  in the planned SKA1 Wide, Deep, and Ultra Deep surveys, respectively. Selecting sources with $S_\mathrm{1.4GHz}<10~\mu$Jy 
(corresponding to $S_\mathrm{3GHz}<6~\mu$Jy, assuming $\alpha=-0.7$)
will result in samples consisting of about 76\%, 83\%, 89\% SFGs in the three SKA1 surveys, respectively.  Considering only objects with no radio-excess these numbers are about 92\%, 97\%, 99\% and 96\%, 98\%, 99\% for upper flux limits of 80 and 10~$\mu$Jy, respectively.  Hence, it can be expected that at the faint flux limits to be reached by the SKA1 surveys, a selection based only on radio flux limits can provide a simple tool to efficiently identify samples highly dominated by SFGs.

\section{Summary}
\label{sec:summary}

Using one of the deepest radio datasets so far, the VLA-COSMOS 3~GHz Large Project, down to a flux limit of $\sim$11~$\mu$Jy at 5$\sigma$, we associated optical/NIR/MIR counterparts to radio sources. Countepart candidates were drawn from three different catalogs, COSMOS2015, $i$-band and IRAC, prioritizing them in that order. A false match probability was calculated for each counterpart candidate using a background model designed to reproduce realistic magnitude and spatial distribution of counterparts to radio sources. We select as reliable counterparts those counterpart candidates with false match probability below 20\% in at least one of the three catalogs. In total, we find \nTotCpts \ counterparts to our 3~GHz radio sources  over a common unmasked area of 1.77 deg$^2$, with an estimated fraction of spurious identifications of $\sim2\%$.

We analyzed the multi-wavelength star formation- and AGN-related properties of our radio-selected population,  reaching out to a redshift of $z\sim6$. Based on the multi-wavelength properties, including the estimated bolometric (i.e. radiative) AGN luminosities, we separate our radio sources into subsamples of clean star forming galaxies, and those of radio AGN  into low-to-moderate, and moderate-to-high radiative luminosity radio AGN, and consider these AGN subpopulations to be candidates for high-redshift analogues to local low- and high-excitation emission line AGN, respectively. An alternative separation is that into subsamples of galaxies with and without radio-excess, where the radio-excess threshold is defined as a $3\sigma$ excess in radio emission relative to the IR-based star formation rates in their host galaxies.

An analysis of fractional contributions of various galaxy types as a function of radio flux shows that below $\sim$100~$\mu$Jy at 1.4~GHz  star-forming (no radio-excess) galaxies constitute the dominant fraction of the radio population, with a $\sim60$\% (80\%) contribution at $S_\mathrm{1.4GHz}\sim50~\mu$Jy. At higher flux density levels the fractional contribution of star-forming (no radio-excess) galaxies diminishes, and AGN (radio-excess galaxies) constitute the dominant fraction.

The Euclidean normalized number counts of different observed populations are generally in agreement with the number counts of similar (observed or simulated) populations from the literature. Using simple extrapolations of our faint-end ($S_\mathrm{1.4GHz}<80~\mu$Jy) counts we estimate the differential and cumulative fractions of star forming galaxies and galaxies with no radio-excess. These suggest that at the faint flux limits the planned (Wide, Deep and UltraDeep) SKA1 surveys will reach, a selection based only on radio flux limits can provide a simple tool to efficiently identify samples highly dominated by SFGs.

\begin{acknowledgements}
	Zgal group: European Union's Seventh Framework program under grant agreement 333654 (CIG, 'AGN feedback'; N.B., V.S.) and grant agreement 337595 (ERC Starting Grant, 'CoSMass'; V.S., J.D., M. N.). CL acknowledges the funding by a Discovery Early Career Researcher Award (DE150100618). Partially based on data obtained with the European Southern Observatory Very Large Telescope, Paranal, Chile, under Large Programs 175.A-0839 (the zCOSMOS Spectroscopic Redshift Survey) and  185.A-0791 (the VIMOS Ultra Deep Survey). AK and EV acknowledge support from the collaborative research center 956 (sub-project A1) funded
by the Deutsche Forschungsgemeinschaft (DFG)
\end{acknowledgements}

	\bibliographystyle{aa}
	\bibliography{biblio}

\appendix

\section{Associating counterparts to single-component radio sources}
\label{sec:method}

We here describe in detail the procedure used to associate optical-MIR counterparts to our 3~GHz, single-component radio sources.

\subsection{Separation between positions of radio and optical-MIR sources}
\label{sec:methodRadius}

\begin{figure}
	\centering
	\includegraphics[bb = 140 150 800 700, width=0.495\linewidth]{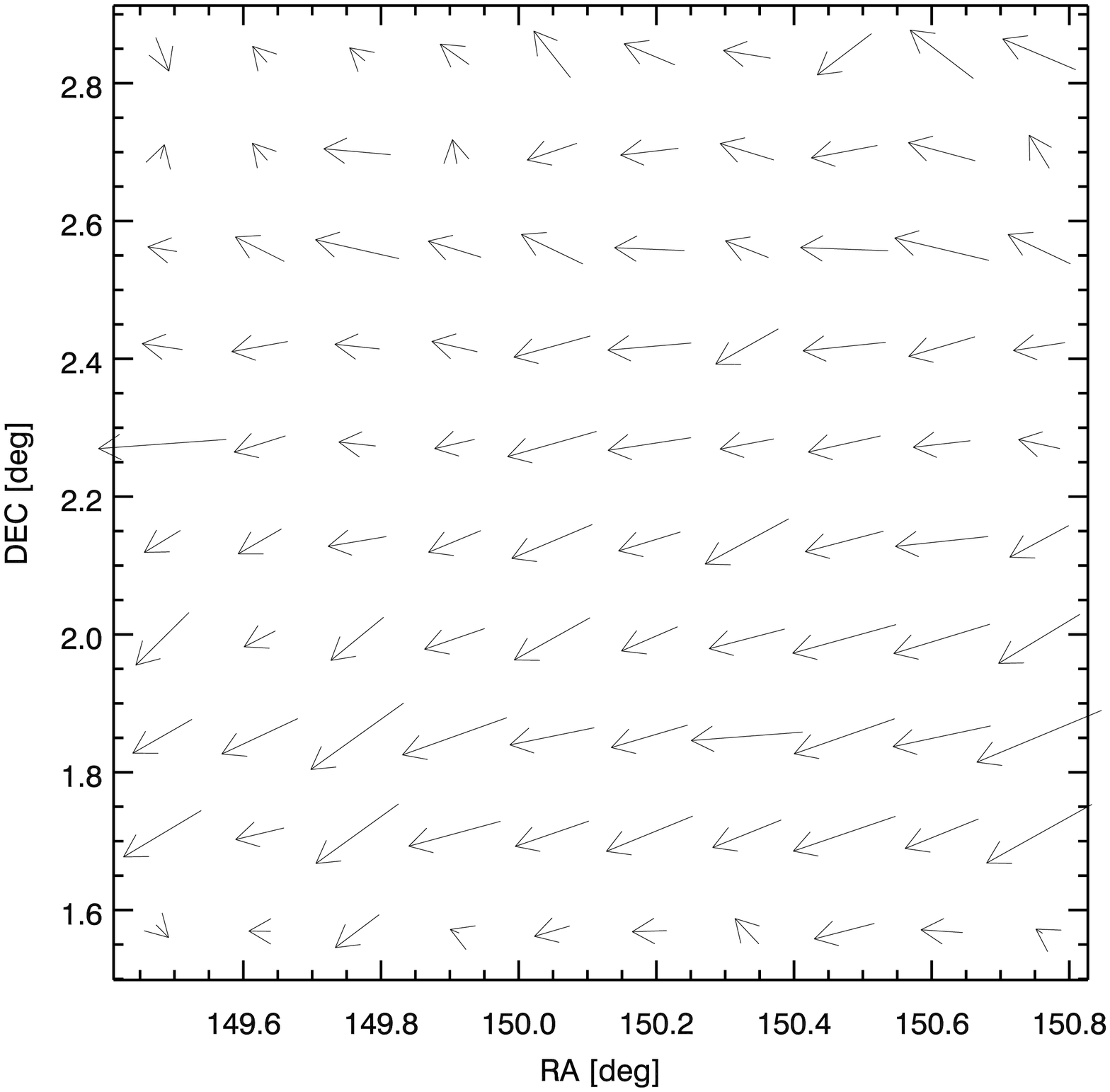}
	\includegraphics[bb = 140 150 800 700, width=0.495\linewidth]{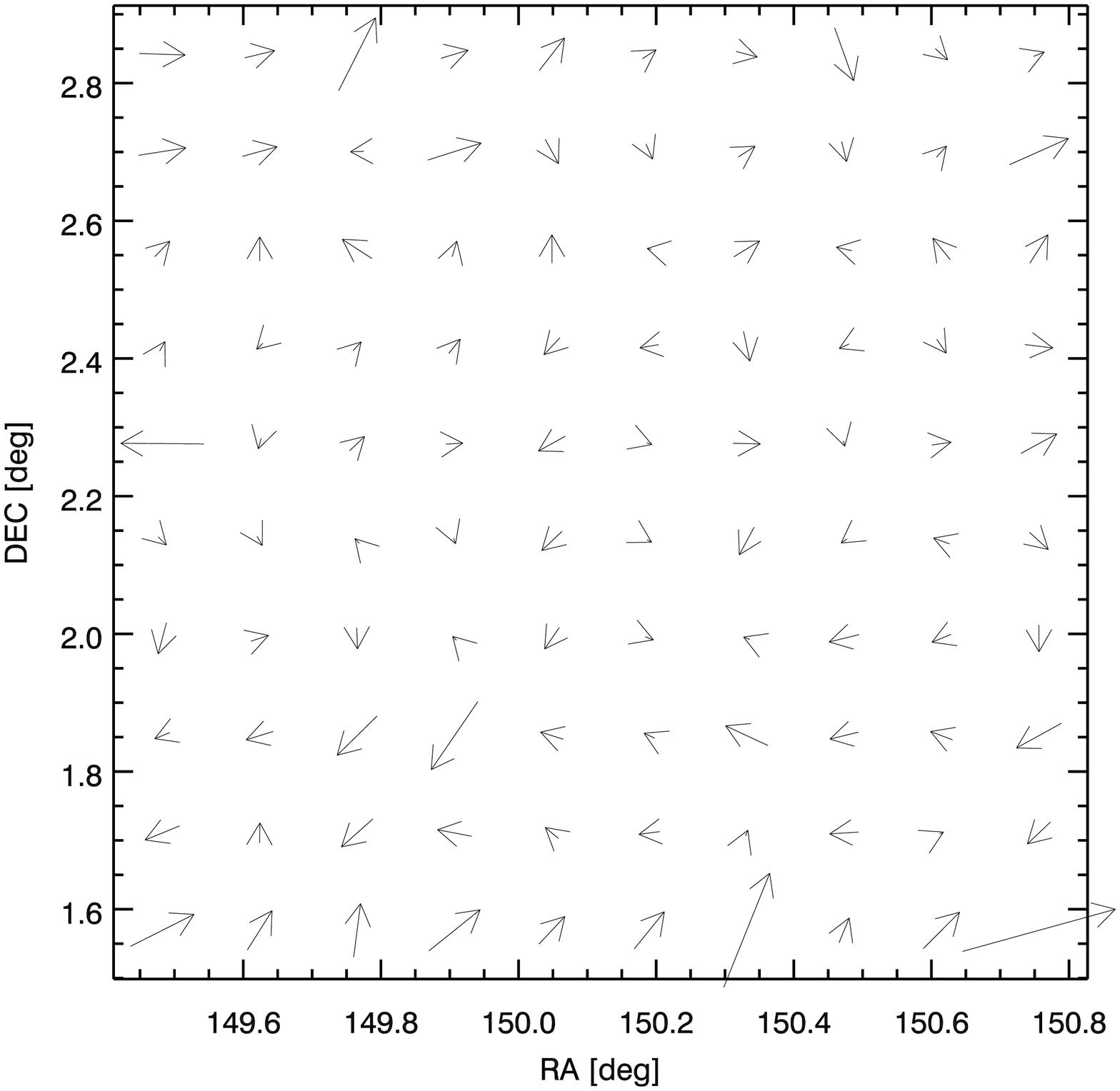}\\
	\includegraphics[bb = 140 150 800 800, width=0.495\linewidth]{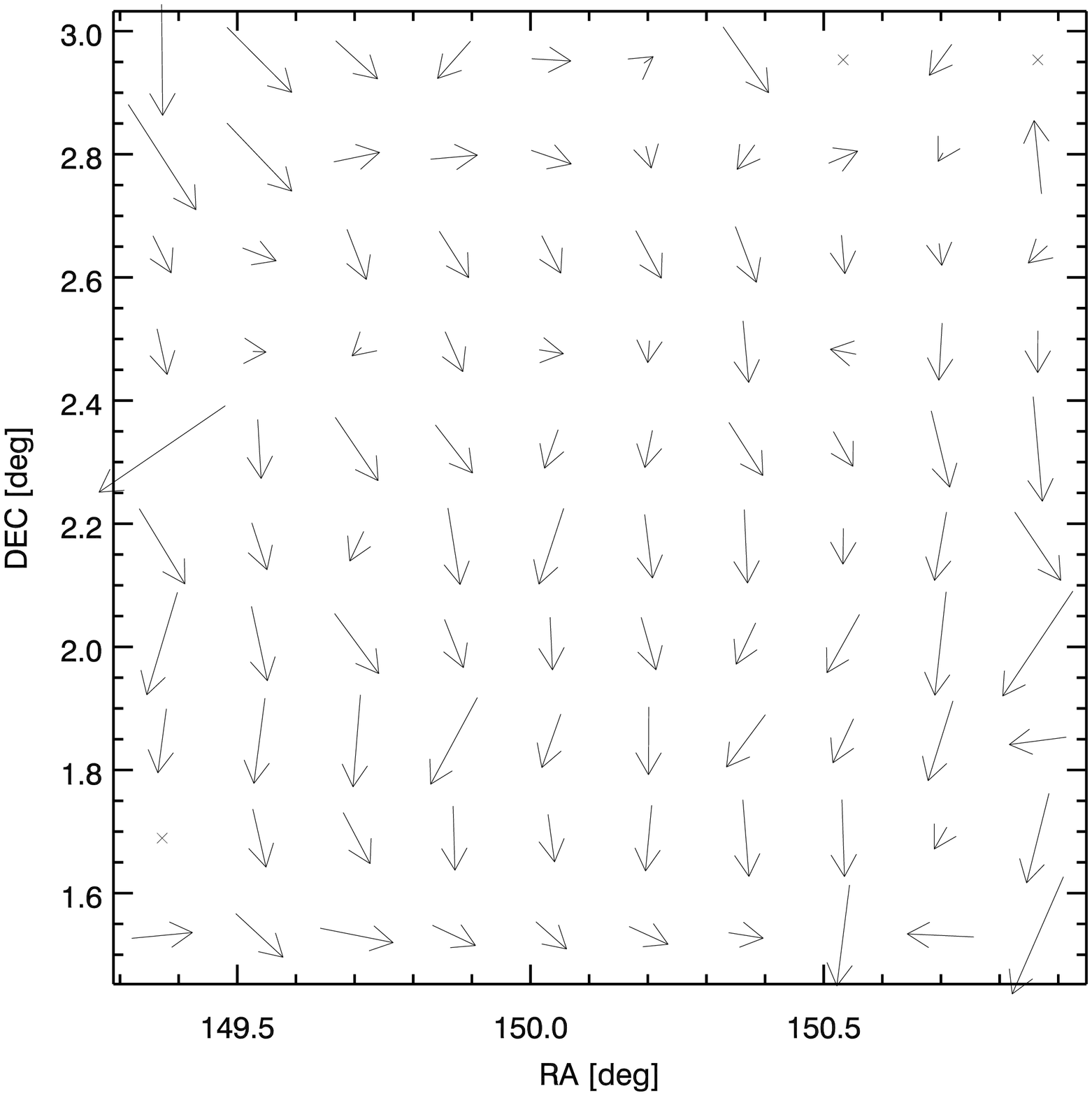}
	\includegraphics[bb = 140 150 800 800, width=0.495\linewidth]{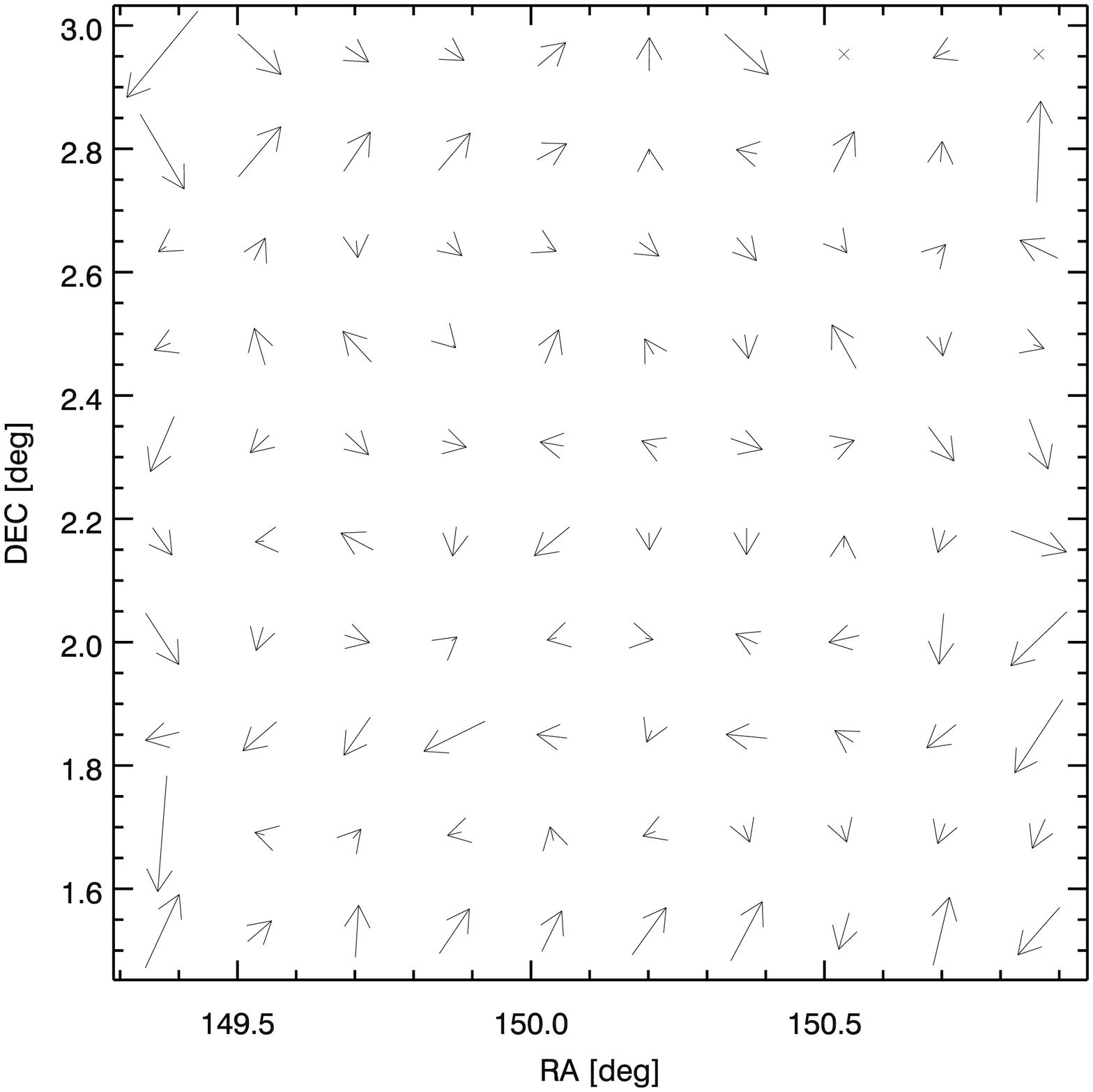}\\
	\includegraphics[bb = 140 150 800 800, width=0.495\linewidth]{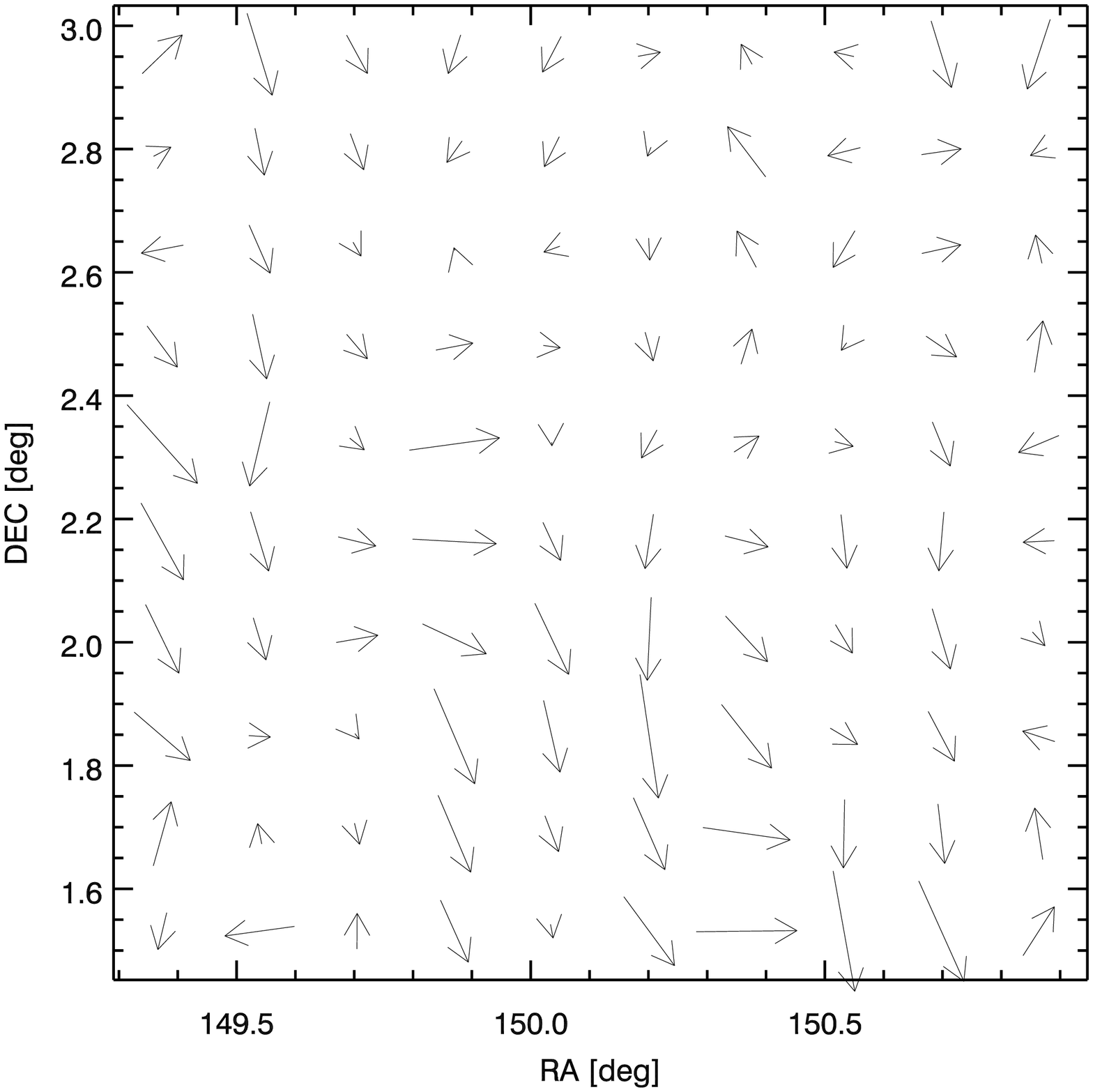}
	\includegraphics[bb = 140 150 800 800, width=0.495\linewidth]{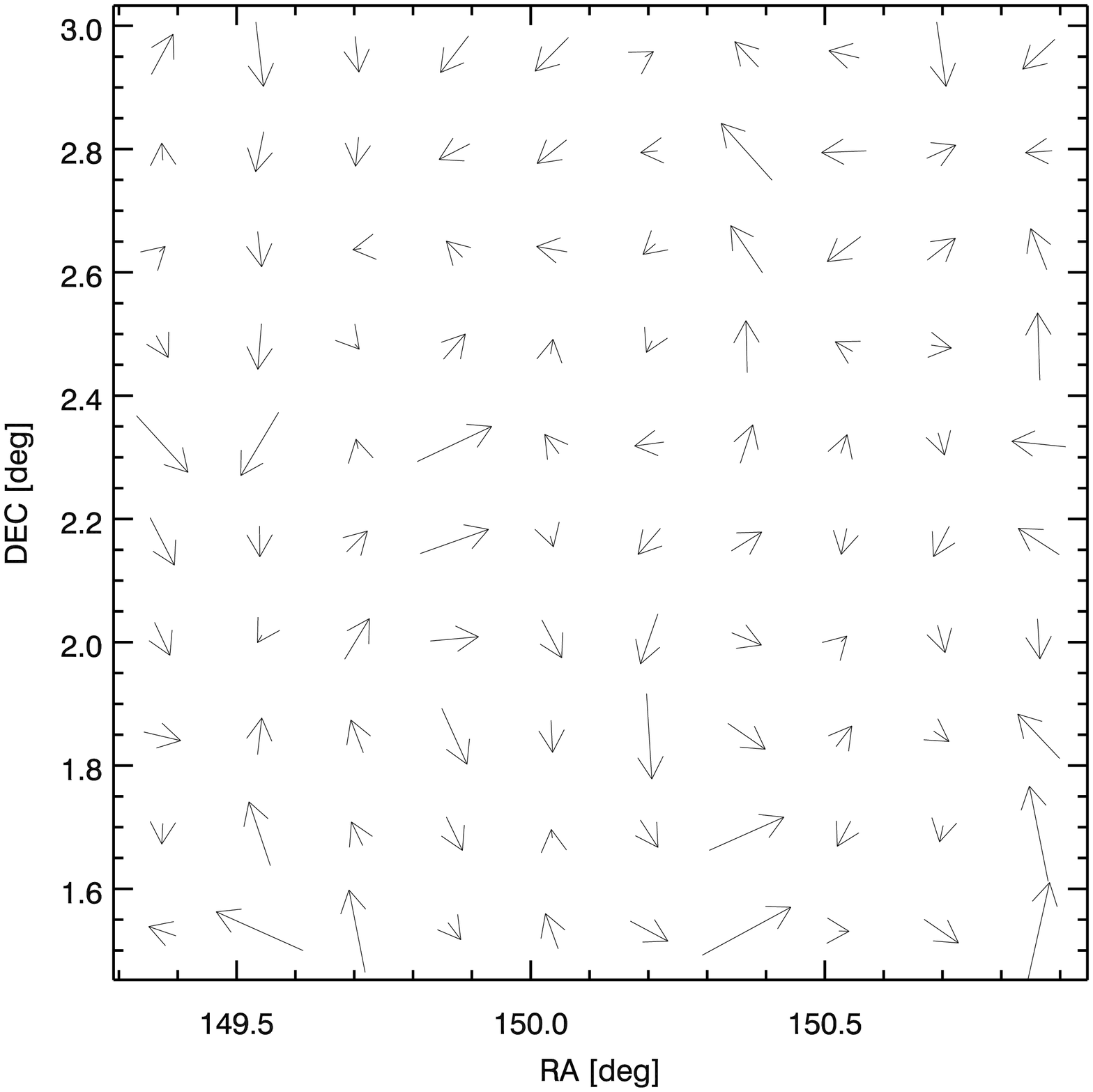}
	\caption{Systematic positional offsets in the (RA, Dec) plane between the VLA-COSMOS 3~GHz Large Project  and the COSMOS2015 (top panels), i-band (middle panels), and IRAC (bottom panels)  sources before (left panels), and after corrections (right panels). To obtain the corrections the $1.4^\circ\times1.4^\circ$ COSMOS field was divided into 100 bins 0.14$^\circ$ on the side. In each bin, a Gaussian was fitted to the separation distribution in right ascension and declination. The means of these two Gaussians are shown as the vectors shown in the plot. For visibility, the sizes of the vectors are multiplied by a factor 3,600. }
	\label{fig:sepVectors}
\end{figure}

Positional matching of the VLA-COSMOS 3~GHz Large Project (\smo et al., accepted) sources with those in the COSMOS2015 \citep{laigle16},  $i$-band \citep{capak07}, and 3.6~$\mu$m IRAC \citep{sanders07} selected catalogs reveals small systematic positional offsets, listed in Table~\ref{tab:meanOffsets}, which also depend on the position on the sky (see left panels of Fig.~\ref{fig:sepVectors}). Thus, prior to associating optical-MIR counterparts to our radio sources, we correct the multi-wavelength catalog positions for these astrometric offsets via a linear fit to the mean (radio-optical, radio-NIR or radio-MIR) astrometric offsets as a function of RA and Dec (see the right panels of Fig.~\ref{fig:sepVectors}).
\begin{table}
\begin{center}
\begin{tabular}{|c|c|c|}
	\hline 
	Catalogue    & $\left< \Delta RA \right>$ 	& $\left< \Delta Dec \right>$ \\ 
	\hline 
	COSMOS2015 & $\sim-$0.09\arcsec  & $\sim-$0.014\arcsec\\ 
	$i$-band   & $\sim-$0.06\arcsec  & $\sim-$0.011\arcsec\\ 
	IRAC 3.6~$\mu$m  & $\sim$0.02\arcsec   & $\sim-$0.04\arcsec\\ 
	\hline 
\end{tabular}
	\caption{The mean offsets in RA and Dec separations for sources matched between the VLA-COSMOS 3~GHz Large Project catalog and the COSMOS2015, the $i$-band and the IRAC 3.6~$\mu$m selected catalogs.}
	\label{tab:meanOffsets}
\end{center}
\end{table}

 The best-fit expressions to determine the corrected optical-MIR positions (RA', Dec') for each counterpart catalog are given below.

 COSMOS2015 catalog:
\begin{eqnarray}
 RA' = RA + ((-0.041 \times RA + 6.059) / 3600.)\\
 Dec' = Dec + ((0.058 \times Dec - 0.147) / 3600.)
\end{eqnarray}

 $i$-band catalog:
\begin{eqnarray}
 RA' = RA + ((-0.043\times RA + 6.408) / 3600.)\\
 Dec' = Dec + ((0.048\times Dec - 0.179) / 3600.)
\end{eqnarray}

 IRAC catalog:
\begin{eqnarray}
 RA' =  RA + ((-0.010\times RA + 1.500) / 3600.)\\
 Dec' = Dec + ((0.055\times Dec - 0.164) / 3600.)
\end{eqnarray}

Next, to choose the limiting counterpart search radius we compare the number of radio -- opt/NIR/MIR nearest neighbors of our radio sources (in the optical, NIR-, and MIR-selected catalogs) as a function of search radius with that obtained from a random position distribution. This is shown in Fig.~\ref{fig:blocking} where the number of matches with random positions is represented by the average of 10 randomizations of each of the optical, NIR-, and MIR-selected catalogs. 

\begin{figure} 
\centering
\includegraphics[ bb=0 40 907 450, width=1\linewidth]{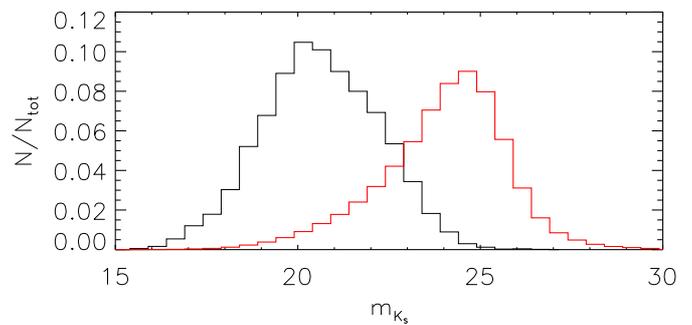}
\caption{ Normalized $K_S$ magnitude distribution of 3~GHz source counterpart candidates from the COSMOS2015 catalog (black), and all sources in the  unmasked regions of COSMOS2015 with $K_S$ band detection (red).  The magnitudes are formal  (positive flux) values derived at the position of objects identified in the $z^{++}YJHK_s$ stack.
}
\label{fig:ksdistro}
\end{figure}

\begin{figure} 
\centering\includegraphics[bb=0 0 432 270, width=\columnwidth]{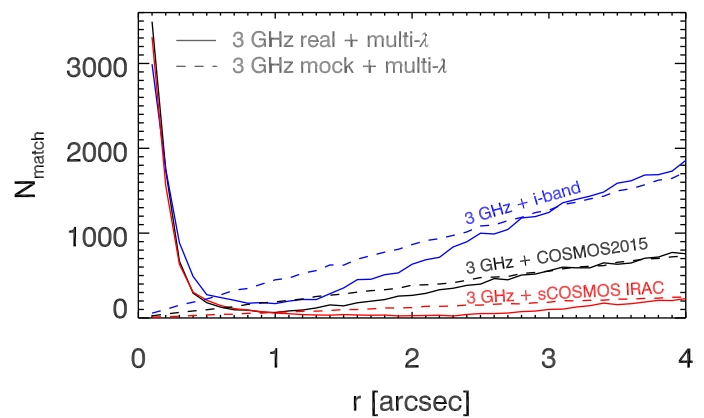}
\caption{Number of all matches within $0.1\arcsec$ wide annuli, between the VLA-COSMOS 3~GHz Large Project sources and the COSMOS2015 (black), the $i$-band selected (blue), and the IRAC 3.6~$\mu$m selected catalogs (red) (solid lines). Dashed lines represent the the average number of matches in the three above-mentioned catalogs around randomized positions. The average was calculated using 10 different randomizations.}
\label{fig:blocking}
\end{figure}

At small separations ($\textless$1\arcsec) the peak in the number of matches around the positions of the radio sources is due to the large number of true counterparts. At large separations ($\textgreater$2\arcsec - 3\arcsec) the numbers of matches around the positions of the radio sources follow a linear relation, consistent with that observed around random positions. Note that the different slopes of these linear relations correspond to the different surface density of sources in the three different catalogs and at these separations there is no excess of counts with respect to what is observed around random positions. At intermediate separations the number of associations around the positions of the radio sources is smaller than that observed around random positions and this number reaches a minimum at separations  of $\sim$1-1.5$\theta$, where  $\theta$ is the FWHM of the PSF ($\theta\sim$0\farcs8 for COSMOS2015/$i$-band, and $\theta\sim$1\farcs66 for IRAC).
This deficiency in the number of  matches with the radio sources, relative to the number of matches with the randomized position catalog (at $r\sim0.7-3\arcsec$ for COSMOS2015 and the $i$-band selected catalogs, spanning up to 4\arcsec\ for the IRAC catalog) is due to the fact that fainter optical/NIR/MIR sources close to the brighter counterparts of radio sources (see Fig.~\ref{fig:ksdistro}) at smaller separations can not be easily detected and most of them are therefore missing in the 
catalogs (i.e., a blocking effect).
Although Fig.~\ref{fig:blocking} appears to suggest, at face value, that most of the counterparts between 0\farcs4 and 0\farcs8 are not associated to radio sources, the magnitude distribution of these counterparts as a function of separation suggests instead that many of those sources, especially the brightest ones, are real associations, as will be discussed in the next section.

\subsection{False match probability}
\label{sec:methodFMP}
To estimate the reliability of the radio-optical/NIR/MIR matches and the false match probability ($p_\mathrm{false}$), we have created a background model that reproduces the blocking effect of optical, NIR and MIR sources around radio sources. To simulate the $i$-band and IRAC background, their selection band was used ($i$, and 3.6~$\mu$m respectively), while for the COSMOS2015 background model the 3.6~$\mu$m  band, measured for a large majority of the sources, was chosen.
First, we divide each sample of $N$ counterpart candidates (COSMOS2015, $i$-band and IRAC) into $N=N_m+N_u$, where $N_m$ ($N_u$) is the number of closest counterparts found within 0.5$\theta$ ($\langle0.5,1.5]\theta$) from the radio source position. For all three catalogs $N_m$ and $N_u$ are $\sim$85-90\% and $\sim$10-15\% of $N$,  respectively. As said above, it is this large number of relatively bright counterparts close to the radio positions that artificially reduces the number of sources in the optical, NIR and MIR catalogs at separations larger than 0.5$\theta$. To reproduce this effect, from each catalog, we select $N_m$ sources requiring them to follow the magnitude distribution of real counterparts matched using a 0.5$\theta$ search radius; then, starting from the positions of these randomly selected sources, we set the centre of our search region by moving it in a random direction following the separation distribution of real counterpart candidates matched using a 0.5$\theta$ search 
radius. Thus, we simulate $N_
m$ sources whose magnitudes and separations fully reflect those of the real sample of counterparts within the 0.5$\theta$ search radius. Finally, we randomly distribute the remaining $N_u$ sources over the entire radio field. We generate 100 different mock (background) catalogs following this recipe, and extract the expected background magnitude distribution from the counterparts to these mock sources. 

The false match probability as a function of magnitude was then calculated in  three separation bins as the ratio between the distributions of the simulated background and real magnitudes. We then linearly interpolate the histogram of this ratio, and assign to each source the false match probability that this interpolation has at the value of the magnitude of the source. This is illustrated in Fig.~\ref{fig:falseMatch}, and summarized in Fig.~\ref{fig:fmp}. Details on matching with the COSMOS2015, the $i$-band and the IRAC sources, as well as the false match probability estimation, are given in the following sections. For sources without the measured magnitude in the band used for the background estimation, false match probabilities were not calculated, and their values were set to 1 (see Sec.~\ref{sec:crossmatching}, and \ref{sec:crossmatching_iband}). 

\subsection{Cross-matching with the COSMOS2015 sources: NIR counterpart candidates}
\label{sec:crossmatching}
\begin{figure} 
\centering
	\includegraphics[bb =  0 -60 432 432, scale=0.23]{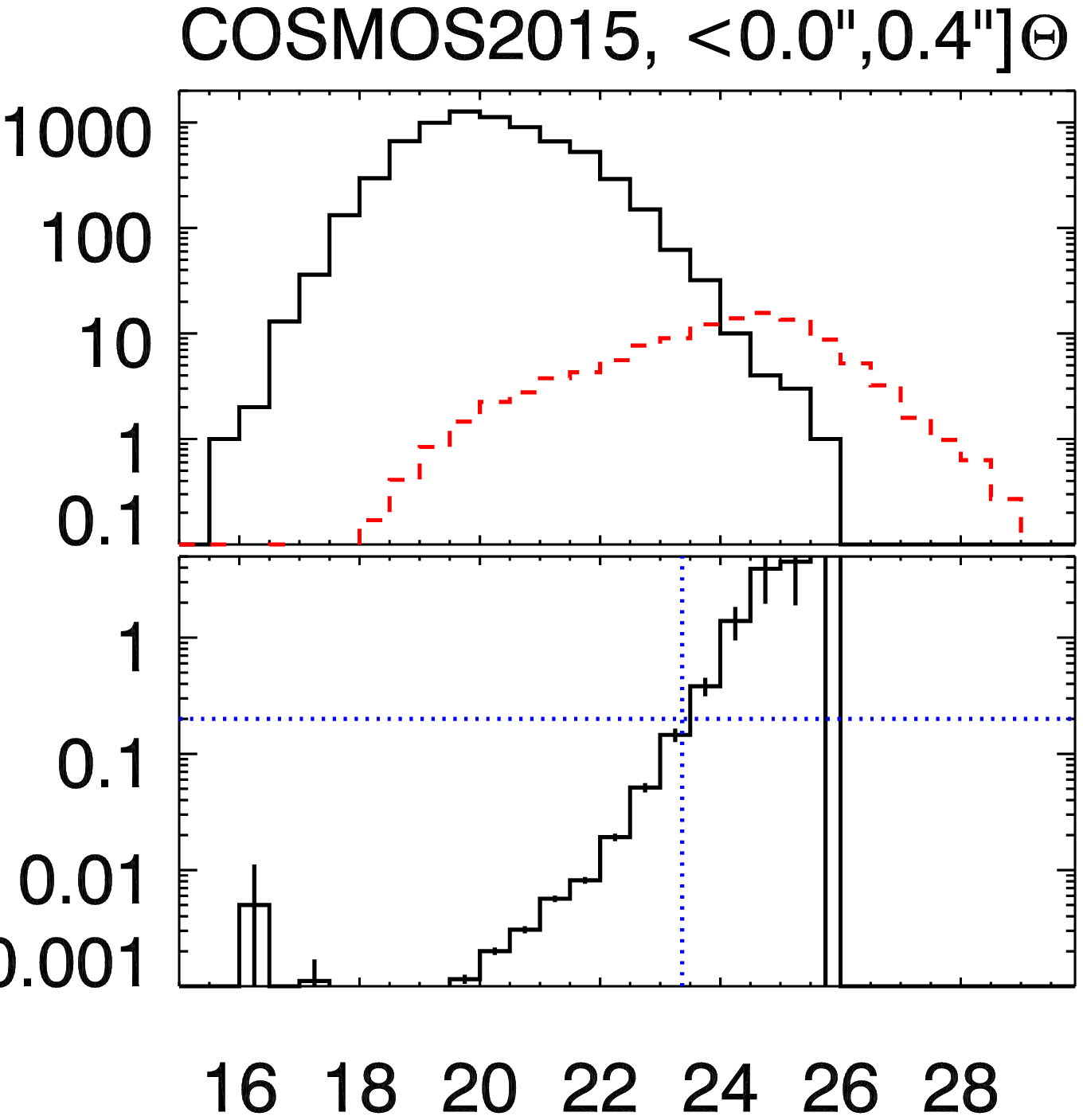}	
	\includegraphics[bb =  108 -60 432 432, scale=0.23]{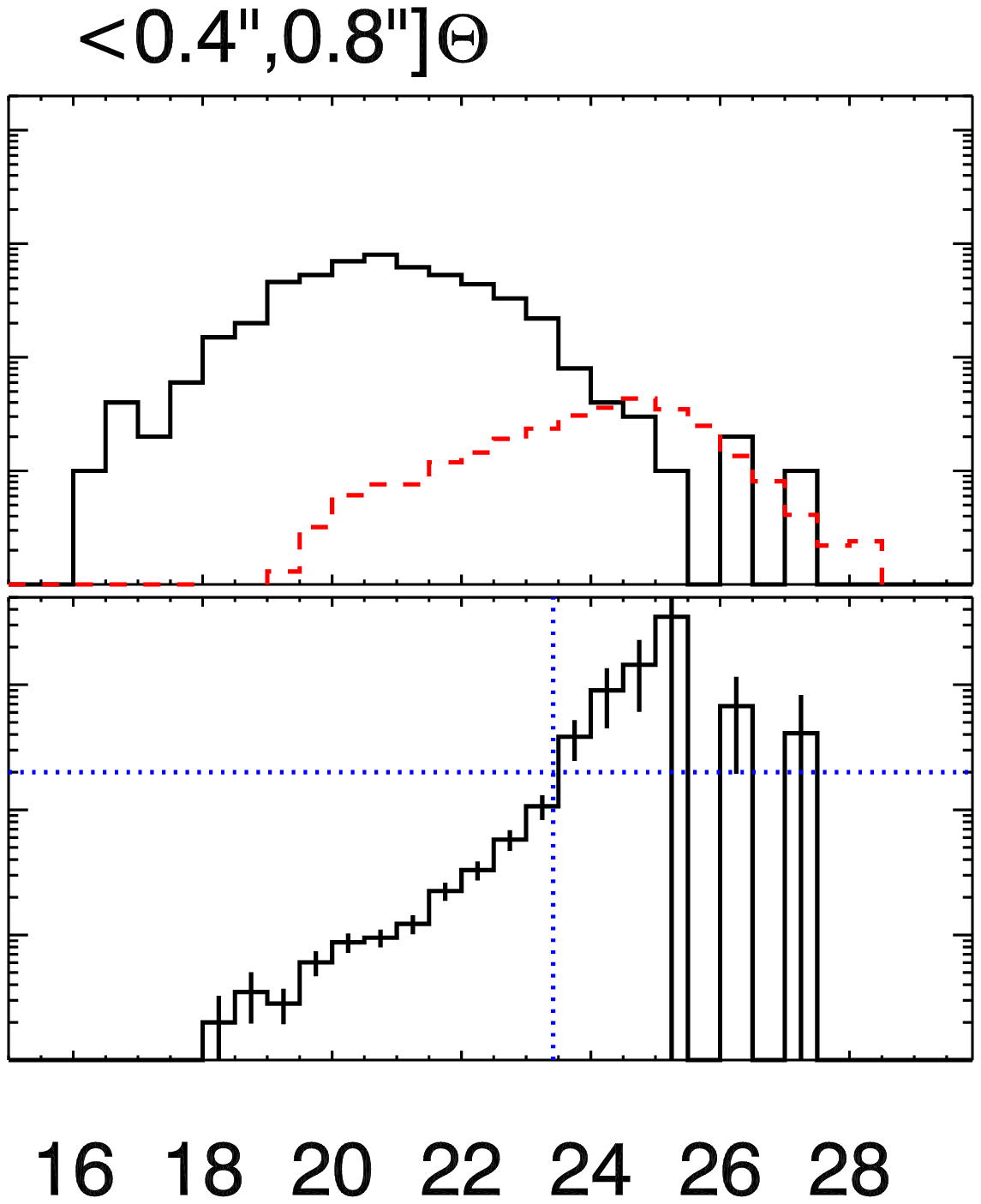}	\\
	\includegraphics[bb =  0 -60 432 432, scale=0.23]{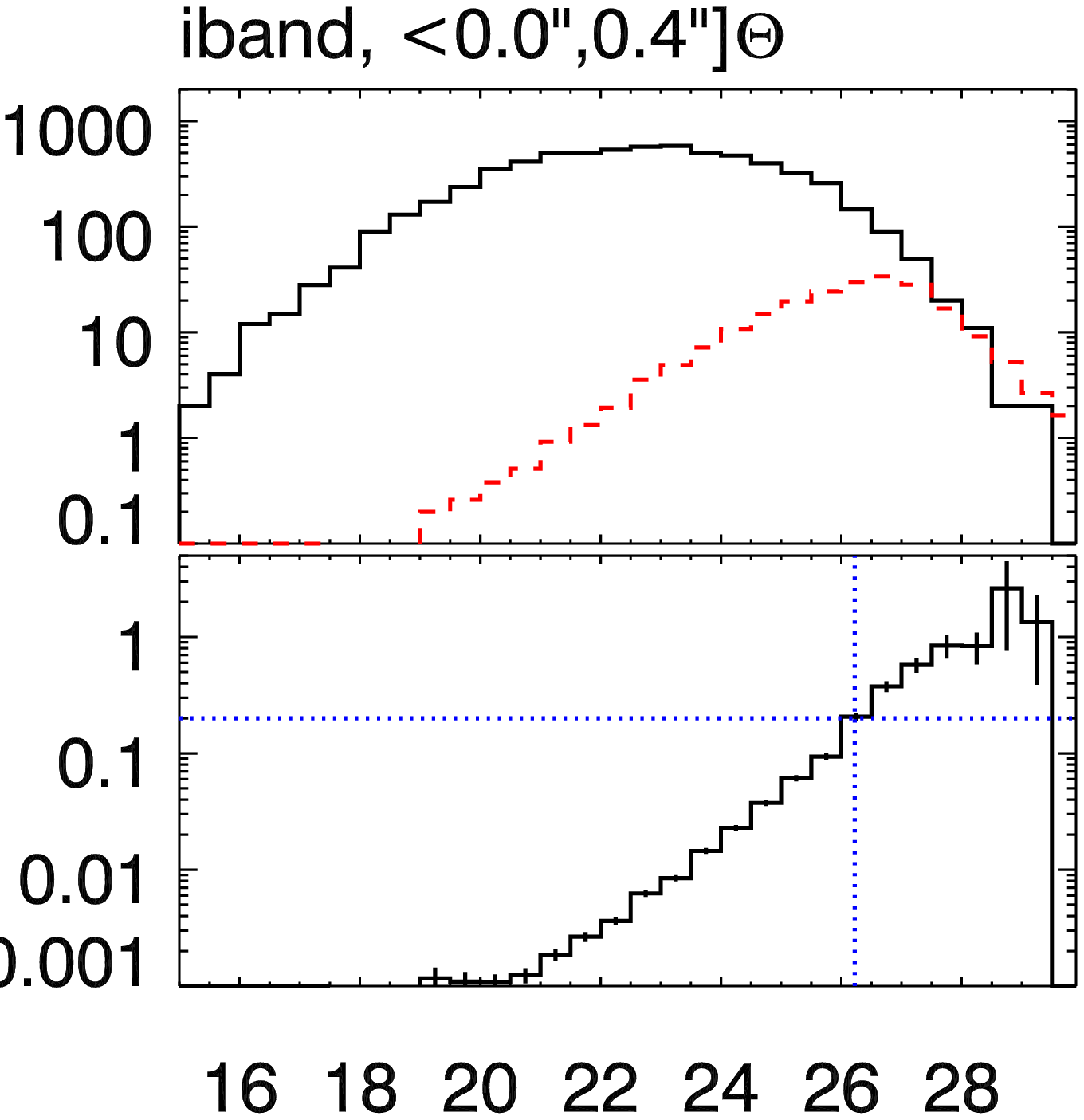}	
	\includegraphics[bb =  108 -60 432 432, scale=0.23]{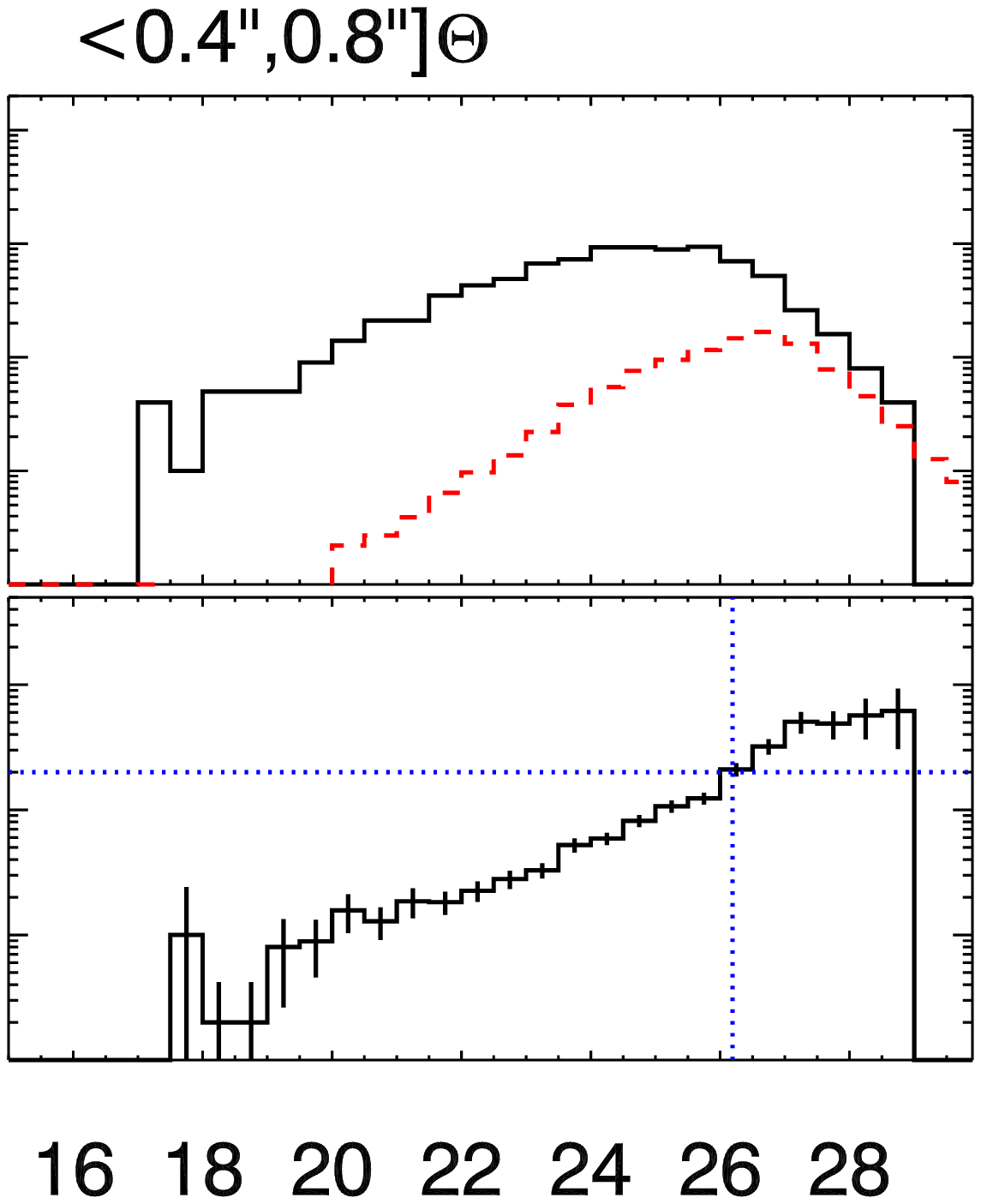}	\\
	\includegraphics[bb =  0 -60 432 432, scale=0.23]{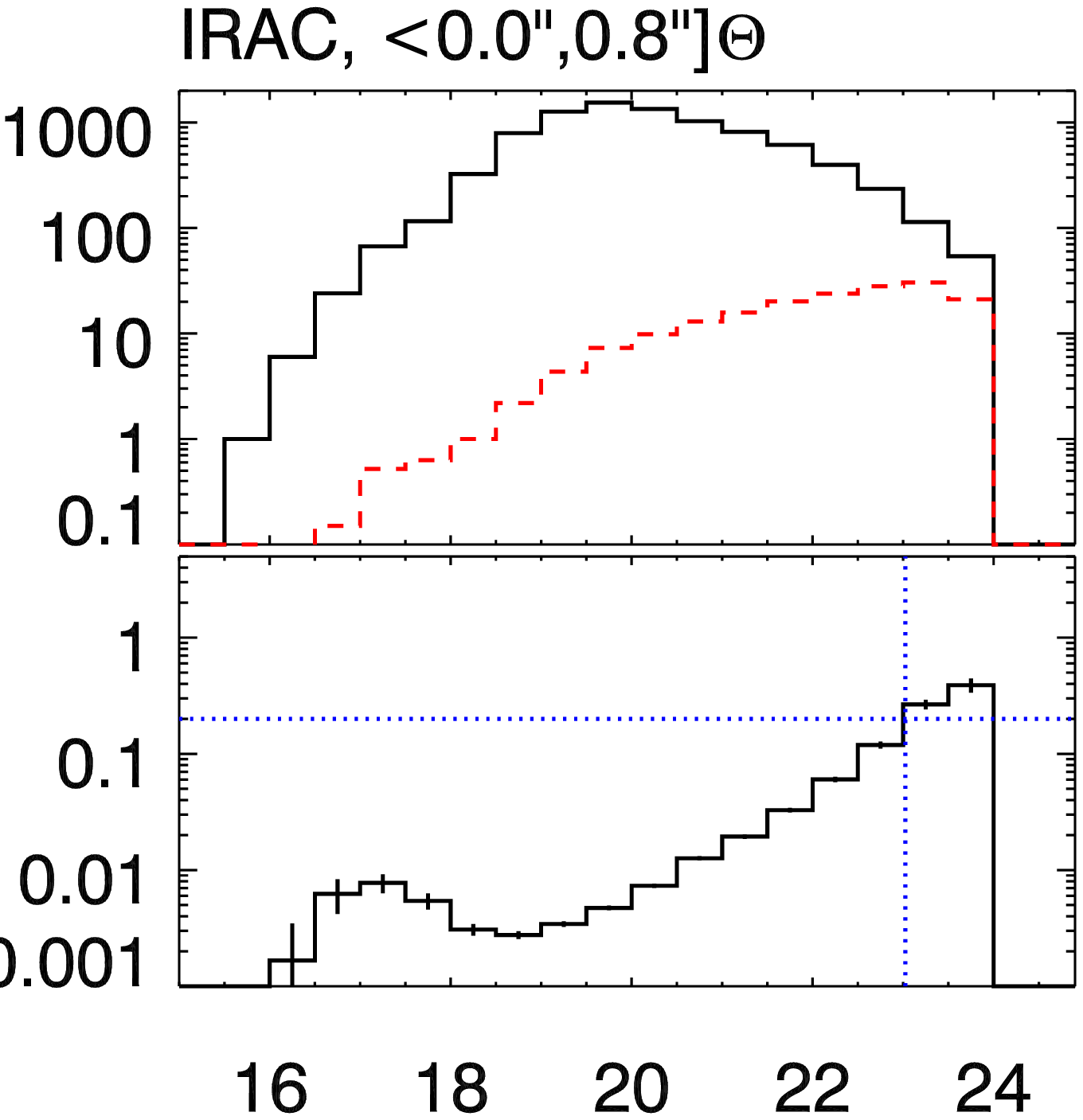}	
	\includegraphics[bb =  108 -60 432 432, scale=0.23]{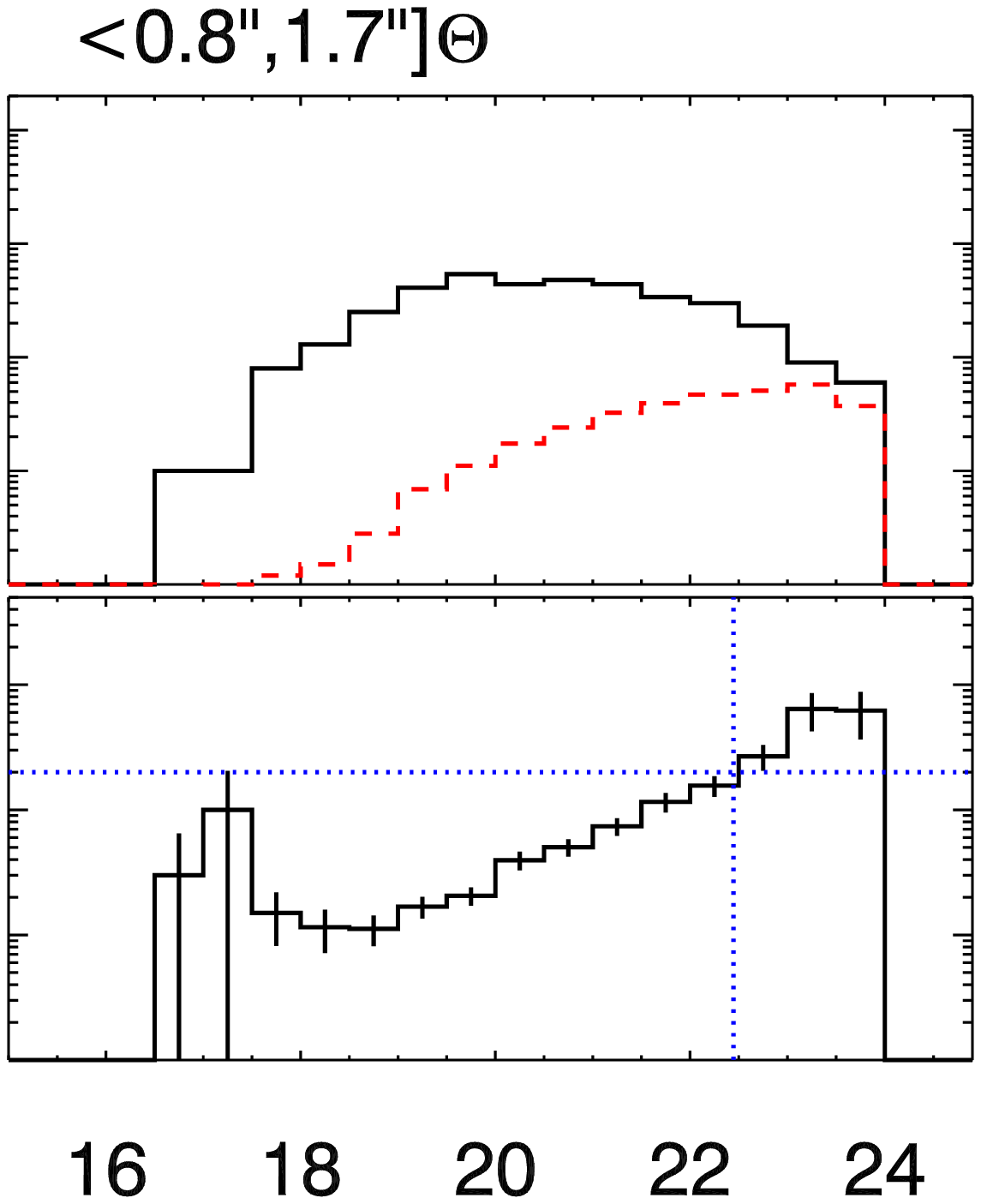}	\\
	\caption{\textit{Top panels in each row:} Magnitude distribution of the real VLA-COSMOS 3~GHz counterparts (solid black line), and the simulated background (red dashed line) for the COSMOS2015 (top row), i-band (middle row), and IRAC (bottom row) catalogs. The range of allowed separations between the radio sources and their counterparts is indicated above each panel (see text for details).
	\textit{Bottom panels in each row}: The ratio between  the background and the real magnitude distributions, defined as the false match probability ($p_\mathrm{false}$) as a function of magnitude.
	The dotted horizontal  line shows the $p_\mathrm{false}$ threshold of 0.2 beyond which counterparts are discarded, while the dotted vertical  line shows the effective magnitude cut corresponding to $p_\mathrm{false}= 0.2$. Poisson errors are also indicated.
	}
	\label{fig:falseMatch}
\end{figure}

\begin{figure} 
\centering
	\includegraphics[bb = -20 20 432 410, width=1.0\linewidth]{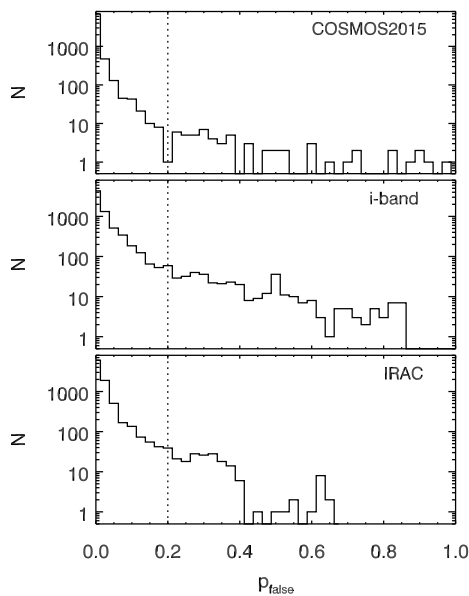}\\
	\caption{Distribution of the false match probability ($p_\mathrm{false}$) for the VLA 3~GHz sources matched to COSMOS2015 selected  (\textit{top}), $i$-band selected sources (\textit{middle}), and IRAC, 3.6~$\mu$m selected sources (\textit{bottom}).  The vertical dotted line shows the $p_\mathrm{false}$ threshold of 20\% used in the counterpart association.}
	\label{fig:fmp}
\end{figure}

Masked regions of the COSMOS2015 catalog illustrated in Fig.~\ref{fig:counterpartsRADEC}, reduce the effective area to $1.77$~deg$^2$. A total of 8,696 ($\sim$80\%) 3~GHz sources fall within this area. COSMOS2015 sources outside the masked regions were positionally matched to the sources in the VLA-COSMOS 3~GHz Large Project catalog, using a search radius of  0\farcs8. Including multiple possible counterpart candidates for a single 3~GHz source, this yielded a total of  7,721 COSMOS2015 counterpart candidates, and  7,701 single radio-NIR source associations. Only  6 counterpart candidates did not have the 3.6~$\mu m$ magnitude listed in the catalog and their false match probabilities were manually set to 1. Magnitude distributions and $p_\mathrm{false}$ (i.e.  ratio between the number of 
 associations with the simulated mock catalogs and the number of associations with the real catalog) vs. magnitude plots are shown in the top panels of Fig.~\ref{fig:falseMatch}. The $p_\mathrm{false}$ distribution of the matched sample is shown in the top panel of Fig~\ref{fig:fmp}.

\subsection{Cross-matching with the $i$-band selected sources}
\label{sec:crossmatching_iband}
From the $i$-band selected catalog \citep{capak07} of 2,017,800 sources, we have excluded those masked (i.e., if saturated or around bright sources) in the $B$, $V$, $i$, and $z$-bands, removing 26\% of the sources, and leaving 1,484,453 reliable detections.
A total of 8,696 ($\sim$80\%) 3~GHz sources fall within the unmasked regions  with an effective area of 1.73 square degrees. Within a search radius of  0\farcs8, we find a total of  7,397 $i$-band selected counterparts of 3~GHz sources (including multiple possible candidates  for the same radio source, and  7,321 for a one-to-one match).
For only  16 sources $i$-band magnitudes were not available, and their false match probabilities were set to 1.
Magnitude distributions and false match probabilities are shown in the middle panels of Fig.~\ref{fig:falseMatch}. The false match probability distribution of the matched sample is shown in the middle panel of Fig~\ref{fig:fmp}. 

\subsection{Cross-matching with the $IRAC$ selected sources}
\label{sec:crossmatching_irac}
From the IRAC 3.6~$\mu$m selected catalog, we have excluded sources masked in any of the four IRAC channels.  This reduced the number of sources from 345,512 to 278,897.
Within a search radius of  1\farcs7, we find  9,158 3.6~$\mu$m sources in total,  all unique IRAC-radio source matches.
All the counterpart candidates have a 3.6~$\mu m$ magnitude. Magnitude distributions and false match probabilities are shown in the bottom panels of Fig.~\ref{fig:falseMatch}.   The false match probability distribution of the matched sample is shown in the bottom panel of Fig~\ref{fig:fmp}.

\subsubsection{Final catalog of optical-MIR counterparts}
\label{sec:finalCat}

\begin{figure} 
\centering
\includegraphics[width=1\linewidth]{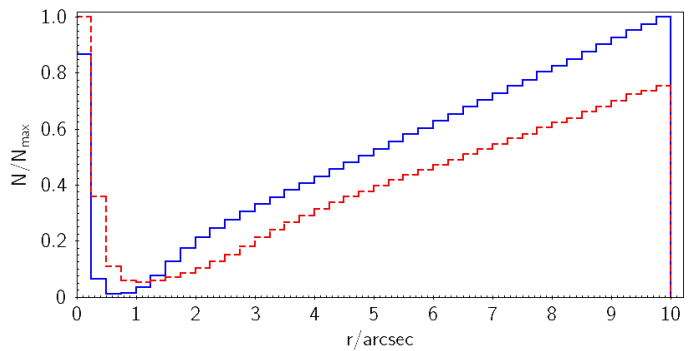}
\caption{Normalized number of $i$-band (IRAC) sources at a given distance from COSMOS2015 sources, shown as a blue solid (red dashed) histogram.}
\label{fig:optCounterparts}
\end{figure}

To find the best possible multiwavelength counterpart of VLA-COSMOS 3~GHz Large Project sources, we compared the three (COSMOS2015, $i$-band, and IRAC) counterpart candidate catalogs described in previous sections. Over 7,000 single-component radio sources have a counterpart candidate in all three catalogs. To verify whether the counterpart candidates found in the three catalogs represent the same physical source we proceed in the following way. 
To find COSMOS2015 counterparts to $i$-band and IRAC sources, and $i$-band counterparts to IRAC sources we perform positional matching between these catalogs. To determine the search radius for the association of sources appearing in two different catalogs we count the number of sources as a function of separation, as shown in Fig.~\ref{fig:optCounterparts}. As the maximum search radius, we use the radius where this number reaches the minimum: above this radius the associations are consistent with being spurious and their number, linearly increasing with separation, is simply due to the surface density of sources in the two analyzed catalogs. This radius is $\sim$0\farcs7 for the match between the COSMOS2015 and $i$-band catalogs, and $1\arcsec$ for the match between the COSMOS2015 and IRAC catalogs. Thus, after selecting a reliable (i.e., one with $p_\mathrm{false}$ less than 20\%) counterpart in the $i$-band and IRAC catalogs, we shift the selection to the COSMOS2015 counterpart if it is within 0\farcs7  
from the $i$-band or $1\arcsec$ from the IRAC counterpart  (but still within a separation of $0\farcs8$ between the radio and COSMOS2015/$i$-band counterpart). Similarly, we shift the selection to the $i$-band counterpart if it is within $1\arcsec$  from a reliable IRAC counterpart. Since it is unlikely that a source detected in two different surveys is spurious, in these cases we associate the COSMOS2015 or $i$-band counterpart regardless of its false match probability. The total number of cases in which COSMOS2015 or $i$-band sources with false match probability $p_\mathrm{false}\textgreater$20\% were selected  was $1.1\%$.

\section{Full 3~GHz radio source counterpart catalog}
\label{sec:fullcat}

The full 3~GHz radio source counterpart catalog contains all 3 GHz sources with counterparts assigned either in the COSMOS2015, $i$-band, or IRAC catalogs, as described in detail in Appendix~\ref{sec:method}, and summarized in Secs.~\ref{sec:cptSingle}, \ref{sec:cptMulti}, and \ref{sec:catalog}. 
The final counterparts assigned to both our multi- and single-component 3~GHz radio sources are summarized in Tab.~\ref{tab:cptsappendix}. 

Overall, 
7,742, 407, and 1,012 
radio  (3~GHz) sources are associated with COSMOS2015,  $i$-band, and IRAC counterparts, respectively.\footnote{ Note that the association is constructed in such a way that the COSMOS2015 sources listed can have IRAC and/or $i$-band counterparts, while the $i$-band sources listed can have IRAC counterparts, but do not have unmasked COSMOS2015 counterparts, and the IRAC sources listed do not have unmasked COSMO2015 nor $i$-band counterparts.} In total, we find 9,161 counterparts for our radio sources. Summing the computed false match probabilities we estimate a total fraction of spurious matches of $\sim2\%$. Since our identification procedure is based on ground-based catalogs, it is likely that this "formal" fraction should be considered as a lower limit because of the limited spatial resolution of these data used in these catalogs.

\begin{figure} 
\centering
\includegraphics[bb= 0 0 432 380, width=1\linewidth]{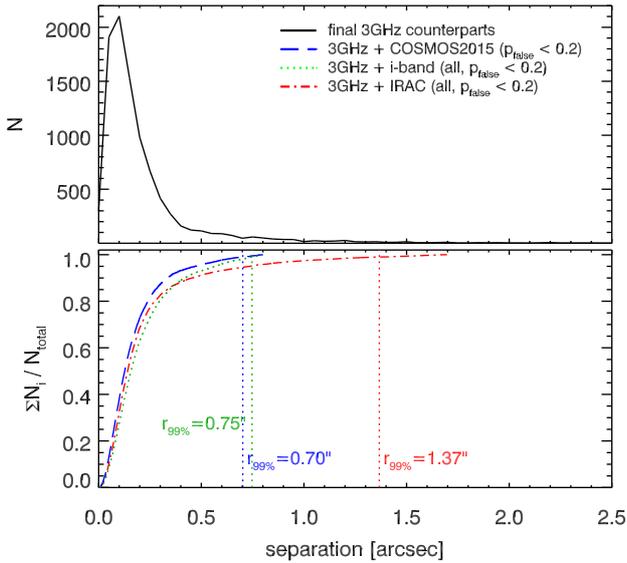}
\caption{Differential (top panel) distribution of the separations of the final counterparts from the (single-component) radio sources,  drawn from the COSMOS2015, $i$-band, or IRAC catalogs (as described in the text).  The bottom panel shows the cumulative distributions of the separations of all counterpart candidates (with $p_\mathrm{false}\leq0.2$), drawn from the COSMOS2015 (blue dashed curve), $i$-band (green dotted curve), and IRAC (red dash-dotted curve) catalogs. Also indicated is the separation  encompassing 99\% of the counterparts for each of the counterpart sample.}
\label{fig:separations}
\end{figure}

In the top panel of Fig.~\ref{fig:separations} we show  the differential  distribution of the separations of final counterparts from the radio sources. The differential distribution is highly peaked at small separations. As evident in the cumulative distribution, separated into all counterpart candidates with $p_\mathrm{false} \leq 0.2$ from the various catalogs, shown in the bottom panel, 99\% of the associations are found within a separation of $\sim0.7\arcsec$ ($1.4\arcsec$) between the 3~GHz and the COSMOS/i-band (IRAC) matches. This further affirms our choice of the limiting matching distance of $0\farcs8$ ($1\farcs7$) for COSMOS2015/$i$-band (IRAC). Extending the chosen limiting search radii would not have significantly increased the counterpart sample,  at the expense of the possible inclusion of a non-negligible fraction of spurious associations.

\begin{figure} 
\centering
\includegraphics[bb = 0 35 865 866, width=\columnwidth]{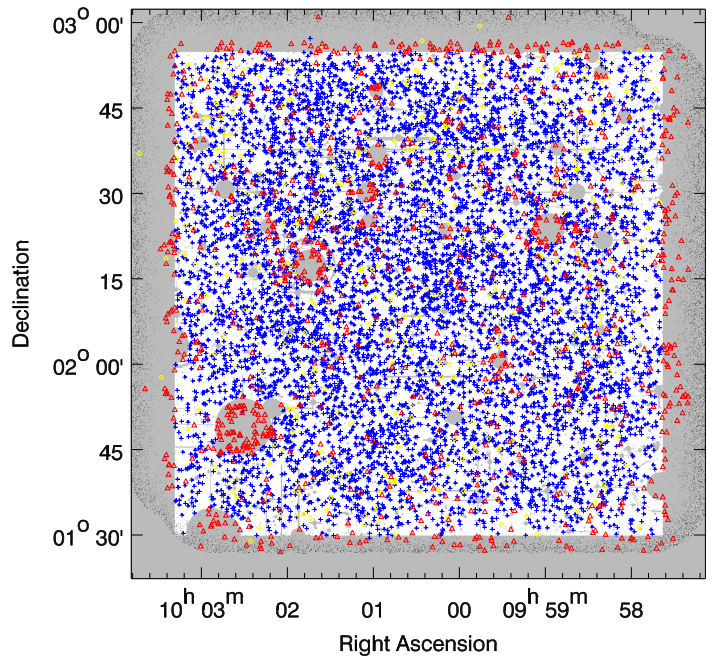}
\caption{
 Positions of counterparts from COSMOS2015 (blue crosses), $i$-band (yellow circles), and IRAC (red triangles) catalogs overlaid on the VLA-COSMOS 3~GHz Large Project mosaic with grayed-out regions masked in COSMOS2015 catalog due to the presence of saturated or bright sources in the optical to NIR bands (see also Fig~\ref{fig:counterpartsRADEC}).
 }
\label{fig:counterpartsRADECappendix}
\end{figure}

In Fig.~\ref{fig:counterpartsRADECappendix} we show the positions of the identified counterparts overlaid on the COSMOS field.
We stress  the difference in effective, unmasked regions of the maps where the three counterpart catalogs were extracted from ($z^{++}JYHK$-stack, $i$-band, $3.6~\mu$m-band, respectively): 1.77 square degrees for COSMOS2015,  1.73 square degree area for the $i$-band, and 2.35 square degree area for the IRAC catalogs.   For this reason the bulk of the IRAC (79\%) and $i$-band (76\%) assigned counterparts, resides either in the outer regions of the field, not covered by the COSMOS2015 catalog or in COSMOS2015 masked regions.

 \begin{table}
 \begin{centering}
\begin{tabular}{lccc}
\hline 
Final  & $A_\mathrm{eff}$ & Total	\\
 counterpart  catalog   & [deg$^2$] \\ 
\hline
COSMOS2015 & 1.77 & 7,742 \\
$i$-band & 1.73 & 407 \\
IRAC & 2.35 & 1,012 \\
\hline
Combined	& 2.35 & 9,161	\\
\hline\\
\end{tabular}
\caption{
Summary of the cross-correlation of the VLA-COSMOS 3~GHz Large Project sources with sources in the multi-wavelength catalogs.  In the last  line we list the combined number of COSMOS2015, $i$-band, and IRAC sources in the full 2.35 square degree area considered here (unmasked in the IRAC catalog). }
\label{tab:cptsappendix}
 \end{centering}
\end{table}

\end{document}